\documentclass[12pt]{article}
\usepackage[marginratio=1:1,height=600pt,width=460pt,tmargin=100pt]{geometry}
\usepackage{authblk}
\usepackage{microtype}
\usepackage{graphicx}
\usepackage{subcaption}
\usepackage{booktabs} 
\usepackage{makecell}
\usepackage{wrapfig}
\usepackage{diagbox}
\usepackage{enumitem}%
\usepackage{amsfonts}
\usepackage{xcolor}

\usepackage{amsthm}
\usepackage{cprotect}
\usepackage{hyperref}
\usepackage{amsmath,amssymb}
\usepackage{multirow}
\usepackage{fancyhdr}
\renewcommand{\headrulewidth}{0pt}
\fancypagestyle{firststyle}
{
   \fancyhf{}
   \fancyfoot[R]{\thepage}
   \renewcommand{\headrulewidth}{0pt} 
}
\usepackage{fancyvrb}
\usepackage{natbib}
\usepackage{algorithm}
\usepackage{algorithmic}
\newcommand{\rev}[1]{{\color{black}#1}}

\newcommand{\magenta}[1]{{\color{black}#1}}

\newcommand{\EnbPI}{\Verb|EnbPI|}

\newcommand{\bigbrac}[1]{\left(#1\right)}
\newcommand{\longbar}{\biggr\rvert}
\newcommand{\betaOracle}{\beta^*}
\newcommand{\betaReal}{\hat{\beta}}
\newcommand{\betaRealBin}{\hat{\beta}_{\rm{line}}}

\newcommand{\R}{\mathbb{R}}
\newtheorem{theorem}{Theorem}
\newtheorem{lemma}{Lemma}
\newtheorem{corollary}{Corollary}

\newtheorem{remark}{Remark}
\newtheorem{assumption}{Assumption}
\usepackage{mathtools}

\DeclarePairedDelimiter\floor{\lfloor}{\rfloor}

\newcolumntype{P}[1]{>{\centering\arraybackslash}p{#1}}
\begin{document}

\title{Conformal prediction for time series
}
\date{}

\author[1]{Chen Xu\footnote{cxu310@gatech.edu}}
\author[1]{Yao Xie\footnote{yao.xie@isye.gatech.edu}}
\affil[1]{{\small H. Milton Stewart School of Industrial and Systems Engineering, Georgia Institute of Technology.}}

\maketitle
\thispagestyle{firststyle}
\vspace{-0.3in}
\begin{abstract}
We develop a general framework for constructing distribution-free prediction intervals for time series. \magenta{Theoretically, we establish explicit bounds on conditional and marginal coverage gaps of estimated prediction intervals, which asymptotically converge to zero under additional assumptions. We obtain similar bounds on the size of set differences between oracle and estimated prediction intervals.} 
Methodologically, we introduce a computationally efficient algorithm called \verb|EnbPI| that wraps around ensemble predictors, which is closely related to conformal prediction (CP) but does not require data exchangeability. \verb|EnbPI| avoids data-splitting and is computationally efficient by avoiding retraining and thus scalable to sequentially producing prediction intervals. We perform extensive simulation and real-data analyses to demonstrate its effectiveness compared with existing methods. We also discuss the extension of \EnbPI \ on various other applications.
\end{abstract}

\section{Introduction}\label{sec:introduction}

Modern applications, including energy and supply chains \citep{diaz2012review,renew_data_challenge}, require sequential prediction with uncertainty quantification for time-series observations with highly complex dependency. In addition to point prediction, it is typical to construct prediction intervals for uncertainty quantification, a fundamental task in statistics and machine learning. 

Constructing accurate prediction intervals for time series is highly challenging yet crucial in many high-stakes applications. In power systems, as outlined in the National Renewable Energy Lab report \citep{renew_data_challenge}, solar and wind power generation data are non-stationary, exhibit significant stochastic variations, and have spatial-temporal correlations among regions. \rev{The inherent randomness of renewable energy sources presents significant challenges for prediction and inference. To overcome these challenges, it is essential to use historical data to accurately predict energy levels from wind farms and solar roof panels and establish prediction intervals. These prediction intervals provide critical information for power network operators, enabling them to understand the uncertainty of the power generation and make necessary arrangements. Incorporating renewable energy into existing power systems requires the prediction of power generation with uncertainty quantification \citep{Gangammanavar2016StochasticOO,7452435}.}
\rev{Although there are various neural-network based quantile prediction models \citep{salinas2020deepar,wen2017multi}, the resulting prediction intervals frequently lack theoretical guarantees, causing concern about their reliability in high-stakes situations. Currently, there is a need for a distribution-free framework that produces prediction intervals for time-series data, along with provable guarantees for interval coverage, which remains an open question in the field.}

\rev{In addition to the difficulties posed by the inherent stochasticity of time-series, constructing prediction intervals for user-specified predictive models also presents further challenges.}
For example, complex prediction models such as random forest \citep{RF} and deep neural networks \citep{DeepReg} are often employed for accurate predictions. \rev{Unlike classical linear regression models, these prediction algorithms do not have straightforward methods for calculating prediction intervals. To construct prediction intervals for such models, practitioners often resort to heuristics like bootstrapping, which lack guarantees.} In practice, ensemble methods \citep{breiman-bagging} are also frequently used to enhance prediction performance by combining multiple prediction algorithms, further complicating the model. Despite this, constructing efficient prediction intervals for time-series data using general prediction methods, which can be arbitrarily complex, remains an under-explored area.

\subsection{Contributions}

In this paper, we develop distribution-free prediction intervals for time series data with a coverage guarantee, inspired by recent works on conformal prediction. Our proposed method, \EnbPI, can provide prediction intervals for ensemble algorithms. The main contributions of this paper are summarized as follows.
\begin{itemize}
\setlength\itemsep{0em}
\item We present a general framework for constructing prediction intervals for time series, which can be asymmetrical. We theoretically upper-bound the conditional and marginal coverage gaps, which converge to zero under mild assumptions on the dependency of stochastic errors and the quality of estimation. We also obtain similar bounds on the size of the set difference between the {\it oracle} and estimated prediction intervals.
\item We develop \EnbPI, a robust and computationally efficient algorithm for constructing prediction intervals around ensemble estimators. The algorithm is designed to avoid expensive model retraining during prediction and requires no data splitting, thanks to a carefully constructed bootstrap procedure. \EnbPI \  is particularly suitable for small-sample problems, and its versatility makes it applicable in various practical settings, such as network prediction and anomaly detection.
\item  We present extensive numerical experiments to study the performance of \EnbPI \ on simulated and real-time series data. The results show that \EnbPI \ can maintain a target coverage when other competing methods fail to do so, and it can yield shorter intervals. Additionally, the experiments demonstrate that \EnbPI \ is robust to missing data. 
\end{itemize}

The rest of this paper is organized as follows. Section \ref{sec:prob+method} describes the problem setup and introduces the oracle prediction interval, which motivates our proposed method. Section \ref{sec:theory} presents asymptotic guarantees for the interval coverage and width and highlights the generality of such guarantees. Section \ref{sec:algo} presents \EnbPI. Section \ref{sec:exper} contains numerical examples with simulated and real data that compare \EnbPI \ with competing methods to demonstrate its good performance in various scenarios. Section \ref{sec:simul_changepts} extends the use of \EnbPI \ when a change point exist. Section \ref{sec:conclu} concludes the paper with discussions. Appendix \ref{app_theory:approx_uni_proof} contains proofs and \ref{app:more_exper} contains experiments. 
Code for this paper can be found at \href{https://github.com/hamrel-cxu/EnbPI}{\Verb|https://github.com/hamrel-cxu/EnbPI|}. 

\subsection{Literature review}

Conformal Prediction (CP) is a popular method for constructing distribution-free prediction intervals. It was formally introduced in \citep{conformaltutorial}, and it assigns "conformity scores" to both training and test data. By inverting hypothesis tests using these scores, prediction intervals can be obtained for the test data. It has been shown that under the assumption of exchangeability in data, this procedure generates valid marginal coverage for the test point.  \rev{Many CP methods have been developed to quantify uncertainty in predictive models. To efficiently compute the conformity scores, a data-splitting method is developed in \citep{inductCP}, which computes the scores on a hold-out set of the training data. \citep{CPquantile} builds on this data-splitting idea for quantile regression models. To avoid data splitting which affects the accuracy of trained predictive model, ``leave-one-out`` (LOO) CP methods are developed to use the entire training samples for computing prediction residuals, a particular choice of conformity scores \citep{jackknife+}. Subsequent works develop more computationally efficient way of training LOO models \citep{j+ab} and generalize the approach to other conformity scores\citep{QOOB}.} Comprehensive surveys and tutorials can be found in \citep{conformaltutorial,conformalreview}. Although no assumption other than data exchangeability is required for marginally exact coverage, the exchangeability assumption is hardly reasonable for time series, \rev{making works above not directly applicable to our setting.}

Adapting CP methods beyond exchangeable data has also been gaining significant interest. \rev{A widely popular type of approach assumes unknown distribution shifts in the test data and weighs the past conformity scores to restore valid coverage. For instance,} the work by \citep{CPcovshift} uses weighted conformal prediction intervals when the test data distribution is proportional to the training distribution. The work by \citep{CProbust} \rev{builds on this idea} when the shifted test distribution lies in an $f$-divergence ball around the training distribution. However, both works still assume {\it i.i.d.} or exchangeable training data, making them not directly applicable for time series. A concurrent work \citep{RinaNonExchange} considers a general set-up for \rev{bounding coverage gap using total variation distances. It then proposes to use fixed weights to correct for the coverage gap.} In retrospect, we consider a more specific setting involving time series, and the upper bounds are captured differently and explicitly using the quality of the estimator and the noise characteristics. \rev{Meanwhile, a recent work for non-exchangeable data sequentially adjusts the significance level $\alpha$ during prediction. For instance,} \citep{Gibbs2021AdaptiveCI} provides approximately valid coverage on sequential data by re-weighting the value $\alpha$ based on online coverage values on test data. \rev{The subsequent work \citep{Zaffran2022AdaptiveCP} proposes more sophisticated re-weighting techniques of $\alpha$.}
However, whether such adjustments are applicable to data with general dependency remain unclear, and \rev{we compare with \citep{Gibbs2021AdaptiveCI} in experiments to show the improved performance of \EnbPI{}.}

Meanwhile, there are many non-CP prediction interval methods. In the traditional time series literature \citep{tsmethod}, there have been abundant work for prediction interval construction, such as ARIMA$(p,d,q)$ \citep{durbin2012time}, exponential smoothing \citep{hyndman2008forecasting}, dynamic factor models \citep{banbura2014maximum} and so on. However, they rely on strong parametric distributional assumptions that limit their applicability.
\rev{On the other hand, recent works have notably leveraged the predictive power of deep neural networks for neural quantile regression. Two of the most popular approaches are MQ-CNN \citep{wen2017multi} and DeepAR \citep{salinas2020deepar}; additional approaches can be found in \citep{makridakis2022m5}. More precisely, MQ-CNN \citep{wen2017multi} leverages the power of sequence-to-sequence neural networks to predict the multi-horizon quantile value of future response variables directly. The framework can also incorporate various temporal and static features and remains scalable to large-scale forecasting. 
Meanwhile, DeepAR \citep{salinas2020deepar} models the conditional distribution of future response using an autoregressive recurrent network. The network is trained by maximizing the log-likelihood of data, assuming Gaussian likelihood for real-valued data and negative-binomial for positive count data. Extensive experiments show its improvement over state-of-the-art methods. 
Although both MQ-CNN and DeepAR have promising performances for a variety of time-series data, they have limitations in requiring special network architecture (not model-free) and providing no theoretical guarantees on coverage. In addition, \citep{salinas2020deepar} imposes distributional assumptions on data through the parametric likelihood models (not distribution-free). In contrast, \EnbPI{} leverages the benefits of conformal prediction to present a general framework for an arbitrary point-prediction model (model-free), with provable guarantees on coverage and without distributional assumption on data (distribution-free).}

Finally, we remark that our assumptions and proof techniques avoid data exchangeability and differ significantly from existing CP works. Most CP methods ensure the finite-sample marginal coverage and distribution-free conditional coverage is impossible at a finite sample size \citep{cond-pi-rina}. In contrast, we achieve an asymptotic conditional coverage guarantee. Such theoretical analyses are inspired by \citep{depend-data,depend-data-new}, yet we refine the proof techniques to improve the convergence rates and extend results under different assumptions. We further analyze the convergence of prediction interval widths. 
We would also like to remark that our work is titled ``conformal prediction'' because \rev{\EnbPI{} builds on the conformal prediction framework in this more general context---in terms of construction, \EnbPI{} intervals closely resemble intervals by existing CP methods (especially J+aB \citep{j+ab}). Meanwhile, the theoretical results in this work can hold for prediction intervals produced by other conformal prediction methods, such as split conformal \citep{inductCP}, J+aB \citep{j+ab}, and so on (see Remark \ref{remark:caveats}). Thus, the theoretical tools presented in this work are general for analyzing CP methods for time series.}

\section{Problem setup } \label{sec:prob+method}

Given an unknown model $f:\mathbb{R}^d\rightarrow \mathbb{R}$, where $d$ is the dimension of the feature vector, we observe data $(x_t, y_t)$ generated according to the following model 
\begin{equation}\label{eq:DGP_model}
    Y_t=f(X_t)+\epsilon_t, \quad t = 1, 2, \ldots
\end{equation}
where $\epsilon_t$ is distributed following a continuous cumulative distribution function (CDF) $F_t$. Note that we do not need  $\epsilon_t$ to be independent and $F_t$ needs not be the same across all $t$. Features $X_t$ can contain exogenous time series sequences that predict $Y_t$ and/or the history of $Y_t$. We assume that the first $T$ samples $\{(x_t,y_t)\}_{t=1}^T$ are training data or initial state of the random process that are observable.  Above, upper case $X_t$, $Y_t$ denote random variables and lower case $x_t,y_t$ denote data.

Our goal is to construct a sequence of prediction intervals as narrow as possible with a certain coverage guarantee. Given a user-specified prediction algorithm, using $T$ training samples, we obtain a trained model represented by $\hat f$. Then we construct $s \geq 1$ prediction intervals $\{\widehat{C}_{T+i}^\alpha\}_{i=1}^s$ for $\{Y_{T+i}\}_{i=1}^s$, where  $\alpha$ is the \textit{significance level}, and the {\it batch size} $s$ is a pre-specified parameter for how many steps we want to look ahead. Once new samples $\{(x_{T+i},y_{T+i})\}_{i=1}^s$ become available, we deploy the pre-trained $\hat f$ on new samples and use the most recent $T$ samples to produce prediction intervals for $Y_{j}, j=T+s+1$ onward without re-training the model on new data. 

The meaning of significance level $\alpha$ is as follows. We consider two types of coverage guarantees. The \textit{conditional} coverage guarantee ensures that each prediction interval $\widehat{C}^{\alpha}_{t}, t>T$  satisfies:
\begin{equation}\label{eq:cond_cov}
    P(Y_t \in \widehat{C}^{\alpha}_{t}|X_t=x_t) \geq 1-\alpha.
\end{equation} 
The second type is the  \textit{marginal} coverage guarantee:
\begin{equation}\label{eq:ave_cov}
    P(Y_t \in \widehat{C}^{\alpha}_{t}) \geq 1-\alpha.
\end{equation} 
Note that (\ref{eq:cond_cov}) is much stronger than (\ref{eq:ave_cov}), which is satisfied whenever data are exchangeable using split conformal prediction \citep{inductCP}. For instance, suppose a doctor reports a prediction interval for one patient's blood pressure. An interval satisfying (\ref{eq:ave_cov}) averages over all patients in different age groups, but may not satisfy (\ref{eq:cond_cov}) for the current patient precisely. In fact, satisfying (\ref{eq:cond_cov}), even for exchangeable data, is impossible without further assumptions \citep{cond-pi-rina}. 
In general, it is challenging to ensure either  (\ref{eq:cond_cov}) or (\ref{eq:ave_cov}) under complex data dependency without distributional assumptions. Despite such difficulty, our theory provides a way to bound the worst-case gap in conditional coverage (\ref{eq:cond_cov}) and marginal coverage \eqref{eq:ave_cov}, under certain assumptions on the error process $\{\epsilon_t\}_{t\geq 1}$ and $\hat f$.  
From now on, we call a prediction interval conditionally or marginally \textit{valid} if it achieves (\ref{eq:cond_cov}) or (\ref{eq:ave_cov}), respectively.

\subsection{Oracle prediction interval}\label{sec:oracle_bound}

To motivate the construction of $\widehat{C}^{\alpha}_{t}$, we first consider the {\it oracle} prediction interval $C^{\alpha}_t$, which contains $Y_t$ with an exact conditional coverage at $1-\alpha$ and is the shortest among all possible conditionally valid prediction intervals. The oracle prediction assumes perfect knowledge of $f$ and $F_t$ in (\ref{eq:DGP_model}). Denote $F_{t,Y}$ as the CDF of $Y_t$ conditioning on $X_t=x_t$, then we have
\begin{align*}
    F_{t,Y}(y)
    =&\mathbb P(Y_t \leq y |X_t=x_t)\\
    =&\mathbb P(\epsilon_t \leq y-f(x_t))
    =F_t(y-f(x_t)).
\end{align*}
For any $\beta \in [0,\alpha]$, we also know that 
\[
  \mathbb P(Y_t \in [F^{-1}_{t,Y}(\beta),F^{-1}_{t,Y}(1-\alpha+\beta)]|X_t=x_t)=1-\alpha,
\]
where $F^{-1}_{t,Y}(\beta):=\inf \{y: F_{t,Y}(y)\geq \beta\}.$ Assume $F^{-1}_{t,Y}(\alpha)$ is attained for each $\alpha \in [0,1]$, and let $y_{\beta}=F^{-1}_{t,Y}(\beta)$. Clearly,
\begin{equation*}
    y_{\beta}=f(x_t)+F_t^{-1}(\beta),
\end{equation*}
which allows us to find $C^{\alpha}_t$ -- the oracle prediction interval with the narrowest width:
\begin{align}
    & C^{\alpha}_t=[f(x_t)+F_t^{-1}(\betaOracle),f(x_t)+F_t^{-1}(1-\alpha+\betaOracle)], \label{eq:oracle_interval}\\
    & \betaOracle:=\underset{\beta \in [0,\alpha]}{\arg\min} \ \bigbrac{F_t^{-1}(1-\alpha+\beta)-F_t^{-1}(\beta)}.\nonumber
\end{align}
A similar oracle construction to \eqref{eq:oracle_interval} appeared in \citep{CPhist}. Thus, if we can approximate unknown  $f(x_t)$, $F_t^{-1}(x)$, $x\in [0,1]$, and $\betaOracle$ reasonably well, the prediction intervals $\widehat{C}^{\alpha}_{t}$ should be close to the oracle $C^{\alpha}_t$.

\subsection{Proposed prediction interval}
We now construct $\widehat{C}^{\alpha}_{t}$ based on ideas above. Recall that the first $T$ data $\{(x_t,y_t)\}_{t=1}^T$ are observable. Denote $\hat{f}_{-i}$ as the $i$-th ``leave-one-out'' (LOO) estimator of  $f$, which is not trained on the $i$-th datum $(x_i,y_i)$ and may include the remaining $T-1$ points. Then,
\begin{align}
 \widehat{C}^{\alpha}_{t}:=[
 &\hat{f}_{-t}(x_t) + \betaReal \text{ quantile of }  \{\hat{\epsilon}_i\}_{i=t-1}^{t-T},\label{CI_form}\\
 &\hat{f}_{-t}(x_t) + (1-\alpha+\betaReal) \text{ quantile of }  \{\hat{\epsilon}_i\}_{i=t-1}^{t-T}],\nonumber
\end{align} 
where the LOO prediction residual $\hat{\epsilon}_i$ and the corresponding $\betaReal$ are defined as
\begin{align*}
     \hat{\epsilon}_i:=&y_i-\hat{f}_{-i}(x_i)\\
     \betaReal:=&
    \underset{\beta \in [0,\alpha]}{\arg\min} \  ((1-\alpha+\beta) \text{ quantile of }  \{\hat{\epsilon}_i\}_{i=t-1}^{t-T} - \\
    & \beta \text{ quantile of }  \{\hat{\epsilon}_i\}_{i=t-1}^{t-T}).
\end{align*}
Thus, the interval centers at the point prediction $\hat{f}_{-t}(x_t)$ and the width is the difference between the $(1-\alpha+\betaReal)$ and $\betaReal$ quantiles over the past $T$ residuals. 

Note that we have to split the training data into two parts: one part is used to estimate $f$, and the second part is used to obtain prediction residuals for the prediction interval. There is a trade-off. On the one hand, we desire the estimator $\hat f$ to be trained on as much data as possible. On the other hand, the quantile of prediction residuals should well approximate the tails of $F_t^{-1}$. These two objectives contradict each other. If we train $\hat{f}$ on all training data, then we overfit; if we train on a subset of training data and obtain prediction residuals on the rest \citep{inductCP}, the approximation $\hat{f}$ to $f$ is poorer. The LOO estimator is known to achieve a good trade-off in this regard. \rev{When obtaining the $i$-th residual, the $i$-th LOO estimator trains on all except the $i$-th training datum so that the LOO estimator is not overfitted on that datum. Then repeating over $T$ training data yields $T$ LOO estimators with good predictive power and $T$ residuals to calibrate the prediction intervals well.} The LOO idea is related to the Jackknife+ procedure \citep{jackknife+}, but it is known to be costly due to the retraining of the model. To address this issue, we will develop a computationally efficient method called \EnbPI \ in Section \ref{sec:algo}, which constructs the LOO estimators as \textit{ensemble} estimators of pre-trained models. 


\section{Theoretical analysis }
\label{sec:theory}

We first present theoretical results for \magenta{bounding the worst-case coverage gap in conditional and marginal coverage}. We then establish \magenta{similar bounds on the difference between estimated and oracle intervals.} The results are general for methods beyond \EnbPI \ (for example, the split conformal method \citep{inductCP}). Without loss of generality and for notation simplicity, we only show guarantees when $t=T+1$, i.e., the one-step-ahead prediction. We will explain how guarantees naturally extend to all prediction intervals from $t = T+2$ onward in Remark \ref{remark:caveats}. In particular, our proof removes the assumptions on data exchangeability by replacing them with general and verifiable assumptions on the error process and estimation quality. All proofs can be found in Appendix \ref{app_theory:approx_uni_proof}.

\subsection{Coverage guarantees}\label{sec:cov_guarantee}

Following notations in Section \ref{sec:oracle_bound}, we first define the empirical $p$-value at $T+1$:
\[\hat{p}_{T+1}:=\frac 1 T\sum_{i=1}^T\textbf{1}\{\hat{\epsilon}_i\leq\hat{\epsilon}_{T+1}\}.\] 
As a result, we see the following equivalence between events:
\begin{align*}
    & \ Y_{T+1} \in \widehat{C}^{\alpha}_{T+1}\bigg|X_{T+1}=x_{T+1} \\
\Longleftrightarrow & \ \hat \epsilon_{T+1} \in  [\betaReal \text{ quantile of }  \{\hat{\epsilon}_i\}_{i=1}^{T}, \\ 
 & (1-\alpha+\betaReal) \text{ quantile of }  \{\hat{\epsilon}_i\}_{i=1}^{T}]\bigg|X_{T+1}=x_{T+1} \\
\Longleftrightarrow & \ \betaReal \leq \hat{p}_{T+1} \leq 1-\alpha+\betaReal,
\end{align*}
where $A|B$ means that the event $A$ conditions on event $B$. 
Therefore, our method covers $Y_{T+1}$ given $X_{T+1}=x_{T+1}$ with probability $1-\alpha$, hence, being conditionally valid if the distribution of $\hat{p}_{T+1}$ is uniform. More precisely, we aim to ensure that $|\mathbb{P}(\beta \leq \hat{p}_{T+1} \leq 1-\alpha+\beta)-(1-\alpha)|$ is small for any $\beta \in [0,\alpha]$. 

Due to the fact that $F_{T+1}(\epsilon_{T+1}) \sim \rm{Unif}[0,1]$ \citep{Casella1990StatisticalI}, $\mathbb{P}(\beta \leq F_{T+1}(\epsilon_{T+1}) \leq 1-\alpha+\beta)=1-\alpha$. Define 
\[
\hat{F}_{T+1}(x):=\frac 1 T\sum_{i=1}^T \textbf{1}\{\hat{\epsilon}_i\leq x\},
\]
whereby we have $\hat{p}_{T+1}=\hat{F}_{T+1}(\hat{\epsilon}_{T+1})$. As a consequence:
\begin{equation*}
\begin{split}
     &|\mathbb{P}(\beta\leq\hat{p}_{T+1}\leq 1-\alpha+\beta)-(1-\alpha)|\\
     =& ~|\mathbb{P}(\beta \leq \hat{F}_{T+1}(\hat{\epsilon}_{T+1})\leq 1-\alpha+\beta)-\\
     & ~\mathbb{P}(\beta \leq F_{T+1}(\epsilon_{T+1})\leq 1-\alpha+\beta)|.
\end{split}
\end{equation*}
Thus, intuitively, we can \magenta{bound gap in conditional coverage} using the \magenta{worst-case difference} between $\hat{F}_{T+1}(\hat{\epsilon}_{T+1})$ and $F_{T+1}(\epsilon_{T+1})$. 
Notice the following coupling between $\hat{\epsilon}_{T+1}$ and $\epsilon_{T+1}$ under model (\ref{eq:DGP_model}) when $X_{T+1}=x_{T+1}$:
\begin{equation}\label{eqn:err_decouple}
    \hat{\epsilon}_{T+1}=\epsilon_{T+1}+(f(x_{T+1})-\hat{f}_{-(T+1)}(x_{T+1})). 
\end{equation}
Therefore, the pointwise function estimation error $f(x_{T+1})-\hat{f}_{-(T+1)}(x_{T+1})$ should be small for $\hat{\epsilon}_{T+1}$ to be a good estimate for $\epsilon_{T+1}$. We will impose this condition when analyzing difference in interval width.

For the analyses, we now introduce another empirical CDF using unknown ``true'' errors $\epsilon_i, i\geq 1$, denoted as $\tilde{F}_{T+1}$:
\[
\tilde{F}_{T+1}(x):=\frac{1}{T}\sum_{i=1}^T \textbf{1}\{\epsilon_i\leq x\}.
\]
Note that $\hat{F}_{T+1}(\hat{\epsilon}_{T+1})$ is close in distribution to $\tilde{F}_{T+1}(\epsilon_{T+1})$ under the same pointwise estimation assumption of $f$ by $\hat{f}$, due to \eqref{eqn:err_decouple}. Meanwhile, the convergence of $\tilde{F}_{T+1}(x)$ to $F_{T+1}(x)$ is well-studied in the literature, which addresses the rate of convergence of an empirical distribution to the actual CDF \citep{DKW,stat_linear,rio2017}. 

Building on notations and ideas above, we now state the precise assumptions with discussions and present the following results: we first bound the worst deviation between $\tilde{F}_{T+1}(x)$ and $F_{T+1}(x)$ in Lemma \ref{regu_lem}. We then bound that between $\hat{F}_{T+1}(x)$ and $\tilde{F}_{T+1}(x)$ in Lemma \ref{consis_lem}. These lemmas are essential to proving our main theoretical results in Theorem \ref{thm:cond_cov}, which has several useful corollaries under slightly modified assumptions on error dependencies.

\newtheorem*{regu_lem}{Lemma \ref{regu_lem}}
\newtheorem*{consis_lem}{Lemma \ref{consis_lem}}
\newtheorem*{regu}{Assumption \ref{regu}}
\newtheorem*{consis}{Assumption \ref{consis}}
\newtheorem*{thm:cond_cov}{Theorem \ref{thm:cond_cov}}

\begin{assumption}[Errors are short-term {\it i.i.d.}]\label{regu}
Assume $\{\epsilon_t\}_{t=1}^{T+1}$ are independent and identically distributed (i.i.d.) according to a common CDF $F_{T+1}$, which is Lipschitz continuous with constant $L_{T+1}>0$.
\end{assumption}


\begin{lemma}\label{regu_lem}
Under Assumption \ref{regu}, for any training size $T$, there is an event $A_T$ which occurs with probability at least $1- \sqrt{\log (16T)/T}$, such that conditioning on $A_T$,  
\[
     \sup_{x} |\tilde{F}_{T+1}(x)-F_{T+1}(x)| \leq \sqrt{\log (16T)/T}.\]
\end{lemma}

\noindent {\it Discussion on Assumption \ref{regu}.} We call it the short-term i.i.d. assumption, since it only requires the past $T+1$ errors to be independent. It is a reasonably mild assumption on the original process $\{(X_t,Y_t)\}_{t\geq 1}$, because the process can exhibit arbitrary dependence and be highly non-stationary but still have i.i.d. errors. 
\magenta{Later on we can relax this assumption for more general cases, for instance, when errors follow linear processes (see Corollary \ref{thm:cond_cov_linear}) or are strongly mixing (see Corollary \ref{thm:cond_cov_stronglymixing}).
\rev{We can empirically examine whether or not the assumptions on residuals hold by using the LOO residuals as surrogates. The procedure is similar to examining the autocorrelation function after fitting a time series model.}
}

\begin{assumption}[Estimation quality]\label{consis} 
\magenta{There exists a real sequence $\{\delta_{T}\}_{T\geq 1}$ such that 
\begin{align*}
    & \frac 1 T \sum_{t=1}^{T} (\hat{f}_{-t}(x_t)-f(x_t))^2 \leq \delta_{T}^2 \mbox{ and }  \\
    & |\hat{f}_{-(T+1)}(x_{T+1})-f(x_{T+1})|\leq \delta_T.
\end{align*}
}
\end{assumption}

\begin{lemma}\label{consis_lem}
Under Assumptions \ref{regu} and \ref{consis}, we have
\begin{align*}
    & \sup_x |\hat{F}_{T+1}(x)-\tilde{F}_{T+1}(x)| \\
    \leq & (L_{T+1}+1)\delta_T^{2/3}+2\sup_x |\tilde{F}_{T+1}(x)-F_{T+1}(x)|.
\end{align*}
\end{lemma}

\noindent {\it Discussion on Assumption \ref{consis}.} \magenta{There are two situations affecting asymptotic guarantees: $\delta_T$ never decays as $T$ grows or converges to zero as $T\rightarrow \infty$. The first situation can happen due to data overfitting, which leads to $\hat{f}_{-t}(x_t) \approx y_t$ and therefore, $\sum_{t=1}^T (\hat{f}_{-t}(x_t)-f(x_t))^2\approx \sum_{t=1}^T \epsilon_t^2$. If $\sum_{t=1}^T\epsilon_t^2 \in \Omega(T)$, the same order holds for the sequence $\{\delta_T\}_{T\geq 1}$, so that the worst-case coverage gap always exists (see Theorem \ref{thm:cond_cov}).} On the other hand, there are examples in the second situation where $\{\delta_{T}\}_{T\geq 1}$ converges to zero. Note that assumptions for estimating  unknown $f$ are necessary due to the well-known \textit{No Free Lunch Theorem} \citep{no-free}. The decay rate of $\delta_T$ is explicit for two classes of $f$ and the following $\mathcal{A}$:
\begin{enumerate}
\item[] (Example 1) If $f$ is sufficiently smooth, $\delta_T=o_P(T^{-1/4})$ for general neural networks sieve estimators \cite[see Corollary 3.2]{nn-consis}. \label{example:smooth_f}
\item[] (Example 2) If $f$ is a sparse high-dimensional linear model, $\delta_T=o_P(T^{-1/2})$ for the Lasso estimator and Dantzig selector. \cite[see Equation 7.7]{high-dim-linear-consis}. \label{example:highdim_f}
\end{enumerate}
In general, one needs to analyze the convergence rate of estimators $\hat f$ to the unknown true $f$. This task is different from analyzing the Mean Squared Error (MSE) of ensemble estimators \citep{breiman-bagging} and likely requires case-by-case analyses, which we leave for future work. 
%


Our main theoretical result is the following Theorem \ref{thm:cond_cov}, which establishes the asymptotic conditional coverage as a consequence of Lemmas \ref{regu_lem} and \ref{consis_lem}.
\begin{theorem}[Conditional coverage gap; errors are short-term i.i.d.]\label{thm:cond_cov}
Under Assumption \ref{regu} and \ref{consis}, for any training size $T$, $\alpha \in (0,1)$, and $\beta \in [0,\alpha]$, we have:
\begin{align}
    & |\mathbb{P}(Y_{T+1}\in \widehat{C}^{\alpha}_{T+1}|X_{T+1}=x_{T+1})-(1-\alpha)| \nonumber \\
    \leq & 12\sqrt{\log (16T)/T} +4 (L_{T+1}+1) (\delta_T^{2/3}+\delta_T). \label{eq:main}
\end{align}
\magenta{Furthermore, if $\{\delta_{T}\}_{T\geq 1}$ converges to zero, the upper bound in \eqref{eq:main} converges to 0 when $T\rightarrow \infty$, and thus the conditional coverage is asymptotically valid.}
\end{theorem}

We briefly comment on the proof techniques and the role of Assumption \ref{regu}. The term $\sqrt{\log (16T)/T}$ on the right-hand side directly relates to how quickly the empirical CDF $\tilde{F}_{T+1}$ converges to the actual CDF $F_{T+1}$. In general, we find sequences $\{s_T\}_{T\geq 1}$ and $\{g(s_T)\}_{T\geq 1}$, both of which converge to zero, such that 
\[
    \mathbb P(\sup_{x} |\tilde{F}_{T+1}(x)-F_{T+1}(x)| > s_T)\leq g(s_T).
\]
The optimal rate of decay reduces to finding $s_T$ such that $s_T=g(s_T)$. Then, the event $A_T$ is chosen to happen with probability at least $1-s_T$, where conditioning on this event, $\sup_{x} |\tilde{F}_{T+1}(x)-F_{T+1}(x)|\leq s_T$. As a result, there are decay rates different from $\sqrt{\log (16T)/T}$ under more relaxed assumptions on $\{\epsilon_t\}_{t=1}^{T+1}$. We summarize two possible results in Corollaries \ref{thm:cond_cov_linear} and \ref{thm:cond_cov_stronglymixing}; certain technical assumptions, precise statements, and definitions are presented in the appendix.  

\begin{corollary}[Conditional coverage gap; errors follow linear processes]\label{thm:cond_cov_linear}
Under Assumption \ref{consis}, suppose that $\{\epsilon_t\}_{t=1}^{T+1}$ satisfy $\epsilon_t=\sum_{j=1}^{\infty} \delta_j z_{t-j}$, with regularity conditions on $\delta_j$ and $z_{t-j}$. There exists a constant $K$ so that for any training size $T$, $\alpha \in (0,1)$, and $\beta \in [0,\alpha]$, we have:
\begin{align}
    & |\mathbb{P}(Y_{T+1}\in \widehat{C}^{\alpha}_{T+1}|X_{T+1}=x_{T+1})-(1-\alpha)| \nonumber \\
    \leq & 12K \log T/\sqrt{T}+4(L_{T+1}+1) (\delta_T^{2/3}+\delta_T). \label{eq:cor1}
\end{align}
\end{corollary}

To introduce the last corollary, we first define the strong mixing coefficient between two $\sigma-$fields $\mathcal{A}$ and $\mathcal{B}$, which measures the dependence between them: 
\[
\alpha(\mathcal{A},\mathcal{B})=2\sup\{|\mathbb{P}(A\cap B)-\mathbb{P}(A)\mathbb{P}(B)|: (A,B)\in \mathcal{A}\times \mathcal{B}\}.
\]
This definition is equivalent to that in \citep{Rosenblatt1956ACL} up to a multiplicative factor of 2. For the sequence $\{\epsilon_t\}_{t\geq 1}$, let $\mathcal{A}_k:=\sigma(\epsilon_t:t\leq k)$ and $\mathcal{B}_{c}:=\sigma(\epsilon_t:t\geq l)$. The coefficients $\{\alpha_n\}_{n\geq 1}$ are defined as
\begin{equation*}
    \alpha_0=1/2 \text{  and  } \alpha_n=\underset{k\in \mathbb{N}}{\sup} \ \alpha(\mathcal{A}_k,\mathcal{B}_{k+n}) \text{ for any  } n>0.
\end{equation*}
The sequence is said to be \textit{strongly mixing} if $\underset{n\rightarrow \infty}{\lim} \alpha_n=0$.

\begin{corollary}[Conditional coverage gap; errors are strongly mixing]\label{thm:cond_cov_stronglymixing}
Under Assumption \ref{consis}, suppose $\{\epsilon_t\}_{t=1}^{T+1}$ are stationary and strongly mixing, where mixing coefficients are summable with $0<\sum_{k\geq 0}\alpha_{k} < M$. For any training size $T$, $\alpha \in (0,1)$, and $\beta \in [0,\alpha]$, we have:
\begin{align}
    & |\mathbb{P}(Y_{T+1}\in \widehat{C}^{\alpha}_{T+1}|X_{T+1}=x_{T+1})-(1-\alpha)| \nonumber \\
    \leq & 12(M/2)^{1/3}(\log T)^{2/3}/T^{1/3}+4(L_{T+1}+1)(\delta_T^{2/3}+\delta_T). \label{eq:cor2}
\end{align}
\end{corollary}

Lastly, the following asymptotic marginal validity guarantee holds as a consequence of earlier results by the tower law property (proof omitted):
\begin{theorem}[Marginal coverage gap]\label{thm:marg_cov}
Under Assumption \ref{regu} and \ref{consis}, for any training size $T$, $\alpha \in (0,1)$, and $\beta \in [0,\alpha]$, we have:
\begin{align}
    & |\mathbb{P}(Y_{T+1}\in \widehat{C}^{\alpha}_{T+1})-(1-\alpha)| \nonumber \\ 
    \leq & 12\sqrt{\log (16T)/T}+4(L_{T+1}+1) (\delta_T^{2/3}+\delta_T). \label{eq:marginal}
\end{align}
Moreover, the right-hand side decay rate in \eqref{eq:marginal} is $\mathcal{O}(\log T/\sqrt{T}+\delta_T^{2/3})$ if $\{\epsilon_t\}$ follow a linear process as in Corollary \ref{thm:cond_cov_linear}, and $\mathcal{O}((\log T)^{2/3}/T^{1/3}+\delta_T^{2/3})$ if $\{\epsilon_t\}$ are strongly mixing with summable mixing coefficients as in Corollary \ref{thm:cond_cov_stronglymixing}.
\end{theorem}




We make two final comments for the above theorems and corollaries. Firstly, to build prediction intervals that have at least $1-\alpha$ coverage, one needs to incorporate the upper bounds on the right-hand side of \eqref{eq:main}---\eqref{eq:marginal} into the prediction interval construction. However, we will not do so in \EnbPI \ (our proposed algorithm), which is a \textit{general wrapper} that can be applied to most regression models $\mathcal{A}$. Secondly, The rate $\mathcal O(\sqrt{\log (16T)/T}+\delta_T^{2/3}$) is a worst-case analysis for both marginal and conditional coverage; empirical results show that even at a small training data size $T$, \EnbPI \ can achieve both marginal and conditional validity.

\subsection{Width guarantees}

Our next goal is to bound the gap between the estimated prediction interval $\widehat{C}^{\alpha}_{T+1}$ and the oracle $C^{\alpha}_{T+1}$ in (\ref{eq:oracle_interval}). Define set difference $\Delta:\mathbb N \rightarrow \mathbb R$ such that $\Delta(T)=\widehat{C}^{\alpha}_{T+1}\triangle C^{\alpha}_{T+1}$, where for any two subsets $A, B \subset \mathbb R$ under the Lebesgue measure $\mu$, $A\triangle B:=\mu(\{x\in \R:x\in A, x\notin B\})
+\mu(\{x\in \R:x\in B, x\notin A\}).$
Theorem \ref{thm:width} below bounds $\Delta(T)$ under Assumptions \ref{regu}, \ref{consis}, and other regularity conditions;
the bound is similar to that in Theorem \ref{thm:cond_cov}.
\begin{theorem}[Width gap bound; errors are i.i.d.]\label{thm:width}
Under Assumption \ref{regu} and \ref{consis}, further assume $F_{T+1}^{-1}$ is Lipschitz continuous with constant $K_{T+1}$. \magenta{With probability at least $1-\sqrt{\log (16T)/T}$,}
\begin{align*}
\Delta(T) \leq & \delta_T+ \alpha K'_{T+1}/m+ 2(K_{T+1}+M_{T+1}) \cdot\\
    & \left(3\sqrt{\log (16T)/T} +(L_{T+1}+1)(\delta_T^{2/3}+\delta_T) \right),
\end{align*}
where $m$ is the number of grids for line-search of $\betaReal$ based on the past $T$ LOO residuals, \magenta{$K'_{T+1}:=\underset{j=1,\ldots,T-1}{\max} \hat{\epsilon}_{j+1}-\hat{\epsilon}_j$ using sorted LOO residuals indexed from the smallest to the largest,} and $M_{T+1}$ is a constant that depends only on $L_{T+1}$, $K_{T+1}$, and $K^{'}_{T+1}$.
\end{theorem}

When $\{\epsilon_t\}_{t=1}^T$ are not i.i.d., results similar to Corollaries \ref{thm:cond_cov_linear} and \ref{thm:cond_cov_stronglymixing} can be established for Theorem \ref{thm:width} using similar proof techniques. More precisely, the rate $\sqrt{\log (16T)/T}$ will be replaced by $\log T/\sqrt{T}$ when errors follow linear processes, and by $(\log T)^{2/3}/T^{1/3}$ when errors are strongly mixing with summable mixing coefficients. 

\begin{remark}[Theorem applicability and caveats]\label{remark:caveats} 
All theoretical results hold for $t>T+1$, as long as Assumptions \ref{regu} and \ref{consis} hold at indices $t-T,\ldots,t$. The same proof techniques apply. Meanwhile, as long as the same assumptions hold, all previous results apply to other conformal prediction methods, such as split conformal \citep{inductCP}. However, unlike our \EnbPI{} that requires no data-splitting, split conformal and its variants require data splitting by treating a subset of training data as the ``calibration data.'' As a result, the value $T$ on the right-hand side of Theorem \ref{thm:cond_cov} and all subsequent corollaries become the size of the calibration data, not that of the full training data. This is because prediction residuals $\hat{\epsilon}$ are only computed on calibration data, whose empirical distribution is used to approximate that of the true distribution of errors $\epsilon$. In such cases, the worst-case coverage gap becomes larger. 
\end{remark}

\section{EnbPI Algorithm} \label{sec:algo}

We now present a general conformal prediction algorithm for time series in Algorithm \ref{algo:hypo-test-ensemble}, which is named \EnbPI. \magenta{On a high-level, \EnbPI \ has a training phase and the prediction phase. In the training phase, \EnbPI \ first fits a fixed number of bootstrap estimators from subsets of the training data. Then, it aggregates predictions from these bootstrap estimators on the training data in an efficient leave-one-out (LOO) fashion, resulting in \textit{both} LOO predictors and LOO residuals for prediction. In the prediction phase, \EnbPI \ aggregates predictions from LOO predictors on each test datum to compute the center of the prediction interval. Then, it builds the prediction interval using the past LOO residuals, where the interval width is also optimized through a simple one-dimensional line search. Lastly, residuals are slid forward as soon as actual response variables in test data are observed to ensure adaptiveness in the prediction intervals.}

In the algorithm description, $\hat{f}^b$ is the $b$-th bootstrap estimator, the superscript $\phi$ denotes variables with dependence on the aggregation function $\phi$. The \textit{block bootstrap} with $T$ non-overlapping blocks is used in line 2, which is a popular method for bootstrapping dependent data \citep{Kreiss2011BootstrapMF}. The basic idea is to split the $T$ training samples into $l$ (non-)overlapping blocks, each with a size $\floor*{T/l}$. Then, sample from $l$ blocks randomly with replacement. 

We comment on the choice of hyperparameters as follows. (1) In general, $\mathcal{A}$ can be a family of (parametric and non-parametric) prediction algorithms. (2) Different choices of aggregation functions $\phi$ bring different benefits, such as reducing the MSE under mean, avoiding sensitivity to outliers under median, or achieving both under trimmed mean. (3) As the number of pre-trained bootstrap models $B$ increases, interval widths may be narrower. Empirically, we find that choosing $B$ between 20 and 50 is sufficient, especially for computationally intensive methods such as neural networks. 
(4) Larger $s$ requires prediction further in the future without feedback; however, as $s$ increases, the prediction becomes harder, which is reflected in that intervals become wider and the coverage deteriorates; how large $s$ can be is determined by the dynamics of the data.

\begin{algorithm}[h!]
\cprotect\caption{Ensemble batch prediction intervals (\EnbPI)}
\label{algo:hypo-test-ensemble}
\begin{algorithmic}[1]
\REQUIRE{Training data $\{(x_i, y_i)\}_{i=1}^T$, prediction algorithm $\mathcal{A}$, significance level $\alpha$, aggregation function $\phi$, number of bootstrap models $B$, batch size $s$, and test data $\{(x_t,y_t)\}_{t=T+1}^{T+T_1}$; $y_t$ is revealed as feedback only after prediction at $t$ is done.
}
\ENSURE{Ensemble prediction intervals $\{C^{\phi,\alpha}_{t}(x_t)\}_{t=T+1}^{T+T_1}$}
\FOR {$b = 1, \dots, B$}
\STATE Sample with replacement an index set $S_b=(i_1,\ldots,i_T)$ from indices $(1,\ldots,T)$.
\STATE Compute $\hat{f}^b=\mathcal{A}((x_i,y_i), i \in S_b)$.
\ENDFOR
\STATE Initialize $\widehat{\boldsymbol \epsilon}=\{\}$ \magenta{as an ordered set.}
\FOR {$i=1,\dots,T$}
\STATE $\hat{f}_{-i}^{\phi}(x_i)=\phi(\hat{f}^b(x_i), i \notin S_b)$
\STATE Compute $\hat{\epsilon}_i^{\phi}=y_i-\hat{f}_{-i}^{\phi}(x_i)$
\STATE $\widehat{\boldsymbol \epsilon}=\widehat{\boldsymbol \epsilon} \cup \{\hat{\epsilon}_i^{\phi}\}$
\ENDFOR
\FOR {$t=T+1,\dots,T+T_1$}
\STATE $\hat{f}^{\phi}_{-t}(x_t)=\phi(\hat{f}^{\phi}_{-i}(x_t), i=1, \ldots, T)$
\STATE Compute $\betaReal$ as 
\vspace{-0.1in}
\[{\arg\min}_{\beta\in[0,\alpha]} (1-\alpha+\beta) \text{ quantile of \ } \widehat{\boldsymbol \epsilon}-\beta \text{ quantile of \ } \widehat{\boldsymbol \epsilon})\]
\vspace{-0.2in}
\STATE  $w_{t,\rm{lower}}^{\phi,\alpha}= \betaReal \text{ quantile of } \widehat{\boldsymbol \epsilon}$
\STATE  $w_{t,\rm{upper}}^{\phi,\alpha}= (1-\alpha+\betaReal) \text{ quantile of } \widehat{\boldsymbol \epsilon}$.
\STATE Return $C^{\phi,\alpha}_{t}(x_t)=[\hat{f}^{\phi}_{-t}(x_t) +w_{t,\rm{lower}}^{\phi,\alpha},\hat{f}^{\phi}_{-t}(x_t) +w_{t,\rm{upper}}^{\phi,\alpha}]$
\IF {$t-T \equiv 0$ mod $s$} 
\FOR {$j=t-s,\ldots,t-1$}
\STATE Compute $\hat{\epsilon}_j^{\phi}=y_j-\hat{f}^{\phi}_{-j}(x_t)$  
\STATE $\widehat{\boldsymbol \epsilon}=(\widehat{\boldsymbol \epsilon}- \{\hat{\epsilon}_1^{\phi}\}) \cup \{\hat{\epsilon}_j^{\phi}\}$ and reset index of $\widehat{\boldsymbol \epsilon}$.
\ENDFOR
\ENDIF
\ENDFOR
\end{algorithmic}
\end{algorithm}

\subsection{Properties of \protect \EnbPI } \label{sec:enbpi_properties}

\noindent
{\it Computational efficiency.} Note that in \EnbPI, the prediction models in the ensemble are pre-trained once and stored; when deploying \EnbPI \ for prediction, residuals are computed from $T$ pre-trained models on the fly, and the interval is constructed based on quantile values of $T$ residuals. Thus, the main computation of \EnbPI \ for obtaining the prediction interval is tolerable in calling the prediction algorithm $\mathcal{A}$ $B$ times. In comparison, the Jackknife+ approach \citep{jackknife+} requires requires $B$ times training of $\mathcal{A}$ on \textit{each} leave-$i$-out sample $\{(x_j,y_j)\}_{j=1, j\neq i}^T$. This requires $BT$ training of $\mathcal{A}$, which can be computationally intensive for complex prediction algorithms such as deep neural networks.

\vspace{.1in}
\noindent
{\it No overfitting or data splitting.} Traditional CP methods such as split conformal \citep{inductCP} use data-splitting to avoid overfitting. In contrast, inspired by the J+aB procedure in \citep{j+ab}, \EnbPI \ trains LOO ensemble models on full data and avoids overfitting through thoughtful aggregations in lines 6-10. In particular, to construct the $i$-th LOO ensemble predictor, \EnbPI \ aggregates all $B$ bootstrap models that are \textit{not} trained on the training datum $(x_i,y_i)$. Thus, the actual number of aggregated models is a Binomial random variable with parameters $B$ and $(1-1/T)^T$; the Chernoff bound ensures that each ensemble predictor aggregates a balanced number of pre-trained models. 

\vspace{.1in}
\noindent
{\it Leverage new data without model retraining.} \EnbPI \ constructs sequential prediction intervals without retraining $\mathcal{A}$. Instead, it leverages feedback by updating past residuals through a sliding window of size $T$, which adapts the interval widths to data and can better adapt to data non-stationarity. \rev{In practice, we acknowledge the benefits of retraining, especially in reducing the widths of the prediction intervals. However, retraining can be costly for certain models, and one should consider the trade-off between interval widths and computation involved in retraining.}

\subsection{\protect \EnbPI \ on challenging tasks }

We comment that \EnbPI \ is flexible and can handle various challenging tasks. \rev{In appendix \ref{app_expr:CA_solar_wind}, we also discuss how \EnbPI{} can construct prediction intervals for outputs from each node of a network.}

\vspace{.1in}
\noindent
{\it Handle missing data.} We suggest a heuristic approach to handle missing data by \EnbPI, which is verified in Section \ref{expr:second_part}. When training and/or test data have missing entries, we can properly increase the size of bootstrap samples being drawn from the rest available training data--- this is appropriate since a common data model $f$ is assumed. On test data, when \EnbPI \ encounters a missing index $t^{\prime}$, we impute the feature $x_{t^{\prime}}$ if it is missing to compute $\hat{f}_t(x_{t^{\prime}})$, the interval center, and use the most recent $T$ residuals to compute the interval width. The sliding window would skip over the residual $\epsilon^{\phi}_{t^{\prime}}$ when $y_{t^{\prime}}$ is unobserved. Section \ref{expr:second_part} considers the solar dataset with missing data.

\vspace{0.1in}
\noindent
{\it Unsupervised anomaly detection.} Suppose there is an anomalous point $y_{t^*}$ at time $t^*$, due to either a change in model $f$ at $t^*$ or an unusually large stochastic error $\epsilon_{t^*}$. As a result, $y_{t^*}$ tends to lie far outside the interval (equivalently, $\epsilon^{\phi}_{t^*}$ is well below or above the $\betaReal$ or $(1-\alpha+\betaReal)$ quantile of past $T$ residuals) and thus can be detected using the prediction interval. An example applying \EnbPI \ to detect anomalous traffic flows appears in Section \ref{sec:anomaly_detection}.

\section{Experimental results} \label{sec:exper}

The experiments are organized as follows. In Section \ref{exp:simul}, we provide extensive simulations to examine the coverage and width of \EnbPI \ intervals. In Section \ref{expr:first_part}, we show that \EnbPI \ attains valid marginal coverage on real data, whereas competing methods may fail. In Section \ref{expr:second_part}, we present real-data experiments to examine the conditional coverage of \EnbPI \ against other methods when missing data are present. \rev{In Section \ref{sec:anomaly_detection}, we present an example for anomaly detection in  traffic flow using \EnbPI{}.} In Appendix \ref{app_expr:CA_solar_wind} and \ref{app_expr:other_data}, we present more time-series data examples to demonstrate that \EnbPI \ has valid coverage and shorter intervals than the competing methods.

\subsection{Simulation results}\label{exp:simul}

\rev{We first conduct three simulated examples based on the assumption $Y_t=f(X_t)+\epsilon_t$ to examine the performance of \EnbPI{}. We then consider a more complex example based on a noisy helix trajectory.

\vspace{0.1in}
\noindent \textit{Three simulated examples.} We construct these examples with increasing levels of model sophistication in the design of $f(X_t)$ and under more complex error dependency in $\epsilon_t$. The detailed data-generating procedures and additional details are described in Appendix \ref{append_simul}. The results shown in Figure \ref{fig:simulation_nochangepts} of Appendix \ref{append_simul} indicate the satisfactory performance of \EnbPI{} to maintain valid coverage.} The interval widths also converge to the oracle width as the training sample size grows, validating Theorem \ref{thm:width}.

\vspace{0.1in}
\noindent \textit{Simulation with a noisy helix trajectory.} Consider $Y_t$ given by a nonlinear mapping of components of a helix in three-dimensional space contaminated by noise:  $X_t=[r\cos(\theta_t),r\sin(\theta_t),H\theta_t]$, $f(X_t)=r\cos(\theta_t)\cdot (|r\sin(\theta_t)|)^{1/2} \cdot (H\theta_t+\varepsilon)^{-1/2}$, $\varepsilon=10^{-3}$, and $\epsilon_t=\rho \epsilon_{t-1}+e_t$ where $\rho=0.6$ and $e_t$ are {\it i.i.d.} normal random variables with zero mean and unit variance. The color map of the helix is proportional to $Y_t$. We fix $H=3$, $r=10$ and generate 1000 samples parametrized by $\theta_t$, which are uniformly spaced between 0 and $8\pi$. The first 500 data points are used for training \EnbPI \ with random forest regression (RF) and the rest 500 are used for testing. The RF setup is described in Appendix \ref{app_expr:data_des}. In Figure \ref{fig:simul_helix}, we see that in the test phase intervals by \EnbPI \ tightly cover the unknown response $Y_t$. Moreover, the blue and orange curves corresponding to $\hat{Y_t}$ and $Y_t$ are very close, which indicates that LOO ensemble predictors approximate the unknown model $f$ very well.

\begin{figure*}[!t]
    \centering
    \includegraphics[width=\linewidth]{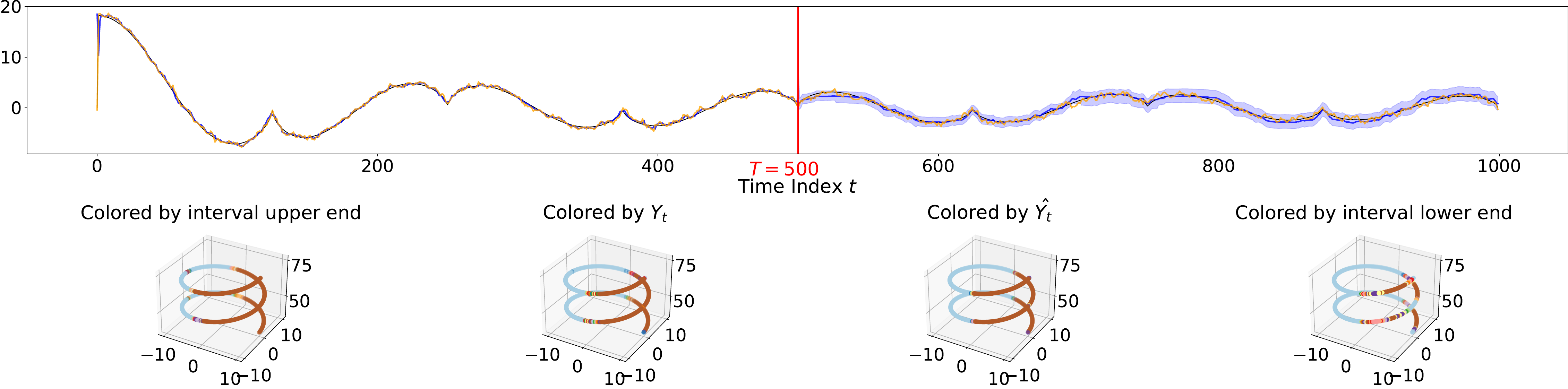}
    \caption{
    Helix colored by $Y_t$. We observe that the predicted colors closely match the actual color on the bottom row because values of $Y_t$ colored in orange are contained in prediction intervals colored in shaded blue with high probability, and intervals are very narrow, as shown on the top row.}
    \label{fig:simul_helix}
\end{figure*}


\subsection{Real-data: Marginal validity and the interval width} \label{expr:first_part}

In this section, we consider predictions for renewable energy generation. In this setting, the prediction and uncertainty quantification is critical due to their high stochasticity and non-stationarity.

\vspace{0.1in}
\noindent {\it Data description.} The renewable energy data are from the National Solar Radiation Database and the Hackberry wind farm in Austin\footnote{NSRDB: https://nsrdb.nrel.gov/. Wind farm: https://github.com/Duvey314/austin-green-energy-predictor}. We use 2018 hourly solar radiation data from Atlanta and nine cities in California and 2019 hourly wind energy data. \magenta{We remove recordings before 6 a.m. and after 8 p.m. for the solar radiation data due to zero radiation levels during the period.} In total, there are 11 time series from 11 sensors (one from each sensor), and each time series contain other features such as temperature, humidity, wind speed, etc. In particular, California solar data constitute a network, where each node is a sensor. From now on, we call $X_t$ \textit{univariate} if it is the history of $Y_t$ and \textit{multivariate} if it contains other features that predict $Y_t$.

\vspace{0.1in}
\noindent {\it Comparison methods.} We compare \Verb|EnbPI| with traditional time series and other conformal prediction methods. 
The time series methods are ARIMA(10,1,10), Exponential Smoothing (ExpSmoothing), and Dynamic Factor model (DynamicFactor). \magenta{The CP methods are split/inductive conformal predictor (ICP) \citep{inductCP} and, weighted ICP (WeightedICP) \citep{CPcovshift}, quantile out-of-bag method (QOOB) \citep{QOOB}, adaptive conformal inference (AdaptCI) \citep{Gibbs2021AdaptiveCI}, and jackknife+-after-bootstrap (J+aB) \citep{j+ab}. For the former two CP methods (resp. AdaptCI),  we split the training data into $50\%$ (resp. $75\%$) proper training set for training a predictor and $50\%$ (resp. $25\%$) calibration set for computing non-conformity scores.} Appendix \ref{app_expr:data_des} describes more detailed setup. 
\vspace{0.1in}

\noindent\textit{Prediction algorithm $\mathcal{A}$.} We choose four prediction algorithms: ridge regression, random forest (RF), neural networks (NN), and recurrent neural networks (RNN) with LSTM layers. The first two are implemented in the Python \Verb|sklearn| library, and the last two are built using the \Verb|keras| library. See Appendix \ref{app_expr:data_des} for their specifications. 
\vspace{0.1in}

\noindent\textit{Other hyperparameters.} Since the three CP methods are trained on random subsets of training data, we repeat all experiments below for ten trials with an independent random split in each trial. The time series methods are only applied once on training data because they do not use random subsets. Throughout this subsection, we fix $s=1$. Let $\alpha=0.1$ and use the first 20$\%$ of the total hourly data for training unless otherwise specified. This creates small training samples for a challenging long-term predictive inference task. We use \EnbPI \ under $B=25$ and $\phi$ as taking the sample mean. 

\begin{table*}[!b]
    \centering
    \cprotect \caption{Solar power prediction in Atlanta, comparison of \EnbPI{} with AdaptCI, J+aB, QOOB, ICP, and Weighted ICP. We vary the percentage of total data as training data at $\alpha=0.1$. Cells in brackets for CP methods indicate standard deviation over ten trials.} \label{tab:marginal_cov_solar_QOOB_AdaptCI}
    
    \resizebox{\textwidth}{!}{%
    \def\arraystretch{1.5}
    \begin{tabular}{p{1.6cm} p{1.6cm}p{1.6cm}p{1.6cm}p{1.6cm}p{1.6cm}p{1.6cm}|p{1.6cm}p{1.6cm}p{1.6cm}p{1.6cm}p{1.6cm}p{1.6cm}|p{1.6cm}p{1.6cm}p{1.6cm}p{1.6cm}p{1.6cm}p{1.6cm}}
    \toprule
       Train ratio & \multicolumn{6}{c}{0.10} & \multicolumn{6}{c}{0.19} & \multicolumn{6}{c}{0.28} \\
       \hline
    CP method &           EnbPI &            AdaptCI &                J+aB &               QOOB &                ICP &       Weighted ICP &           EnbPI &            AdaptCI &                J+aB &               QOOB &                ICP &        Weighted ICP &           EnbPI &            AdaptCI &                J+aB &               QOOB &                ICP &        Weighted ICP \\
    \midrule
    Coverage &    0.893 (1.8e-03) &    0.828 (2.4e-02) &    0.747 (2.7e-03) &    0.684 (1.1e-02) &    0.646 (1.2e-01) &   0.608 (1.4e-01) &    0.897 (5.9e-04) &    0.891 (6.1e-03) &    0.777 (3.0e-03) &    0.783 (2.8e-03) &    0.703 (1.2e-01) &    0.698 (1.2e-01) &    0.905 (7.0e-04) &    0.909 (1.5e-03) &    0.819 (1.9e-03) &    0.850 (3.5e-03) &    0.760 (1.1e-01) &    0.746 (1.3e-01) \\
    Width    &  204.597 (1.8e+00) &  178.870 (1.7e+01) &  116.129 (1.3e+00) &  106.199 (2.1e+00) &  104.745 (4.0e+01) &  96.728 (4.2e+01) &  215.442 (4.4e-01) &  222.328 (2.2e+00) &  148.174 (1.6e+00) &  140.723 (1.9e+00) &  132.888 (4.8e+01) &  131.247 (4.9e+01) &  227.286 (6.8e-01) &  211.686 (2.6e+00) &  180.081 (7.0e-01) &  160.231 (1.4e+00) &  165.545 (6.5e+01) &  163.855 (6.8e+01) \\
    \bottomrule
    \end{tabular}
    }
\end{table*}

\vspace{0.1in}
\noindent{\it Results.} All results in Section \ref{expr:first_part} and \ref{expr:second_part} come from using the Atlanta solar data. Similar results using California solar data and Hackberry wind data are in Appendix \ref{app_expr:CA_solar_wind}. \rev{We first compare \EnbPI{} with the conformal prediction methods at a fixed $\alpha=0.1$. \EnbPI{} results are based on one of the four prediction algorithms that yield the narrowest interval when reaching valid $1-\alpha$ coverage. Table \ref{tab:marginal_cov_solar_QOOB_AdaptCI} shows that out of all the CP methods, \EnbPI{} is the only choice that consistently yields valid coverage at 0.9 regardless of the amount of training data. In contrast, the baseline CP methods may yield narrower intervals than \EnbPI{}, yet their intervals often have a high coverage gap with respect to the 0.9 target level. Hence, this indicates that \EnbPI{} is the most suitable method for this dataset. To better compare \EnbPI{} with the baselines, we adjust the $\alpha$ parameter for each baseline method so that they yield approximately the same interval widths as \EnbPI{}. Table \ref{tab:marginal_cov_solar_QOOB_AdaptCI_2} compares the performance of all methods under adjusted $\alpha$, where we see that baseline methods often fail to reach valid $1-\alpha$ coverage as \EnbPI{}. In addition, we often need to use extremely conservative values of $\alpha$ to reach the same interval widths as \EnbPI{} (e.g., reduce to 0.03 for QOOB under 0.28 train ratio). Furthermore, \EnbPI{} intervals also have the smallest standard deviation in width, indicating more stable interval construction by our proposed method.

In addition, Table \ref{tab:marginal_cov_solar_tseries} compares \EnbPI{} with commonly used time-series methods, where we also include AdaptCI as the best-performing CP baseline method. Compared to \EnbPI{}, the time-series baseline methods either yield conservative intervals under valid coverage or narrower intervals which nevertheless fail to cover at target $1-\alpha$ levels.}

\begin{table*}[!t]
    \centering
    \cprotect \caption{\rev{Solar power prediction in Atlanta. We adjust the hyper-parameter $\alpha$ for baseline methods to ensure they yield intervals with nearly the same width as \EnbPI{}, under identical setup to Table \ref{tab:marginal_cov_solar_QOOB_AdaptCI}.}} \label{tab:marginal_cov_solar_QOOB_AdaptCI_2}
    
    \resizebox{\textwidth}{!}{%
    \def\arraystretch{1.5}
    \begin{tabular}{p{1.6cm} p{1.6cm}p{1.6cm}p{1.6cm}p{1.6cm}p{1.6cm}p{1.6cm}|p{1.6cm}p{1.6cm}p{1.6cm}p{1.6cm}p{1.6cm}p{1.6cm}|p{1.6cm}p{1.6cm}p{1.6cm}p{1.6cm}p{1.6cm}p{1.6cm}}
    \toprule
       Train ratio & \multicolumn{6}{c}{0.10} & \multicolumn{6}{c}{0.19} & \multicolumn{6}{c}{0.28} \\
       \hline
    CP method &           EnbPI &            AdaptCI &                J+aB &               QOOB &                ICP &       Weighted ICP &           EnbPI &            AdaptCI &                J+aB &               QOOB &                ICP &        Weighted ICP &           EnbPI &            AdaptCI &                J+aB &               QOOB &                ICP &        Weighted ICP \\
    \midrule
    $\alpha$ value &                0.1 &               0.05 &             0.0115 &             0.0075 &              0.018 &             0.0125 &                0.1 &               0.13 &               0.04 &              0.025 &               0.04 &               0.04 &                0.1 &              0.125 &              0.055 &               0.03 &               0.07 &               0.06 \\
    Coverage      &    0.893 (1.8e-3) &    0.855 (4.5e-3) &    0.844 (2.5e-3) &    0.840 (9.8e-3) &    0.848 (1.4e-2) &    0.828 (3.4e-2) &    0.897 (5.9e-4) &    0.869 (1.0e-3) &    0.850 (2.4e-3) &    0.876 (2.4e-3) &    0.843 (1.8e-2) &    0.844 (1.2e-2) &    0.905 (7.0e-4) &    0.883 (6.1e-4) &    0.879 (2.3e-3) &    0.931 (1.7e-3) &    0.859 (7.4e-3) &    0.869 (7.1e-3) \\
    Width         &  204.597 (1.8e+0) &  210.124 (4.7e+0) &  210.747 (2.7e+0) &  203.400 (1.0e+1) &  214.308 (1.3e+1) &  203.511 (3.2e+1) &  215.442 (4.4e-1) &  215.896 (1.1e+0) &  214.418 (1.9e+0) &  212.158 (2.0e+0) &  218.597 (1.2e+1) &  217.359 (8.7e+0) &  227.286 (6.8e-1) &  224.981 (9.7e-1) &  224.418 (1.7e+0) &  223.890 (1.6e+0) &  220.091 (6.1e+0) &  225.935 (5.2e+0) \\
    \bottomrule
    \end{tabular}
    }
\end{table*}

\begin{table*}[!b]
    \centering
    \cprotect \caption{\rev{Solar power prediction in Atlanta, comparison of \EnbPI{} with AdaptCI, ARIMA, Exponential Smoothing, and Dynamic Factor Models. We vary $\alpha \in [0.05,0.10,0.15,0.20]$ and use the first 20\% data as training data. 
    }}
    \label{tab:marginal_cov_solar_tseries}
    \resizebox{\textwidth}{!}{%
    \def\arraystretch{1.5}
    \begin{tabular}{p{1.2cm} p{1.05cm}p{1.2cm}p{1.2cm}p{2cm}p{1.5cm}|p{1.05cm}p{1.2cm}p{1.2cm}p{2cm}p{1.5cm}|p{1.05cm}p{1.2cm}p{1.2cm}p{2cm}p{1.5cm}|p{1.05cm}p{1.2cm}p{1.2cm}p{2cm}p{1.5cm}}
    \toprule
       $\alpha$ & \multicolumn{5}{c}{0.05} & \multicolumn{5}{c}{0.10} & \multicolumn{5}{c}{0.15} & \multicolumn{5}{c}{0.20} \\
       \hline
Method & EnbPI &  AdaptCI &    ARIMA & Exp Smoothing & Dynamic Factor & EnbPI &  AdaptCI &    ARIMA & Exp Smoothing & Dynamic Factor & EnbPI &  AdaptCI &    ARIMA & Exp Smoothing & Dynamic Factor & EnbPI &  AdaptCI &    ARIMA & Exp Smoothing & Dynamic Factor \\
\midrule
Coverage &    0.950 &    0.863 &    0.839 &        0.900 &          0.917 &    0.896 &    0.831 &    0.784 &        0.868 &          0.887 &    0.846 &    0.806 &    0.743 &        0.852 &          0.855 &    0.798 &    0.776 &    0.711 &        0.840 &          0.832 \\
Width    &  288.581 &  215.258 &  158.581 &      351.181 &        262.006 &  216.989 &  187.504 &  135.404 &      313.185 &        229.151 &  178.140 &  173.079 &  119.870 &      288.428 &        206.448 &  147.297 &  154.322 &  107.652 &      269.379 &        187.840 \\
        \bottomrule
    \end{tabular}
    }
\end{table*}

\magenta{\begin{remark}[Computational challenges of quantile-based conformal inference methods]
    Quantile regression models aim to predict quantiles of the response distribution accurately and capture the unknown distribution during inference. Such benefit can be reflected in the narrow prediction intervals by quantile-based conformal inference methods \citep{CPquantile,QOOB,Gibbs2021AdaptiveCI}. However, one should be cautious with the following subtle computational concern. 
    
    To fit a quantile regression model, one uses the empirical risk minimization under the following loss, which depends on the quantile $\alpha$ and the sign of the residual $\hat{\epsilon}_i:=y_i-\hat{f}(x_i)$:
\begin{equation}\label{quantile_loss}
    \mathcal{L}(\hat{\epsilon}_i,\alpha)=
    \begin{cases}
    \alpha \hat{\epsilon}_i & \text{if } \hat{\epsilon}_i\geq 0,\\
    (\alpha-1) \hat{\epsilon}_i & \text{if } \hat{\epsilon}_i< 0.
    \end{cases}
\end{equation}
Therefore, producing intervals at different desired $1-\alpha$ coverage levels requires fitting the baseline algorithm $\mathcal{A}$ inside a quantile-based conformal method multiple times. 

In comparison, \EnbPI \ trains the LOO estimators only once to compute all LOO residuals, during which one needs not to specify the desired $\alpha$ value (see Algorithm \ref{algo:hypo-test-ensemble}, line 1-10). Then, constructing intervals at a particular $1-\alpha$ only requires making a point prediction using fitted LOO estimators and evaluating the empirical quantiles of LOO residuals. The whole procedure is computationally efficient when different target coverage levels are specified.
\end{remark}
}
 
\subsection{Real-data: Missing data, conditional coverage} \label{expr:second_part}

In this section, we move beyond marginal coverage with two particular goals. Firstly, we aim to show conditional validity of \EnbPI \ as it looks ahead beyond one step to construct multiple prediction intervals before receiving feedback (that is, $s>1$). Secondly, we show that \EnbPI \ can handle time series with missing data, which commonly exist in reality. We compare \EnbPI \ against QOOB and AdaptCI in this setting.

\vspace{0.1in}
\noindent\textit{Setup.} \magenta{The same setup applies to all three conformal inference methods, so we only describe the general setup.} All hyperparameters except choices of $s$ are kept the same unless otherwise specified. We fit each CP method separately on subsets of hourly data, given that radiation data exhibit significant periodic variations (for example, recordings near noon have much larger magnitudes than the rest). More precisely, we fit each CP method once on data between 10 AM ---2 PM and once on data from the rest 5 hours. Then, we let $s=5$ hours, so \EnbPI \ constructs five-hour ahead prediction intervals every day, after which the conditional coverage is computed separately at each hour. To create a more challenging missing data situation, we randomly drop 25$\%$ of both training and test data. As $X_t$ may contain the history of $Y_t$ for prediction, we impute missing entries as independent random samples from a normal distribution, whose mean and variance parameters are empirical mean and standard error of the most recent $s$ observations. We assume exogenous features (temperature, humidity, wind speed, etc.) are readily available and perform no imputation on them. The training data come from the first 92 days of observation (January-March), and intervals always lie within $[0,\infty)$, as solar radiation value cannot be negative. For clarity, we only show results under one typical trial.

\begin{figure}[!t]
    \centering
    \begin{minipage}[t]{\linewidth}
        \includegraphics[width=\textwidth]{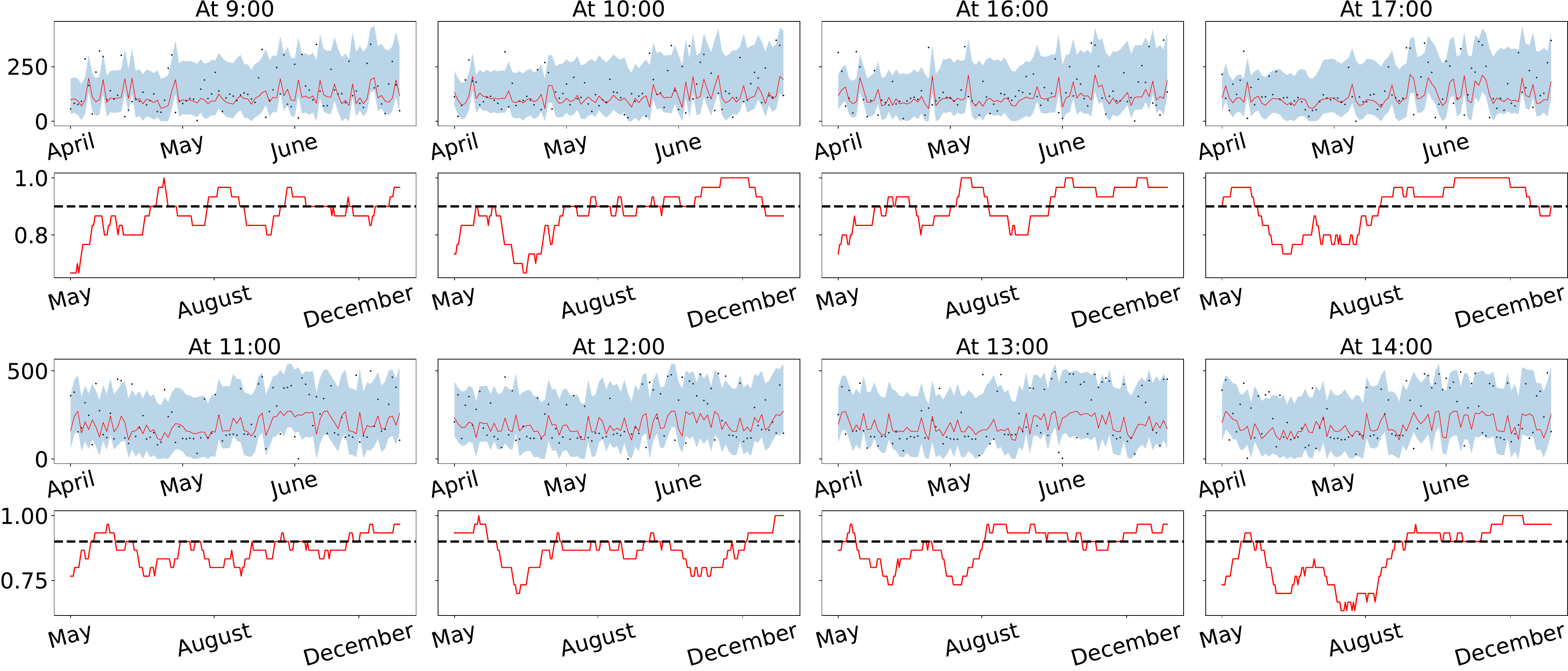}
    \end{minipage}
    \cprotect \caption{Solar power prediction in Atlanta, when \EnbPI{} looks ahead beyond one step. At each hour (i.e., a two-row subfigure), the top figure visualizes observations in black, estimates in red, and prediction intervals in blue for three months (April-June). The bottom subfigures compute coverage using a sliding window of 30 days. The sliding coverage is much poorer near summertime (for example, August), when the data distribution may differ. Conditional coverage at each hour is always near 0.9 (cf. Table \ref{tab:cond_cov_QOOB_AdaptiveCI}).}
    \label{fig:plts_daily_slide_nn}
\end{figure}

\vspace{0.1in}
\noindent\textit{Results. } 
Figure \ref{fig:plts_daily_slide_nn} shows conditional coverage of \EnbPI \ under RF. We title each subfigure by the hour, in which the bottom row visualizes the coverage over a sliding window to illustrate how \EnbPI \ performance evolves. Several things are noticeable. 
Firstly, despite not being shown, empirical distributions of LOO residuals in the rightmost figures are asymmetric around 0, justifying the need to build asymmetric intervals in \EnbPI.
Secondly, \EnbPI \ can nearly obtain conditional coverage at all these hours (see the first row of Table \ref{tab:cond_cov_QOOB_AdaptiveCI}) even with missing data. We note that the sliding coverage can be much poorer near the summer (for example, in August), likely because radiation data near the summer experience unknown shifts in the model $f$ and violate our assumption for the data-generating process. Lastly, applying \EnbPI \ separately onto group training data that are more ``similar'' (for example, by morning and afternoon) can be essential, especially when the data-generating processes are heterogeneous over subgroups. 
In general, we believe \EnbPI \ can obtain conditionally valid coverage on real data even in missing data. In Appendix \ref{app_expr:add_results}, we show more results when no feedback is available to \EnbPI \ (that is, $s=\infty$), illustrating the necessity to slide past residuals for a dynamic interval calibration. Table \ref{tab:cond_cov_QOOB_AdaptiveCI} \rev{in Appendix \ref{app_expr:add_results}} reports the conditional coverage and width for \EnbPI, QOOB, and AdaptCI. We see that QOOB can lose coverage at all hours, but AdaptCI can maintain conditional validity. In particular, AdaptCI prediction intervals for radiation levels in the morning are almost identical in width to those by \EnbPI. However, those for radiation levels in the afternoon are wider than those by \EnbPI. In Appendix \ref{app_expr:add_results}, we also visualize the sliding coverage and prediction intervals by QOOB and AdaptCI as in Figure \ref{fig:plts_daily_slide_nn} for \EnbPI.

\subsection{\rev{Real data: Unsupervised anomaly detection}} \label{sec:anomaly_detection}

In this section, we use \EnbPI \ to detect anomalies in traffic flow observations with missing data. In this setting, it is important to dynamically update decision thresholds (for example, upper and lower ends of prediction intervals) based on spatial and temporal information in the traffic sensor network because traffic data are highly correlated and non-stationary. Data description, setup, and comparison methods are described in Appendix \ref{append:anomaly_detection}

\begin{figure}[!b]
    \centering    
    \includegraphics[width=\linewidth]{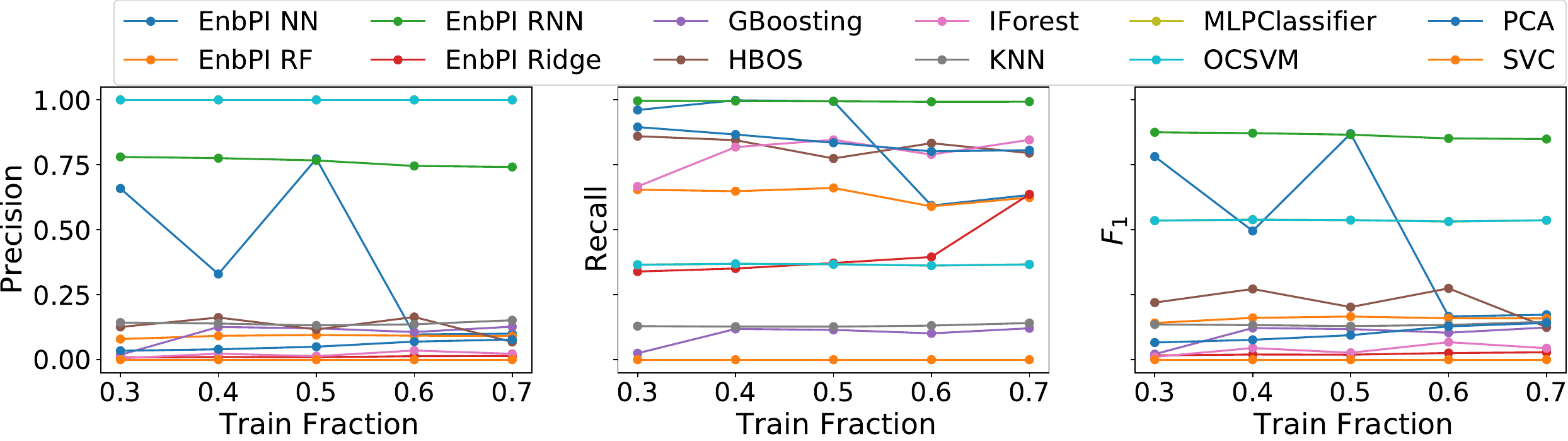}
    \vspace{-0.4cm}
    \cprotect\caption{Traffic flow anomaly detection. Precision, Recall, and $F_1$ scores versus different amounts of training data (as percentages of total data) for different detectors. \EnbPI \ under RNN and NN outperforms the other methods.}
    \label{fig:kaggle}
    \vspace{-0.4cm}
\end{figure}

\vspace{0.1in}
\noindent {\it Results.} Figure \ref{fig:kaggle} compares all methods on a particular traffic sensor as we vary the size of training data. It is clear that \EnbPI \ consistently obtains the highest $F_1$ scores when RNN is used as the prediction model; $F_1$ scores by \EnbPI \ also are consistent across over training sample sizes. In addition, Table \ref{tab:f1_tab} shows the results with more sensors, from which \EnbPI \ under NN or RNN still outperforms the other competitors by a large margin. In the future, we will consider multiple testing corrections to improve the performance \citep{candes_anomaly,contextual-fdr,LORD++}, where the critical step is to examine the dependency of $p$-value as a correction step.

\begin{table*}[!t]
\cprotect\caption{Traffic flow anomaly detection. $F_1$ scores, Precision, and Recall by 12 methods on selected sensors. Bold cells indicate the highest scores. \EnbPI \ RNN or NN are better on this task in terms of $F_1$ scores.}
\label{tab:f1_tab}
\centering
\resizebox{0.9\linewidth}{!}{%
\begin{tabular}{ccccc|cccccccc}
\multicolumn{13}{c}{$F_1$ score}\\
\hline
Sensor ID  & EnbPI  Ridge & EnbPI  RF   & EnbPI  NN   & EnbPI  RNN  & HBOS & IForest & OCSVM & PCA  & SVC & GBoosting & KNN  & MLPClassifier \\
282 & 0.13  & 0.14 & 0.88 & \textbf{0.88} & 0.16 & 0.02    & 0.51  & 0.09 & 0.0 & 0.04      & 0.07 & 0.51          \\
248 & 0.02  & 0.17 & 0.87 &\textbf{0.87} & 0.20 & 0.03    & 0.54  & 0.09 & 0.0 & 0.12      & 0.13 & 0.54          \\
151 & 0.02  & 0.14 & \textbf{0.81} & 0.80 & 0.11 & 0.04    & 0.39  & 0.08 & 0.0 & 0.08      & 0.12 & 0.39          \\
235 & 0.57  & 0.59 & 0.77 & \textbf{0.77} & 0.01 & 0.00    & 0.45  & 0.00 & 0.0 & 0.23      & 0.24 & 0.45          \\
\bottomrule
\end{tabular}%
}
\vspace{0.05in}

\centering
\resizebox{0.9\linewidth}{!}{%
\begin{tabular}{ccccc|cccccccc}
\multicolumn{13}{c}{Precision}\\
\hline
Sensor ID  & EnbPI  Ridge & EnbPI  RF   & EnbPI  NN   & EnbPI  RNN  & HBOS & IForest & OCSVM & PCA  & SVC & GBoosting & KNN  & MLPClassifier \\
282 & 0.46  & 0.59 & 0.96 & \textbf{0.96} & 0.58 & 0.71    & 0.34  & 0.75 & 0.0 & 0.04      & 0.07 & 0.34          \\
248 & 0.37  & 0.66 & 0.99 &\textbf{0.99} & 0.77 & 0.85    & 0.37  & 0.84 & 0.0 & 0.11      & 0.13 & 0.37          \\
151 & 0.24  & 0.61 & 0.96 & \textbf{0.96} & 0.30 & 0.47    & 0.24  & 0.46 & 0.0 & 0.08      & 0.11 & 0.24          \\
235 & 0.60  & 0.60 & 0.70 & \textbf{0.70} & 0.04 & 0.03    & 0.29  & 0.00 & 0.0 & 0.23      & 0.24 & 0.29          \\
\bottomrule
\end{tabular}%
}
\vspace{0.05in}

\centering
\resizebox{0.9\linewidth}{!}{%
\begin{tabular}{ccccc|cccccccc}
\multicolumn{13}{c}{Recall}\\
\hline
Sensor ID  & EnbPI  Ridge & EnbPI  RF   & EnbPI  NN   & EnbPI  RNN  & HBOS & IForest & OCSVM & PCA  & SVC & GBoosting & KNN  & MLPClassifier \\
282   & 0.07 & 0.08 & 0.81 & 0.81 & 0.10    & 0.01  & \textbf{1.0} & 0.05 & 0.0       & 0.04 & 0.07          & 1.0 \\
248   & 0.01 & 0.09 & 0.77 & 0.77 & 0.12    & 0.01  & \textbf{1.0} & 0.05 & 0.0       & 0.12 & 0.13          & 1.0 \\
151   & 0.01 & 0.08 & 0.69 & 0.68 & 0.07    & 0.02  & \textbf{1.0} & 0.04 & 0.0       & 0.09 & 0.12          & 1.0 \\
235   & 0.55 & 0.59 & 0.87 & 0.87 & 0.01    & 0.00  & \textbf{1.0} & 0.00 & 0.0       & 0.24 & 0.24          & 1.0 \\
\bottomrule
\end{tabular}%
}
\vspace{-0.3cm}
\end{table*}

\section{\texttt{EnbPI} under change points}\label{sec:simul_changepts}

In real applications, there can exist abrupt changes in the underlying data distribution, which are called {\it change points} \citep{xie2021sequential,aminikhanghahi2017survey}. In this section, we present numerical experiments to demonstrate the performance of  \EnbPI \ in the presence of change points. We also discuss the potential adaption of \EnbPI \ for change point detection.

We conider a change point happening during the testing phase and follow the setup in Section \ref{exp:simul}. Assume a change point at $T^*=0.6(T+T_1)$, which alters the underlying model $f$ for the last 40$\%$ test data. As a result, the post-change responses $Y_t$ are very different from the pre-change ones. We call the post-change model $f_1$. For the linear model, let $f_1(X_t)=\beta_1 X_t$ and $\beta_1$ be entry-wise {\it i.i.d.} $U[0,5]$. Recall the pre-change $\beta$ is entry-wise {\it i.i.d.} $U[0,1]$. For the high-dimensional sparse linear model, $\beta_1$ has twice many non-zero components as that of $\beta$ and the components are drawn from $U[0,1]$ independently. For the nonlinear model, we keep the same $\beta$ but square the value $f(X_t)$. Choices of $X_t$ and $\epsilon_t$ remain the same in each case.

Recall $T$ is the length of the pre-change training data; let $T=0.3(T+T_1)=600$. To adapt to post-change dynamics as quickly as possible, we retrain the prediction algorithm on $0.1T$ data after the change point $T^*$. We assume the $T^*$ is known to us (for instance, we can be detected and estimated using a change point detection algorithm \citep{xie2021sequential}). To quickly detect change points that highly correlate with differences in interval widths, we only take the empirical quantile of the most recent $T', T'<T$ residuals and fix $T'=100$.

\begin{figure}[!t]
    \centering
    \begin{minipage}[t]{0.32\linewidth}
        \includegraphics[width=\textwidth]{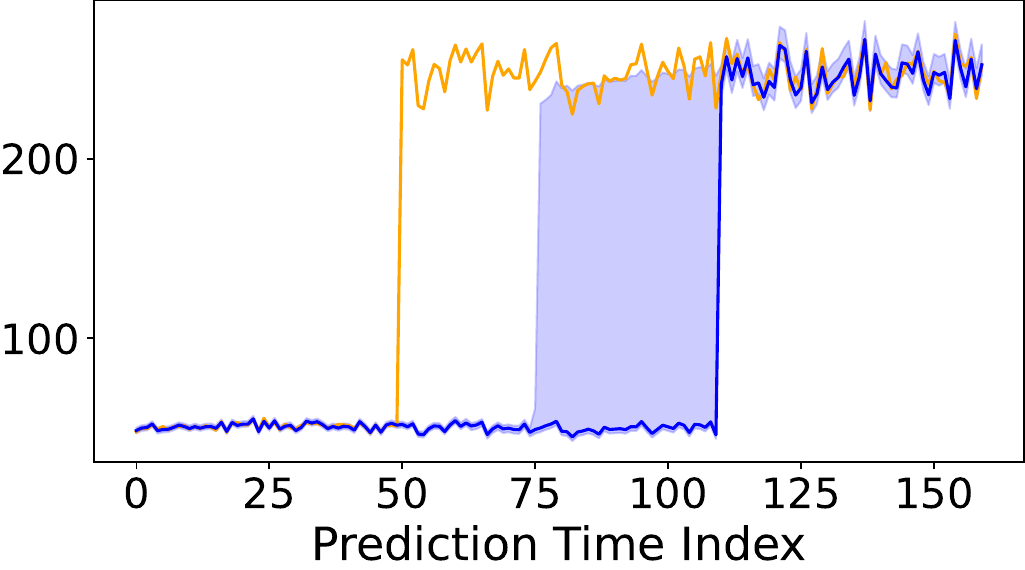}
        \subcaption{Case 1, EnbPI}
    \end{minipage}
    \begin{minipage}[t]{0.32\linewidth}
        \includegraphics[width=\textwidth]{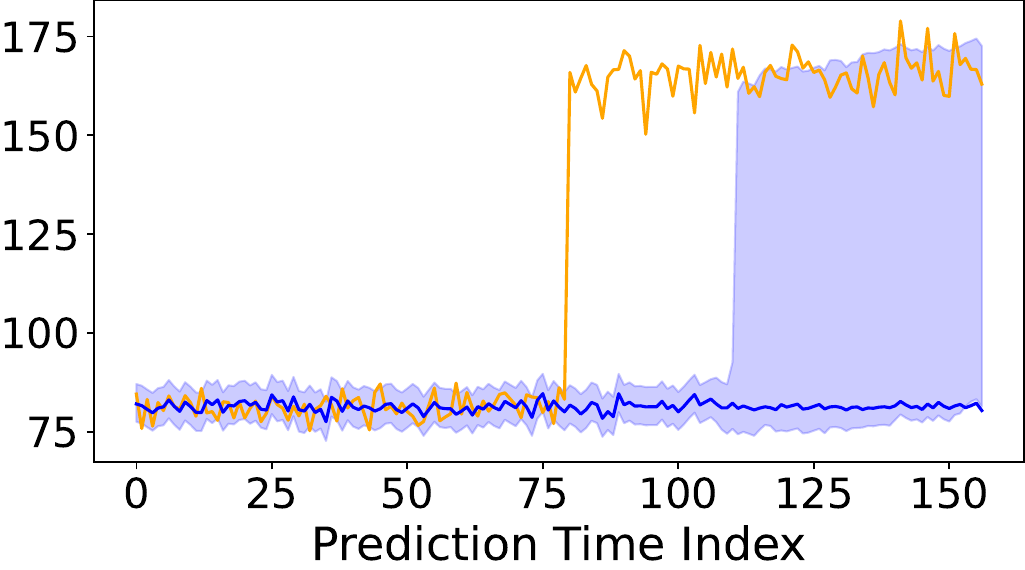}
        \subcaption{Case 2, EnbPI}
    \end{minipage}
    \begin{minipage}[t]{0.32\linewidth}
        \includegraphics[width=\textwidth]{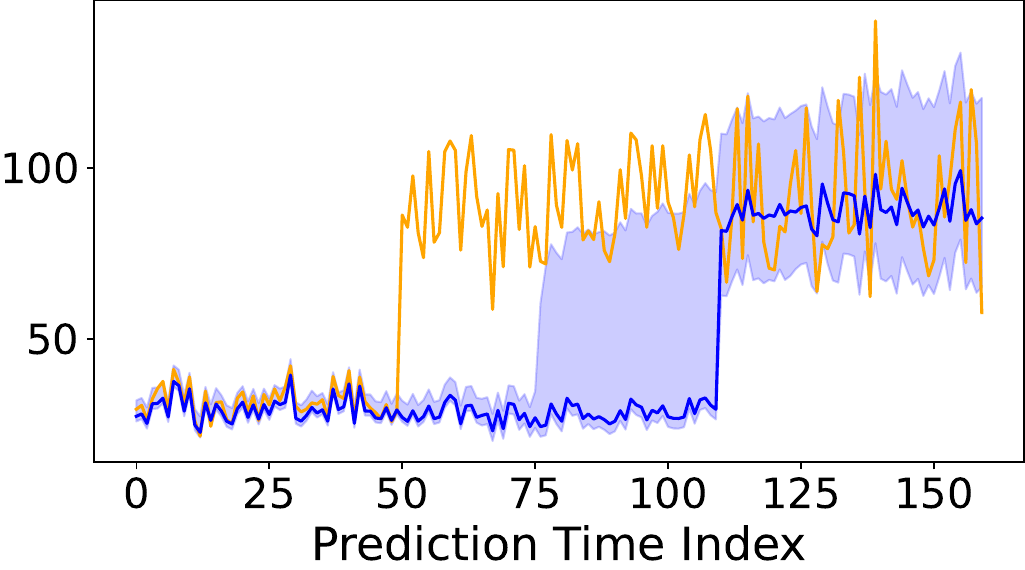}
        \subcaption{Case 3, EnbPI}
    \end{minipage}
    \begin{minipage}[t]{0.32\linewidth}
        \includegraphics[width=\textwidth]{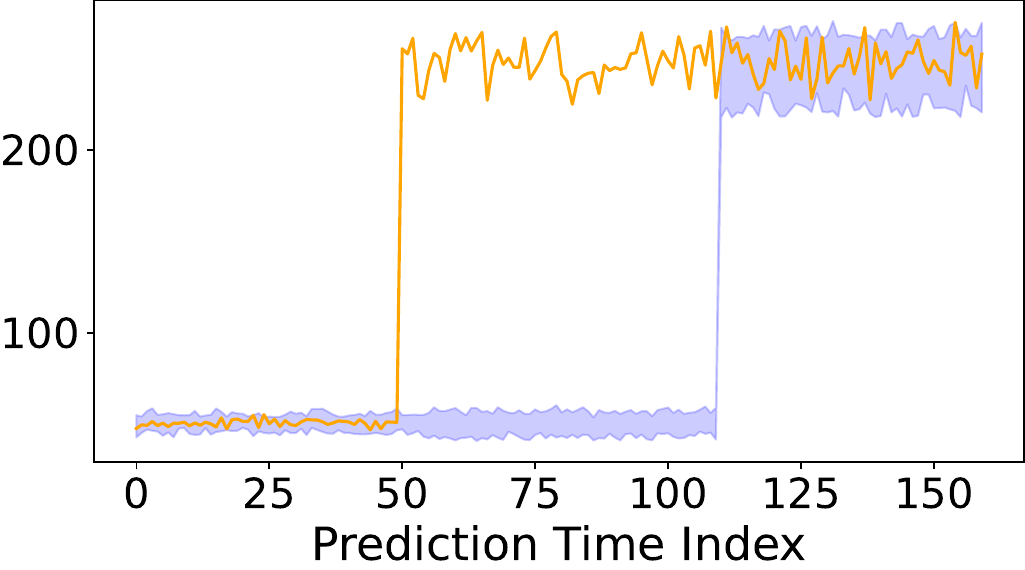}
        \subcaption{\rev{Case 1, AdaptCI}}
    \end{minipage}
    \begin{minipage}[t]{0.32\linewidth}
        \includegraphics[width=\textwidth]{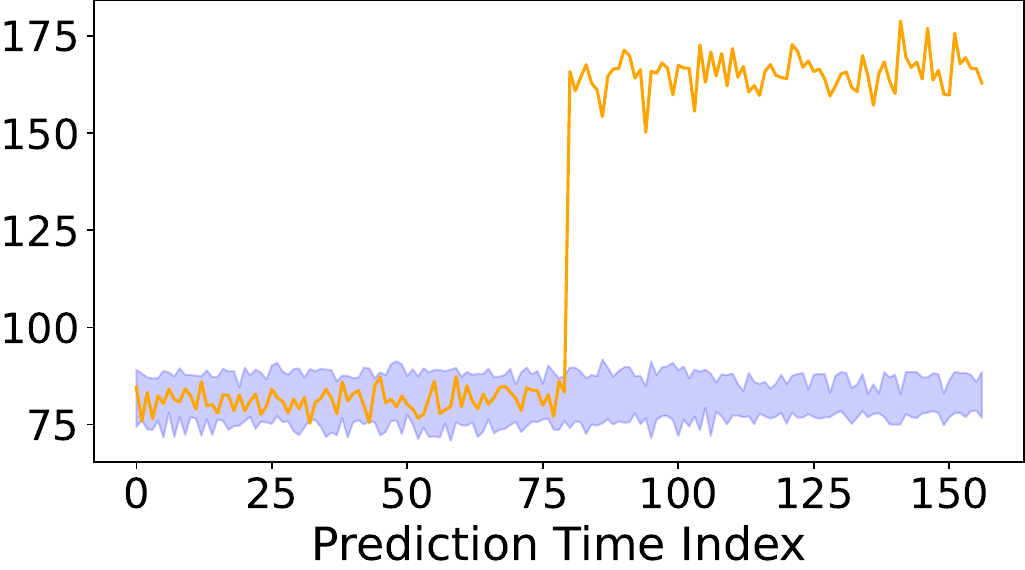}
        \subcaption{\rev{Case 2, AdaptCI}}
    \end{minipage}
    \begin{minipage}[t]{0.32\linewidth}
        \includegraphics[width=\textwidth]{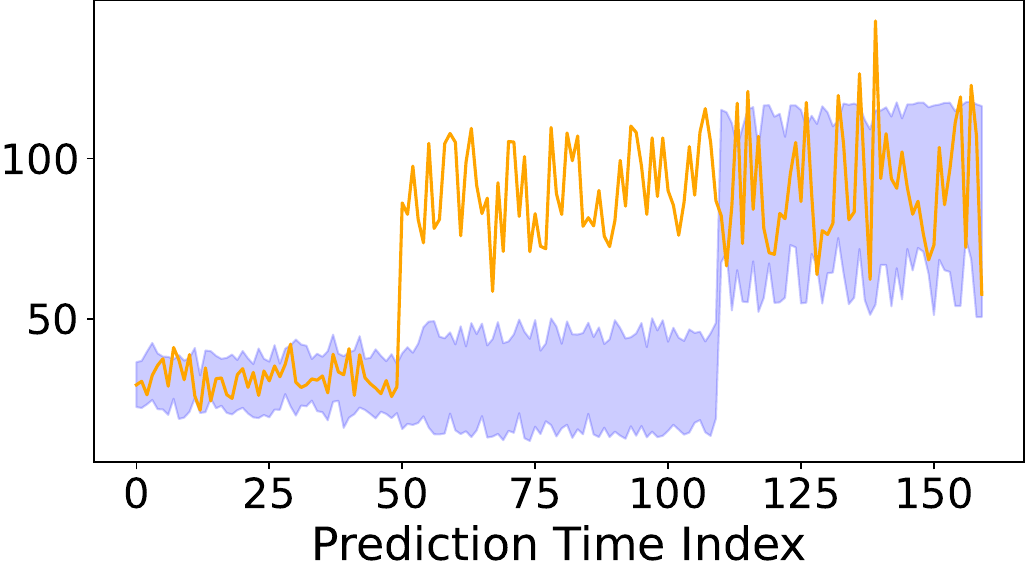}
        \subcaption{\rev{Case 3, AdaptCI}}
    \end{minipage}
    \cprotect \caption{Simulation with a change point at index 50. We overlay prediction intervals in shaded blue on top of the actual data. In particular, we collect 60 post-change data points to refit $\mathcal{A}$ at index 110. We expect that collecting more post-change data to fit \EnbPI \ will yet better estimation with tighter prediction intervals.}
    \label{fig:simul_changepts_result}
    \vspace{-0.3cm}
\end{figure}

Figure \ref{fig:simul_changepts_result} plots prediction intervals on top of actual data for three cases. Firstly, except for data indexed between $T^*$ and $T^*+0.1T$ (that is, between index 50 and 110 in the figure), most prediction data from both pre-change and post-change models are covered by \EnbPI \ intervals. Secondly, prediction intervals built with pre-change models on post-change data tend to have much wider widths than others, reflecting a poor estimation of $\hat f$ by the pre-change models. Nevertheless, such a dramatic increase in width can enable change point detection, as we elaborate on below. \rev{Thirdly, we observe that AdaptCI intervals are non-adaptive in this setting, as they fail to contain the true observations before retraining the predictive model. In Figure \ref{fig:simul_changepts_result_ETS}, we further compare \EnbPI{} with the ETS model \citep{Hyndman2013ForecastingPA}, which shows similar performance as AdaptCI.}

One can potentially adapt \EnbPI \ to detect change points as follows. From Figure \ref{fig:simul_changepts_result}, we observe that the change point leads to unusually wide post-change prediction intervals. As a result, one should monitor \textit{both} the evolution of interval widths \textit{and} coverage performances. On the one hand, when only $f$ changes but the distribution of errors remains the same, the interval tends to be wider, but the coverage is worse. On the other hand, if $f$ remains the same but the distribution changes, intervals may also become wider. However, coverage may not be as greatly affected because estimators by \EnbPI \ can approximate $f$ well. Due to a sliding window over residuals, one can adapt to the post-change distribution. These ideas resonate with several other works: \citep{Gibbs2021AdaptiveCI} construct prediction sets under distribution shifts sequentially and prove that when shifts are small, the marginal coverage is approximately maintained. As a result, when coverage is significantly less than $1-\alpha$, it can indicate an abrupt shift in distribution. Such ideas may also be used to test whether the test distribution lies in an $f$-divergence ball of the training distribution, given i.i.d. training and test data from the corresponding distribution \citep{CProbust}; extensions to time series remain unexplored. On the other hand, a line of works \citep{a6e703ae932a433c8ceedb6ae3131ef7,vovk2021conformal,vovk2021retrain} builds martingales to detect change points which however, violates data exchangeability. The lower bound for the average-run-length is established for the Shiryaev–Roberts procedure using such martingale \cite[Proposition 4.1]{a6e703ae932a433c8ceedb6ae3131ef7}. How to extend the ideas beyond testing exchangeable data remains an open question.

\section{Conclusions and Discussions} \label{sec:conclu}

In this paper, we present a predictive inference method for time series. Theoretically, we can show that the constructed intervals are asymptotically valid without assuming data exchangeability: relaxing this requirement is crucial for time series data, and the interval width converges to the oracle one.  We also present a simple, computationally friendly, and interpretable algorithm called \Verb|EnbPI|, which is an efficient ensemble-based wrapper for many prediction algorithms, including deep neural networks. Empirically, it works well on time series from various applications, including network data and data with missing entries, and maintains validity when other predictive inference methods fail. Furthermore, one can use \Verb|EnbPI| for unsupervised sequential anomaly detection. While the theoretical guarantee of \Verb|EnbPI| requires consistent estimation of the true model, empirical results are valid even under potentially misspecified models, and coverage is almost always valid.

Future work includes several possible directions. We may adapt \EnbPI \ for classification problems \citep{MJ_classification,Candes_classification,xuERPAS2022} by defining conformity scores other than residuals. It can also be interesting to further develop \EnbPI\  for online change point detection and adaptation for time series, extending the idea of sequential testing of data exchangeability \citep{vovkinductive} based on the Shiryaev-Roberts procedure. 

\section*{Acknowledgments}
The work is partially supported by NSF CAREER CCF-1650913, CMMI-2015787, DMS-2134037, DMS-1938106, and DMS-1830210. \rev{The method presented in this paper has been implemented in open-source packages MAPIE \citep{taquet2022mapie} and Fortuna \citep{fortuna2022}.}
\bibliographystyle{plainnat}
\bibliography{icml_paper}
\appendix
\section{Proofs }\label{app_theory:approx_uni_proof}
For notation simplicity, we remove subscripts $T+1$ for $\hat{F}, \tilde{F},$ and $F$.

\begin{proof}[Proof of Lemma \ref{regu_lem}] 
When the error process is i.i.d., the famous Dvoretzky–Kiefer–Wolfowitz inequality \cite[p.210]{iidbound} implies that 
\[
\mathbb P(\sup_{x} |\tilde{F}(x)-F(x)| > s_T) \leq 2e^{-2Ts_T^2}.
\]
Pick $s_T=\frac{\sqrt{W(16T)}}{2\sqrt{T}}$, where $W(T)$ is the Lambert $W$ function that satisfies $W(T)e^{W(T)}=T$. We see that $s_T\leq \sqrt{\log(16T)/T}$. Thus, define the event $A_T$ on which $\sup_x |\tilde{F}(x)-F(x)|\leq \sqrt{\log(16T)/T}$, whereby we have 
\begin{align*}
     \sup_{x} |\tilde{F}(x)-F(x)|\longbar A_T & \leq \sqrt{\log(16T)/T}\\
     P(A_T) & > 1-\sqrt{\log(16T)/T}
\end{align*}

\end{proof}

\begin{proof}[Proof of Lemma \ref{consis_lem}]
Note that Assumption \ref{consis} is equivalent to requiring $\sum_{i=1}^T(\hat \epsilon_i-\epsilon_i)^2\leq \delta_T^2$. Thus, let $S:=\{i\in [T] : |\hat \epsilon_i-\epsilon_i|\geq \delta_T^{2/3}\}$. It follows that 

$$
|S|\delta_T^{4/3} \leq \sum_{i=1}^T (\hat \epsilon_i-\epsilon_i)^2 \leq T\delta_T^2,
$$
where the second inequality follows by Assumption \ref{consis}. As a result, $|S|\leq T\delta_T^{2/3}$ and we see that for any $x \in \mathbb{R}$,
\begin{align}
    & |\hat{F}(x)-\tilde{F}(x)| 
     \leq \frac{1}{T} \sum_i |\textbf{1}\{\hat{\epsilon}_i \leq x\}-\textbf{1}\{\epsilon_i \leq x\}| \nonumber\\
     \overset{(i)}{\leq} & \frac{1}{T} (|S|+\sum_{i\notin S} |\textbf{1}\{\hat{\epsilon}_i \leq x\}-\textbf{1}\{\epsilon_i \leq x\}|) \nonumber\\
     \overset{(ii)}{\leq}&  \frac{1}{T} (|S|+\sum_{i\notin S} \textbf{1}\{|\epsilon_i - x| \leq \delta_T^{2/3}\})\nonumber\\
     \leq & \delta_T^{2/3} + \mathbb P(|\epsilon_{T+1}-x|\leq \delta_T^{2/3}) + \nonumber\\ \qquad & \sup_{x}|\frac{1}{T}\sum_{i\notin S} \textbf{1}\{|\epsilon_i - x| \leq \delta_T^{2/3}\} -\mathbb P(|\epsilon_{T+1}-x|\leq \delta_T^{2/3}) |\nonumber\\
     \overset{(iii)}{=} & \delta_T^{2/3} + [F(x+\delta_T^{2/3})-F(x-\delta_T^{2/3})] + 
     \sup_{x}|[\tilde{F}(x+\delta_T^{2/3})\nonumber\\  
     \qquad & - \tilde{F}(x-\delta_T^{2/3})]-[F(x+\delta_T^{2/3})-F(x-\delta_T^{2/3})]|\nonumber\\
     \leq & (L_{T+1}+1)\delta_T^{2/3}+2\sup_x |\tilde{F}(x)-F(x)|. \label{close2}
\end{align}

We remark that (i) follows as we analyze $i\in S$ and $i \notin S$ separately, (ii) follows since $|\textbf{1}\{a\leq x\}-\textbf{1}\{b\leq x\}|\leq \textbf{1}\{ |b-x| \leq |a-b|\}$ for any constant $a,b$ and univariate $x$ and $|\hat{\epsilon}_i - \epsilon_i|\leq \delta_T^{2/3}$ for $i\notin S$, and (iii) follows since we assume $\{\epsilon_t\}_{t=1}^T$ have the common CDF $F$.

\end{proof} 

\begin{proof}[Proof of Theorem \ref{thm:cond_cov}]
Recall the following definitions:
\begin{itemize}[noitemsep,topsep=1pt]
    \item $\hat{p}_{T+1}:=\frac{1}{T}\sum_i \textbf{1}\{\hat{\epsilon}_i \leq \hat{\epsilon}_{T+1}\}$, which is the empirical $p$-value defined using residuals.
    \item $\tilde{F}(x):=\frac{1}{T}\sum_{i=1}^T \textbf{1}\{\epsilon_i\leq x\}$ as the empirical CDF of $F$. Equivalently define $\hat{F}(x)$ using prediction residuals $\hat{\epsilon}_i$.
\end{itemize}
As a consequence, for any $\beta \in [0,\alpha]$, the following are equivalent:
\begin{align} 
    & |\mathbb{P}(Y_{T+1}\in \widehat{C}^{\alpha}_{T+1}|X_{T+1}=x_{T+1})-(1-\alpha)| \nonumber \\ 
    \overset{(i)}{=} & |\mathbb{P}(\beta\leq\hat{p}_{T+1}\leq 1-\alpha+\beta)-(1-\alpha)| \nonumber\\
    = & |\mathbb{P}(\beta \leq \hat{F}(\hat{\epsilon}_{T+1})\leq 1-\alpha+\beta)-\nonumber \\ \qquad &\mathbb{P}(\beta \leq F(\epsilon_{T+1})\leq 1-\alpha+\beta)|, \label{alter}
\end{align}
where (i) follows by the equivalence we pointed out at the beginning of Section \ref{sec:cov_guarantee}.

We can rewrite the right hand side of (\ref{alter}) as follows:
\begin{align}
    & |\mathbb{P}(\beta \leq \hat{F}(\hat{\epsilon}_{T+1})\leq 1-\alpha+\beta)-\nonumber \\ \qquad &\mathbb{P}(\beta \leq F(\epsilon_{T+1})\leq 1-\alpha+\beta)| \nonumber\\
     \leq &\mathbb E|\textbf{1}\{\beta \leq \hat{F}(\hat{\epsilon}_{T+1})\leq 1-\alpha+\beta\} - \nonumber \\ \qquad & \textbf{1}\{\beta \leq F(\epsilon_{T+1})\leq 1-\alpha+\beta\}|\nonumber\\
     \overset{(i)}{\leq} &\mathbb E(|\textbf{1}\{\beta \leq \hat{F}(\hat{\epsilon}_{T+1})\}-\textbf{1}\{\beta \leq F(\epsilon_{T+1})\}|+\nonumber\\ \qquad &|\textbf{1}\{ \hat{F}(\hat{\epsilon}_{T+1})\leq 1-\alpha+\beta\}-\textbf{1}\{ F(\epsilon_{T+1})\leq 1-\alpha+\beta\}|) \nonumber\\
     \overset{(ii)}{\leq} &\mathbb P(|F(\epsilon_{T+1})-\beta| \leq |\hat{F}(\hat{\epsilon}_{T+1})-F(\epsilon_{T+1})|)+\nonumber \\ \qquad &\mathbb P(|F(\epsilon_{T+1})-(1-\alpha+\beta)| \leq |\hat{F}(\hat{\epsilon}_{T+1})-F(\epsilon_{T+1})|)\nonumber,
\end{align}
where inequality (i) follows since for any constants $a,b$ and univariates $x,y$, $|\textbf{1}\{a \leq x \leq b\} - \textbf{1}\{a \leq y \leq b\}|\leq |\textbf{1}\{a \leq x\}-\textbf{1}\{a \leq y\}|+|\textbf{1}\{x \leq b\}-\textbf{1}\{y \leq b\}|$. Moreover, inequality (ii) follows since $|\textbf{1}\{a\leq x\}-\textbf{1}\{b\leq x\}|\leq \textbf{1}\{ |b-x| \leq |a-b|\}$ for any constant $a,b$ and univariate $x$ and $\mathbb E[\textbf{1}\{A\}]=\mathbb P(A)$. 

Recall in Lemma \ref{regu_lem}, we defined $A_T$ as the event on which 
\[
\sup_{x} |\tilde{F}(x)-F(x)|\longbar A_T \leq \sqrt{\log (16T)/T},
\]
where $\mathbb P(A_T)>1- \sqrt{\log (16T)/T}$. Let $A_T^C$ denote the complement of the event $A_T$. For any $\gamma \in [0,1]$, we have that
\begin{align*}
    & \mathbb P(|F(\epsilon_{T+1})-
    \gamma| \leq |\hat{F}(\hat{\epsilon}_{T+1})-F(\epsilon_{T+1})|) \\ 
     \leq &\mathbb P(|F(\epsilon_{T+1})-
    \gamma| \leq |\hat{F}(\hat{\epsilon}_{T+1})-F(\epsilon_{T+1})| \cap A_T) + \mathbb P(A_T^C) \\
     \leq &\mathbb P(|F(\epsilon_{T+1})-
    \gamma| \leq |\hat{F}(\hat{\epsilon}_{T+1})-F(\epsilon_{T+1})| \longbar A_T) + \mathbb P(A_T^C) \\ 
     \leq &\mathbb P(|F(\epsilon_{T+1})-
    \gamma| \leq |\hat{F}(\hat{\epsilon}_{T+1})-F(\hat{\epsilon}_{T+1})|+\nonumber \\ \qquad &|F(\hat{\epsilon}_{T+1})-F(\epsilon_{T+1})| \longbar A_T) + \sqrt{\log (16T)/T}.
\end{align*}
To bound the conditional probability above, we note that conditioning on the event $A_T$, 
\begin{align*}
    & |\hat{F}(\hat{\epsilon}_{T+1})-F(\hat{\epsilon}_{T+1})|+|F(\hat{\epsilon}_{T+1})-F(\epsilon_{T+1})| \longbar A_T\\
     \leq& \sup_x |\hat{F}(x)-F(x)| \longbar A_T+L_{T+1}|\hat{\epsilon}_{T+1}-\epsilon_{T+1}|\\
     \leq& \sup_x |\hat{F}(x)-\tilde{F}(x)|\longbar A_T+\sup_x |\tilde{F}(x)-F(x)| \longbar A_T+\nonumber \\ \qquad &L_{T+1}|\hat{\epsilon}_{T+1}-\epsilon_{T+1}|\\
     \overset{(i)}{\leq}& (L_{T+1}+1) \delta_T^{2/3}+3\sup_x |\tilde{F}(x)-F(x)|\longbar A_T +L_{T+1}\delta_T\\
     \leq& 3\sqrt{\log (16T)/T} + (L_{T+1}+1) (\delta_T^{2/3}+\delta_T), \\
\end{align*}
where (i) holds by Lemma \ref{consis_lem} and Assumption \ref{consis}.

Therefore, because $F(\epsilon_{T+1}) \sim \text{Unif}[0,1]$, we have 
\begin{align*}
    & \mathbb P(|F(\epsilon_{T+1})-
    \gamma| \leq |\hat{F}(\hat{\epsilon}_{T+1})-F(\hat{\epsilon}_{T+1})|+ \nonumber \\ \qquad &|F(\hat{\epsilon}_{T+1})-F(\epsilon_{T+1})| \longbar A_T) \\ 
     \leq & 6\sqrt{\log (16T)/T} +2(L_{T+1}+1) (\delta_T^{2/3}+\delta_T).
\end{align*}
As a result, by letting $\gamma=\beta$ and $1-\alpha+\beta$, we have
\begin{align*}
    & |\mathbb{P}(Y_{T+1}\in \widehat{C}^{\alpha}_{T+1}|X_{T+1}=x_{T+1})-(1-\alpha)| \\ 
\leq & 12\sqrt{\log (16T)/T} +4(L_{T+1}+1) (\delta_T^{2/3}+\delta_T),
\end{align*}
which is the desired result.
\end{proof}

Based on the proof of Theorem \ref{thm:cond_cov}, note that if for certain sequences $\{s_T\}_{T\geq 1}$ and a function $g$ such that $s_T=g(s_T)$, we have the condition 
\[
    \mathbb P(\sup_{x} |\tilde{F}(x)-F(x)| > s_T)\leq g(s_T),
\]
then the following bound always hold under Assumption \ref{consis}:
\begin{align*}
    & |\mathbb{P}(\beta\leq\hat{p}_{T+1}\leq 1-\alpha+\beta)-(1-\alpha)| \\
    \leq & 12s_T +4(L_{T+1}+1) (\delta_T^{2/3}+\delta_T).
\end{align*} 
As a result, proofs of Collaries to Theorem \ref{thm:cond_cov} reduce to finding $s_T$.

\begin{assumption}[Errors follow Linear Processes]
Assume $\{\epsilon_t\}_{t=1}^{T+1}$ satisfy $\epsilon_t=\sum_{j=1}^{\infty} \delta_j z_{t-j}$ for each $i$, under which $z_{i-j}$ are i.i.d. with finite first absolute moment and $\delta_j$ are bounded in absolute value by some function $g$ such that $\sum_{i=1}^{\infty} ig(i)$ converges.
Moreover, assume $\{\epsilon_t\}_{t=1}^{T+1}$ are distributed according to the common CDF $F$, which is Lipschitz continuous with constant $L_{T+1}>0$. 
\end{assumption}

\begin{proof}[Proof of Corollary \ref{thm:cond_cov_linear}]
Following these assumptions, \citep{stat_linear} proves that $\sup_x |\tilde{F}(x)-F(x)|=\mathcal O(\log T/\sqrt{T})$ \cite[see Theorem 3]{stat_linear}. This guarantee yields the desired result by letting $s_T\in\mathcal O(\log T/\sqrt{T})$.
\end{proof}

\begin{proof}[Proof of Corollary \ref{thm:cond_cov_stronglymixing}]
Define $v_T(x):=\sqrt{T}(\tilde{F}(x)-F(x))$. Then, Proposition 7.1 in \citep{rio2017} shows that 
$$ 
\mathbb E(\sup_x |v_T(x)|^2)\leq (1+4\sum_{k=0}^{T} \alpha_{k})(3+\frac{\log T}{2\log 2})^2,
$$
where $\alpha_{k}$ is the $k^{\rm{th}}$ mixing coefficient. Since we assumed that the coefficients are summable with $\sum_{k\geq 0} \alpha_{k} < M$ (for example, $\alpha_{k} = \mathcal{O}(n^{-s}), s>1$), Markov Inequality shows that
\begin{align*}
    \mathbb P(\sup_z |\tilde{F}(x)-F(x)| \geq s_T)
    &\leq \frac{\mathbb E(\sup_x |v_T(x)|^2/T)}{s_T^2}\\
    &\leq \frac{1+4M}{Ts_T^2}(3+\frac{\log T}{2\log 2})^2.
\end{align*}
Thus, we let \[s_T:=\left (\frac{1+4M}{T}(3+\frac{\log T}{2\log 2})^2\right )^{1/3}\approx \left(\frac{M(\log T)^2}{2T}\right)^{1/3}\] and see that 
\begin{align*}
    & \mathbb{P}\bigbrac{\sup_x |\tilde{F}(x)-F(x)|\leq \left(\frac{M(\log T)^2}{2T}\right)^{1/3}} \\
\geq & 1-\left(\frac{M(\log T)^2}{2T}\right)^{1/3}.
\end{align*}

Hence, the event $A_T$ is chosen so that conditioning on $A_T$,  $\sup_x |\tilde{F}(x)-F(x)|\leq (M/2)^{1/3}(\log T)^{2/3}/T^{1/3}$.
\end{proof}

\begin{proof}[Proof of Theorem \ref{thm:width}] The proof has two parts. Firstly, we explicitly define the inverse empirical CDF $\hat{F}^{-1}$ so that it is Lipschitz continuous with an explicit data-driven Lipschitz continuity constant. Secondly, we break $\Delta(T)$ into a sum of multiple terms and bound each term. 
\vspace{0.1in} 

\noindent \textit{(1) Define $\hat{F}^{-1}$.} Recall that 
\begin{align*}
    & \hat{F}(x):=\frac 1 T\sum_{i=1}^T \textbf{1}\{\hat{\epsilon}_i\leq x\}, x\in \R\\
    & \hat{F}^{-1}(\alpha):=\alpha \text{ quantile of } \{\hat \epsilon_i\}_{i=1}^T, \alpha \in [0,1]
\end{align*}
which are the empirical CDF based on LOO residuals and the inverse empirical CDF. Without loss of generality, assume the residuals are sorted so that if $i<j, \hat{\epsilon}_i \leq \hat{\epsilon}_j$. Consider the discrete version $\hat{F}^{-1}_d$:
\[
\hat{F}^{-1}_d(\alpha):=\min\{i^*\in \{1,\ldots,T\}: T^{-1}\sum_i \textbf{1}\{\hat{\epsilon}_{i^*}\geq \hat{\epsilon}_i\}\geq \alpha\}.
\]
Then, we define $\hat{F}^{-1}$ as follows:
\begin{equation}\label{LipContICDF}
\hat{F}^{-1}(\alpha):=\hat{F}^{-1}_d(\alpha_{-})+(\hat{F}^{-1}_d(\alpha_{+})-\hat{F}^{-1}_d(\alpha_{-}))(\alpha-\alpha_{-}),    
\end{equation}
where $\alpha_{-}:=\max\{1/T,\alpha-\alpha \mod T^{-1}\}, \alpha_{+}:=\alpha_{-}+T^{-1}$. Intuitively, $\alpha_{-}$ (resp. $\alpha_{+}$) is the largest (resp. smallest) integer multiple of $T^{-1}$ that is smaller (resp. larger) than $\alpha$. In this way, we interpolate $\hat{F}^{-1}(\alpha)$ so that it is continuous between each increment of size $T^{-1}$. 

In addition, define $K'_{T+1}:=\underset{j=1,\ldots,T-1}{\max} \hat{\epsilon}_{j+1}-\hat{\epsilon}_j$, which is the largest difference between two consecutive LOO residuals. Based on the definition in \eqref{LipContICDF}, it is easy to see that for any $\alpha_1,\alpha_2 \in [0,1]$, 
\begin{equation}\label{Lipschitz}
    |\hat{F}^{-1}(\alpha_1)-\hat{F}^{-1}(\alpha_2)|\leq K'_{T+1} |\alpha_1-\alpha_2|.
\end{equation}
\vspace{0.1in}

\noindent \textit{(2) Bound $\Delta(T)$.} Call $\betaRealBin$ the estimator of $\betaReal$ after line-search with $m$ grids. By the form of $C_{T+1}^{\alpha}$ and $\widehat{C}^{\alpha}_{T+1}$, we have
\begin{align*}
   \Delta(T)
\leq 
& |f(X_{T+1})-\hat{f}_{-(T+1)}(X_{T+1})|
+ \nonumber \\ \qquad & 2 (\overbrace{|\hat{F}^{-1}(1-\alpha+\betaRealBin)- \hat{F}^{-1}(1-\alpha+\betaReal)|}^{(i)}
+ \\
& \overbrace{|\hat{F}^{-1}(1-\alpha+\betaReal)-F^{-1}(1-\alpha+\betaReal)|}^{(ii)}
+
\nonumber \\ \qquad & \overbrace{|F^{-1}(1-\alpha+\betaReal)-F^{-1}(1-\alpha+\betaOracle)|}^{(iii)}),
\end{align*}
where (i) can be bounded under accurate binning with the Lipschitz continuity condition on the inverse empirical CDF $\hat{F}^{-1}$, (ii) is a consequence of Theorem \ref{thm:cond_cov}, and (iii) applies the bound of (ii) twice.

\vspace{0.1in}

\textit{Proof of (i).} Partition $[0,\alpha]$ into $m+1$ equally spaced values $0,\alpha/m,2\alpha/m,\ldots,\alpha$. It is clear that the estimator $\betaRealBin$ upon we iterating through these $m+1$ values satisfies
\[
|\betaRealBin-\betaReal|
\leq \alpha/2m.
\]
Therefore, the Lipschitz continuity of $\hat{F}^{-1}$ as in \eqref{Lipschitz} guarantees that 
\begin{equation*}
    |\hat{F}^{-1}(1-\alpha+\betaRealBin)- \hat{F}^{-1}(1-\alpha+\betaReal)|\leq \alpha K'_{T+1}/2m.
\end{equation*}

\vspace{0.1in}

\textit{Proof of (ii).} We provide the rate of convergence of $\underset{\alpha \in [0,1]}{\sup} |F^{-1}(\alpha)-\hat{F}^{-1}(\alpha)|$ to zero. For any $y\in [0,1]$, let $x:=\hat{F}^{-1}(y)$. We now have
\begin{align*}
    |F^{-1}(y)-\hat{F}^{-1}(y)| 
    &=  |F^{-1}(\hat{F}(x))-x|
     \\
    &=  |F^{-1}(\hat{F}(x))-F^{-1}(F(x))| \\
    &\overset{(a)}{\leq}  K_{T+1} |\hat{F}(x)-F(x)|\\
    &\leq  K_{T+1} \underset{x'\in \R}{\sup} |\hat{F}(x')-F(x')|, 
\end{align*}
where (a) holds by the Lipschitz continuity assumption of $F^{-1}$. Now, Theorem \ref{thm:cond_cov} implies that with probability at least $1-\sqrt{\log (16T)/T}$,
\[\underset{x'\in \R}{\sup} |\hat{F}(x')-F(x')| \leq 3\sqrt{\log (16T)/T} +C (\delta_T^{2/3}+\delta_T),\]
whereby we thus have
\begin{align*}
   & \underset{\alpha \in [0,1]}{\sup} |F^{-1}(\alpha)-\hat{F}^{-1}(\alpha)|\\
\leq & K_{T+1}(3\sqrt{\log (16T)/T} +C (\delta_T^{2/3}+\delta_T)).
\end{align*}

\vspace{0.1in}

\textit{Proof of (iii).} Intuitively, we know that $\betaReal$ converges to $\betaOracle$, as a consequence of $\hat F^{-1}$ converging to $F^{-1}$. We can indeed bound $\betaReal-\betaOracle$ by upper and lower bounding $|\betaReal-\betaOracle|$ as follows:
\begin{align*}
    \betaReal-\betaOracle 
    & \leq |\betaReal-\betaOracle | \\
    & = \hat F^{-1} (\hat F(|\betaReal-\betaOracle |))-\hat F^{-1} (\hat F(0)) \\
    & \leq K^{'}_{T+1} (\hat F(|\betaReal-\betaOracle |)-\hat F(0)).
\end{align*}
The last inequality follows from Eq. \eqref{Lipschitz}. The quantity $(\hat F(|\betaReal-\betaOracle |)-\hat F(0))$ is non-negative since $\hat F$ is non-decreasing. Similarly,
\begin{align*}
    F(\betaReal-\betaOracle)-F(0)
    & \leq L_{T+1}(\betaReal-\betaOracle) \\
    & \leq L_{T+1}|\betaReal-\betaOracle|,
\end{align*}
which is also non-negative by the property of CDF. As a result, with probability at least $1-\sqrt{\log (16T)/T}$,

\begin{align*}
    |\betaReal-\betaOracle|
    \leq &|1/K^{'}_{T+1}-L_{T+1}|^{-1} \cdot \nonumber \\ \qquad & \bigbrac{|\hat F(|\betaReal-\betaOracle |)-F(|\betaReal-\betaOracle |)|+|\hat F(0)-F(0)|}\\
    \leq & \ 2 \ \underset{x\in \R}{\sup} |\hat{F}(x)-F(x)|/|1/K^{'}_{T+1}-L_{T+1}|\\
    \leq & |1/K^{'}_{T+1}-L_{T+1}|^{-1}  \cdot \nonumber \\ \qquad & \left (6\sqrt{\log (16T)/T} +2C (\delta_T^{2/3}+\delta_T) \right) .
\end{align*}
Thus, by Lipschitz continuity of $F^{-1}$, 
\begin{align*}
    & |F^{-1}(1-\alpha+\betaReal)-F^{-1}(1-\alpha+\betaOracle)| \\
    \leq & |1/K^{'}_{T+1}-L_{T+1}|^{-1} \cdot \nonumber \\ \qquad & \left(6K_{T+1}\sqrt{\log (16T)/T} +2CK_{T+1} (\delta_T^{2/3}+\delta_T) \right).
\end{align*}

Putting (i), (ii), and (iii) together, we have
\begin{align*}
    \Delta(T) \leq & \delta_T+ \alpha K'_{T+1}/m+2(K_{T+1}+M_{T+1})\cdot \nonumber \\ \qquad & \left(3\sqrt{\log (16T)/T} +C(\delta_T^{2/3}+\delta_T) \right),
\end{align*}
where $M_{T+1}:=2K_{T+1}/|1/K^{'}_{T+1}-L_{T+1}|$.
\end{proof}

\section{More experiments} \label{app:more_exper}

We provide a quick overview of how this section is organized: In Appendix \ref{append_simul}, we present detailed setup and additional results for simulated examples. In Appendix \ref{app_expr:data_des}, we describe the renewable energy time series, the competing methods to \EnbPI, and the regression models $\mathcal{A}$. In Appendix \ref{app_expr:add_results}, we show additional results on the Atlanta solar radiation data as mentioned in the main text. In Appendix \ref{app_expr:CA_solar_wind}, we test \EnbPI \ on the network California solar data and Austin wind data, similar to those on Atlanta solar data from the main text. We first show the interval validity of \EnbPI \ and then apply \EnbPI \ on the more challenging multi-step ahead inference task when missing data are present. In Appendix \ref{app_expr:other_data}, we apply \EnbPI \ on other datasets, such as greenhouse gas emission data, air pollution data, and appliances energy data. We observe that \EnbPI \ rarely loses marginal validity and can produce shorter intervals than competing methods. We do not study multi-step ahead inference on these datasets as we primarily aim to demonstrate the applicability of \EnbPI. In Appendix \ref{append:anomaly_detection}, we describe details of performing unsupervised anomaly detection using \EnbPI{}.

\begin{figure}[!t]
    \centering
    \begin{minipage}[t]{0.24\linewidth}
        \centering
        \includegraphics[width=\textwidth]{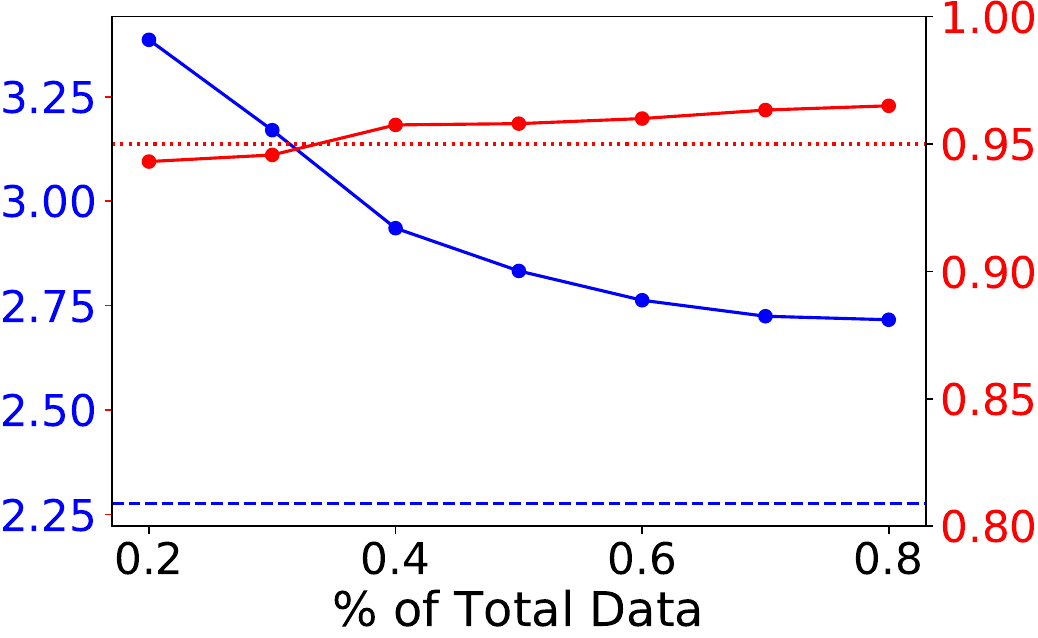}
        \subcaption{Case 1}
    \end{minipage}
    \begin{minipage}[t]{0.24\linewidth}
        \centering
        \includegraphics[width=\textwidth]{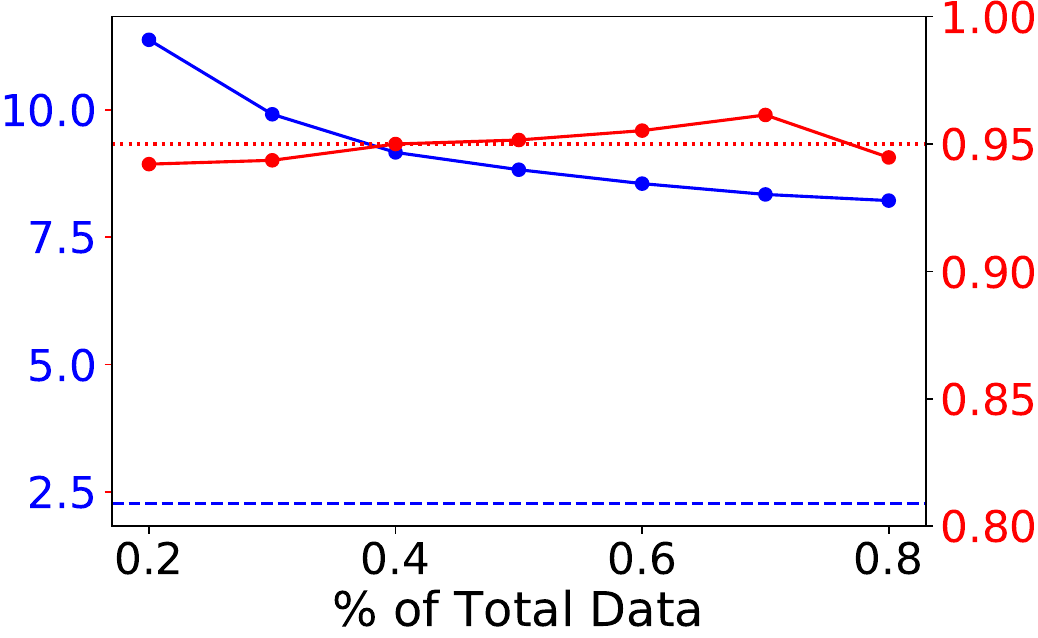}
        \subcaption{Case 2}
    \end{minipage}
     \begin{minipage}[t]{0.24\linewidth}
        \centering
        \includegraphics[width=\textwidth]{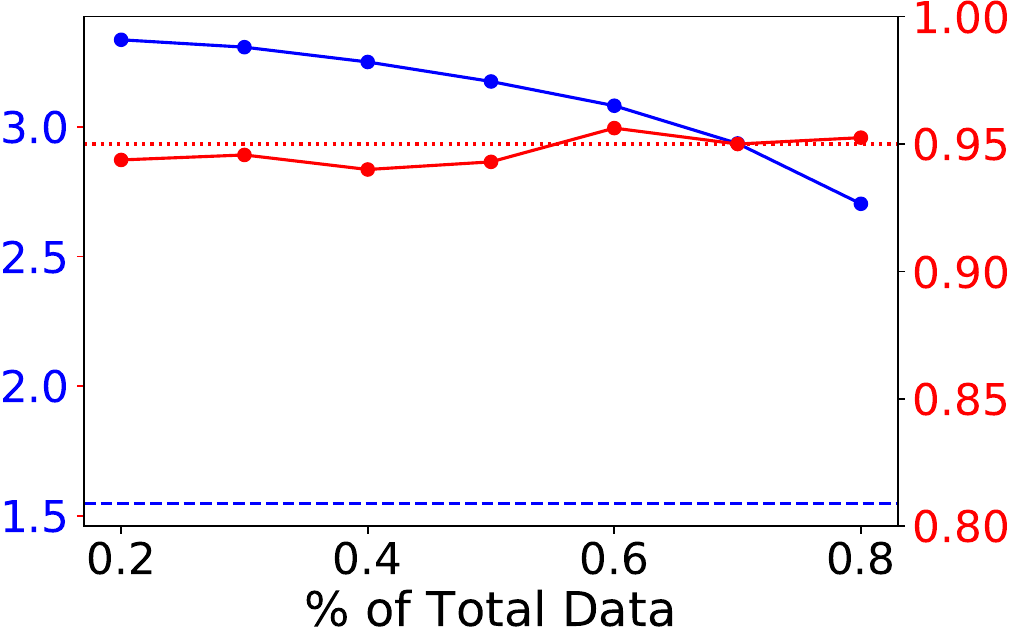}
       \subcaption{Case 3}
    \end{minipage}
    \begin{minipage}[t]{0.24\linewidth}
        \centering
        \includegraphics[width=\textwidth]{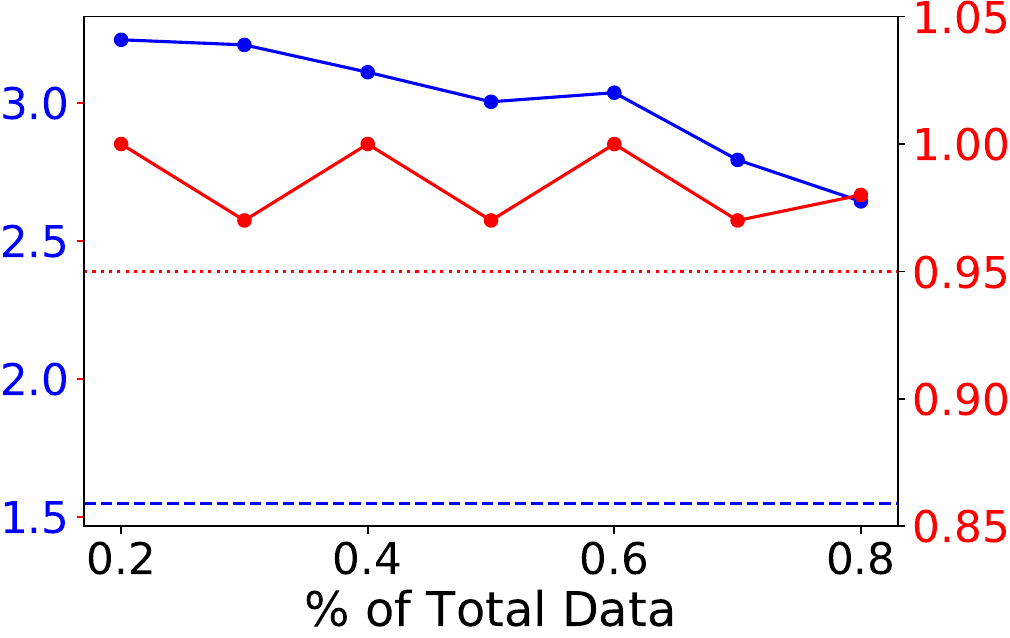}
        \subcaption{Case 3}
    \end{minipage}
    
    \begin{minipage}[b]{0.24\linewidth}
        \centering
        \includegraphics[width=\textwidth]{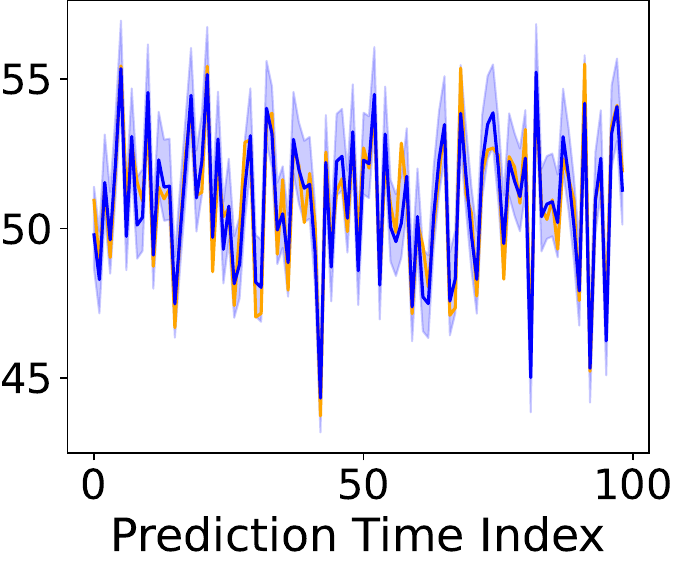}
        \subcaption{Case 1}
    \end{minipage}
    \begin{minipage}[b]{0.24\linewidth}
        \centering
        \includegraphics[width=\textwidth]{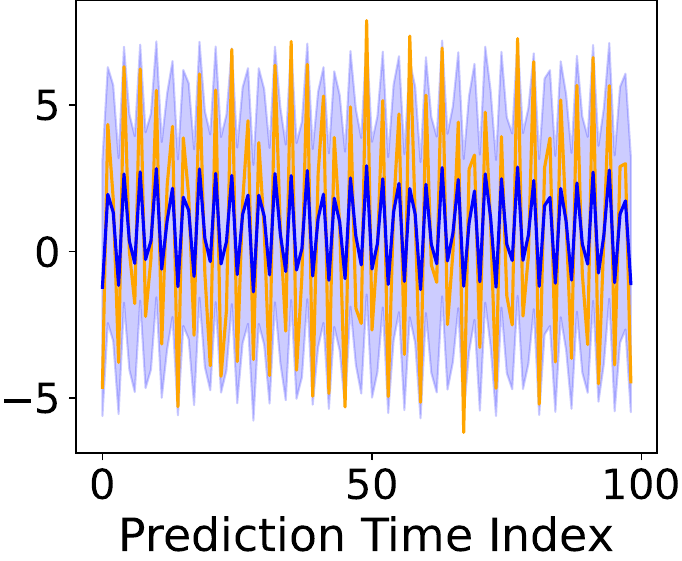}
        \subcaption{Case 2}
    \end{minipage}
    \begin{minipage}[b]{0.23\linewidth}
        \centering
        \includegraphics[width=\textwidth]{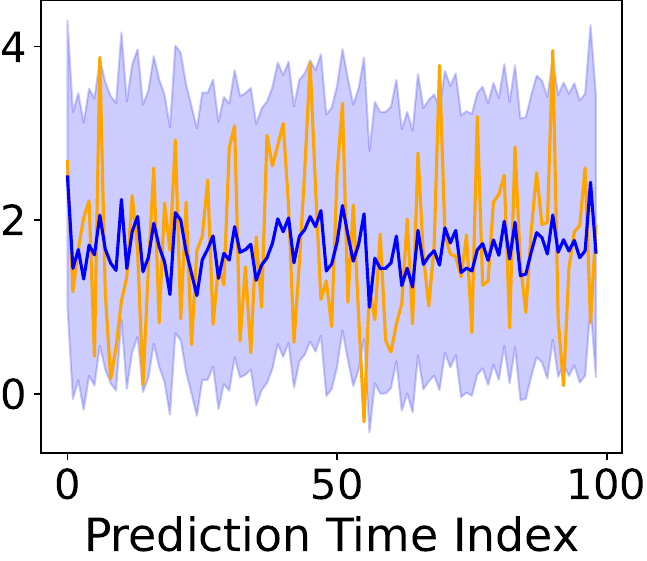}
        \subcaption{Case 3}
    \end{minipage}
    \begin{minipage}[b]{0.23\linewidth}
        \centering
        \includegraphics[width=\textwidth]{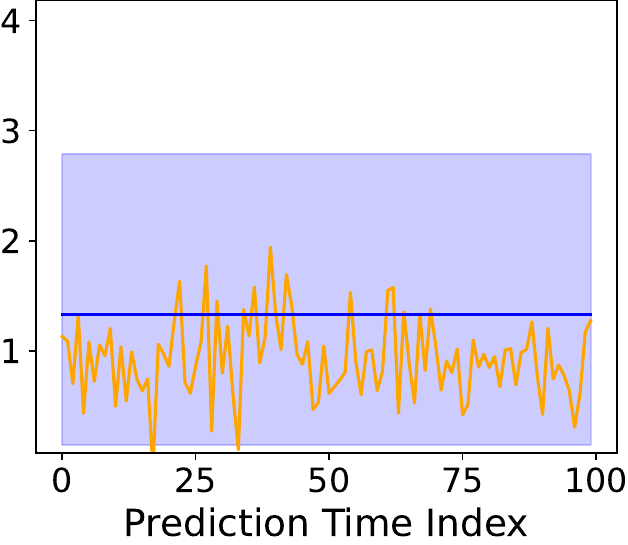}
        \subcaption{Case 3}
    \end{minipage}
    
    \cprotect \caption{\magenta{Simulation results at 95\% target coverage. The first three columns show marginal results and the last column shows conditional results for case 3 on the first test point (that is, $t=T+1$), by repeatedly sampling the error $\epsilon_{T+1}$). Meanwhile, the first row shows coverage in red and width in blue over different training sample sizes. The red (resp. blue) dashed-dotted line indicates the target coverage (resp. oracle interval width). The second row overlays prediction intervals in shaded blue on top of actual response in orange and point estimates in blue.}}
    \label{fig:simulation_nochangepts}
\end{figure}

\subsection{\rev{Simulated examples}}\label{append_simul}
\noindent \textit{Setup.} We examine the performance of \EnbPI \ on different simulated experiments. In particular, for the model $Y_t=f(X_t)+\epsilon_t$ in (\ref{eq:DGP_model}), we consider three cases with increasing levels of model sophistication. We simulate 2000 data and use the first 1000 points to train the LOO ensemble predictors and the remaining 1000 points for testing (that is, $T=T_1=1000$):
\begin{enumerate}[noitemsep,topsep=0.5em]
    \item[(1)] Let $f(X_t)=\beta^TX_t$ be a linear model, where the entries of $\beta$ and $X_t$ are {\it i.i.d.} uniform distributed $U[0,1]$; error $\epsilon_t$ are {\it i.i.d.} following the skewed normal distribution with mean $\mu$, variance $\sigma^2$, and skewness parameter $a$; here let $a=4, \mu=0, \sigma^2=0.1$.
    \item[(2)] Let $f(X_t)=\beta^TX_t$, where 80$\%$ of $\beta$ entries are zeros and observed entries are sampled {\it i.i.d.} from $U[0,1]$. We choose $d=1.6T$ so $f$ is a high-dimensional sparse linear model; this model choice is inspired by Example \ref{example:highdim_f} when discussing Assumption \ref{consis}. Meanwhile, let $X_t=Y_t^{-w}$ be the past $w$ observations, so $Y_t^{-w}=[Y_{t-1},\ldots,Y_{t-w}]$; here $w=100$ and we standardize $X_t$ to have unit $\ell_2$ norm. The errors are drawn from the same skewed normal distribution as above in (1).
    \item[(3)] Let $f(X_t)=(|\beta^TX_t|+(\beta^TX_t)^2+|\beta^TX_t|^3)\magenta{^{1/4}}$, where $\beta$ vector is generated the same as (2). Meanwhile, let $X_t=Y_t^{-w}$ and sample the errors $\epsilon_t$ from an AR(1) process, so that $\epsilon_t=\rho \epsilon_{t-1}+e_t$ and $e_t$ are {\it i.i.d.} normal random variables with zero mean and unit variance with $\rho=0.6$. It is known that  the AR(1) process is strongly mixing, as long as $e_t$ are {\it i.i.d.} with a nontrivial absolutely continuous component and $\mathbb{E}[(\log |e_t|)^+]$ is finite \citep{1986_mixing}.
\end{enumerate}
Hyperparameters to \EnbPI \ are as follows: $\alpha=0.05$, the aggregation $\phi$ takes the mean of the ensemble predictors, $B=50, s=1$. Thus, we build intervals aiming to cover 95$\%$ of test observations using 50 bagging estimators and slide the prediction interval every time we make a prediction. We consider prediction algorithms $\mathcal{A}$ as linear regression without intercept from \Verb|sklearn|, lasso with $\alpha=2$ from \Verb|sklearn|, and a neural network, respectively. The neural network is described in Appendix \ref{app_expr:data_des}. When doing the experiments, we did not optimize the performance of different prediction algorithms to demonstrate that \EnbPI \ works well even under default settings.

\vspace{0.1in}
\noindent \textit{Results.}
Figure \ref{fig:simulation_nochangepts} examines the marginal and conditional validity, as well as the interval width. The marginal coverage is obtained by averaging over $T_1$ test data and the conditional coverage is obtained by averaging over $K$ randomly sampled errors $\{\epsilon_{T+1,i}\}_{i=1}^{K}$ by fixing $f(x_{T+1})$. Here $K=100$. In terms of marginal coverage, despite slight under/over-coverage, we see in the first three columns that the marginal coverage in red is almost always valid at $1-\alpha$ regardless of the sample size $T$, indicating \EnbPI \ is suitable for small-sample problems. Meanwhile, interval width converges to the oracle width as the training sample size grows, validating Theorem \ref{thm:width}. In addition, we see that prediction intervals follow data dynamics and cover the unknown $Y_t$ with high probability. In terms of conditional coverage, the last column illustrates that \EnbPI \ always attains the conditional validity regardless of sample sizes.

\begin{figure}[!t]
    \centering
    \begin{minipage}[t]{0.32\linewidth}
        \includegraphics[width=\textwidth]{EmpvsActual_PtwiseWidth_changepts_LinearRegression_simple.pdf}
        \subcaption{Case 1, EnbPI}
    \end{minipage}
    \begin{minipage}[t]{0.32\linewidth}
        \includegraphics[width=\textwidth]{EmpvsActual_PtwiseWidth_changepts_Lasso_tseries_simple.pdf}
        \subcaption{Case 2, EnbPI}
    \end{minipage}
    \begin{minipage}[t]{0.32\linewidth}
        \includegraphics[width=\textwidth]{EmpvsActual_PtwiseWidth_changepts_Sequential_tseries_mixing_simple.pdf}
        \subcaption{Case 3, EnbPI}
    \end{minipage}
    \begin{minipage}[t]{0.32\linewidth}
        \includegraphics[width=\textwidth]{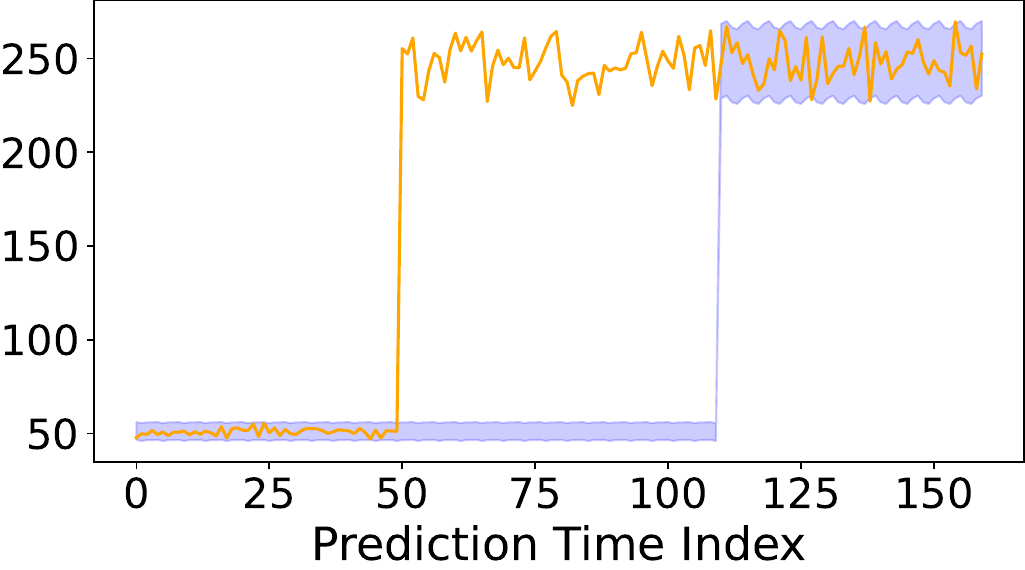}
        \subcaption{\rev{Case 1, ETS}}
    \end{minipage}
    \begin{minipage}[t]{0.32\linewidth}
        \includegraphics[width=\textwidth]{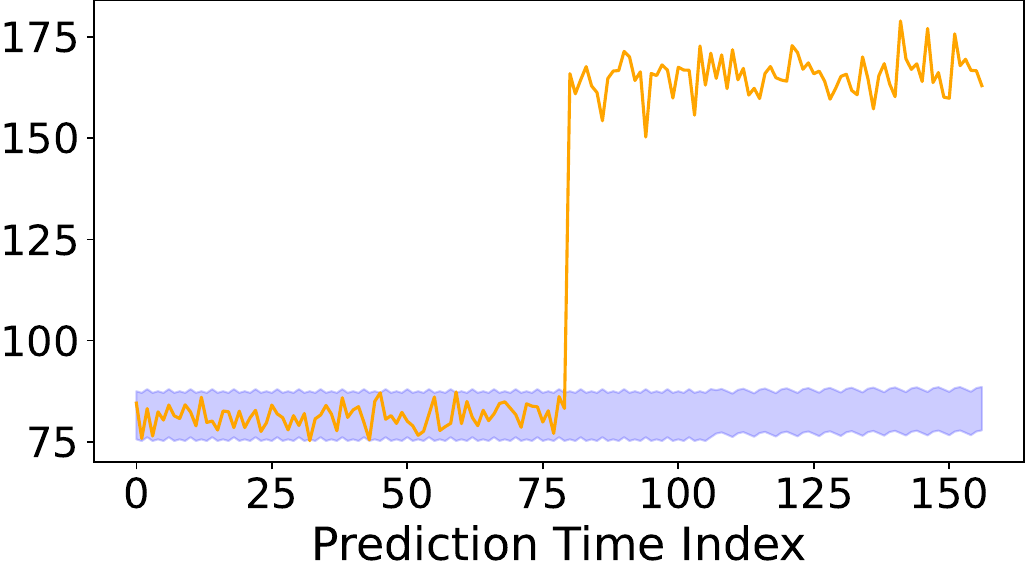}
        \subcaption{\rev{Case 2, ETS}}
    \end{minipage}
    \begin{minipage}[t]{0.32\linewidth}
        \includegraphics[width=\textwidth]{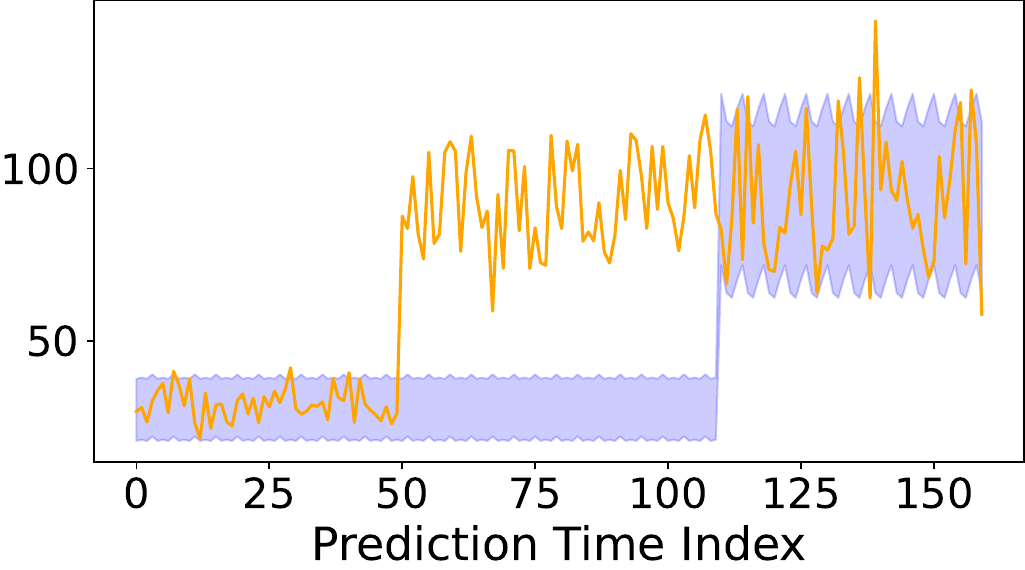}
        \subcaption{\rev{Case 3, ETS}}
    \end{minipage}
    \cprotect \caption{\rev{Simulation with changepoints between \EnbPI{} and ETS model. Same setup as Figure \ref{fig:simul_changepts_result}.}}
    \label{fig:simul_changepts_result_ETS}
    \vspace{-0.3cm}
\end{figure}

\subsection{Setup detail for real-data experiments} \label{app_expr:data_des}

\noindent \textit{A. Competing methods to \EnbPI. }Recall we compete \EnbPI \ against ARIMA(10,1,10), Exponential Smoothing, Dynamic Factor model, split conformal predictor, weighted split conformal predictor, quantile out-of-bag method, adaptive conformal inference, and jackknife+-after-bootstrap:
\begin{itemize}[itemsep=0em]
    \item The first three time series methods are implemented in Python's \Verb|statsmodel| package. The ARIMA model is used under default setting and the other two other methods include damped trend components and seasonal cycles of 24, because most data are updated hourly with daily periodicity.
    \item The WeightedICP \citep{CPcovshift} is proven to work when the test distribution shifts in proportion to the training distribution; it generalizes to more complex settings than ICP. We use logistic regression to estimate the weights for WeightedICP. 
    \item \magenta{The QOOB \citep{QOOB} utilizes a very similar leave-one-out ensemble learning idea, where the main difference lies in computing bootstrap estimators as quantile estimators and consequently adopting a quantile-based non-conformity score. The authors empirically demonstrate that QOOB can yield shorter intervals than other methods. As suggested by the authors, we build quantile random forests as bootstrap estimators, with identical hyper-parameters as (non-quantile) RFs used by \EnbPI.}
    \item \magenta{The AdaptCI \citep{Gibbs2021AdaptiveCI} adaptively updates the significant level $\alpha$ during inference, by examining whether each prediction interval covers the actual response values. It builds on top of conformal quantile regression \citep{CPquantile} whereby the intervals are often empirically shorter and maintain validity on non-exchangeable observations. We use either the quantile linear model or the quantile RF as the predictor and update the significance level according to the simple online update (ibid., Eq (2)).}
\end{itemize}

\noindent \textit{B. Regression models $\mathcal{A}$. }Below are the parameter specifications for the four baseline models $\mathcal{A}$, unless otherwise specified: 
\begin{itemize}[itemsep=0em]
    \item For ridge, the penalty parameter $\alpha$ is chosen with generalized cross-validation over ten uniformly spaced grid points between 0.0001 to 10 (the package default $\alpha$ is 1). Higher $\alpha$ means more robust regularization.
    \item For RF, we build ten trees under the mean-squared-error (MSE) criterion. 
    We only allow each tree to split features rather than samples so that combining RFs trained on subsets of the training data is reasonable for \Verb|EnbPI|.
    \item For NN, we add three hidden layers, each having 100 hidden nodes, and apply 20$\%$ dropout after the second hidden layer to avoid overfitting. We use the Relu activation between hidden layers. The optimizer is Adam with a fixed learning rate of $5\times 10^{-4}$ under the MSE loss. Batch size equals 10\% of training data, and maximum epoch equals 250. We also use early stopping if there is no improvement in training error after ten epochs.
    \item For RNN, we add two hidden LSTM layers, followed by a dense output layer. Each LSTM layer has 100 hidden neurons, so the output from the first hidden layer is fed into the second hidden layer. We use the Tanh activation function for these two hidden layers and the Relu activation function for the dense layer. The optimizer is Adam with a fixed learning rate of $5\times 10^{-4}$ under the MSE loss. Batch size equals 10\% of training data, and maximum epoch equals 10. We use early stopping if there is no improvement in training error after ten epochs.
\end{itemize}


\subsection{Additional results on solar---Atlanta} \label{app_expr:add_results}

The solar dataset is available at \protect\url{https://nsrdb.nrel.gov/}. The 9 cities we chose are Fremont, Milpitas, Mountain View, North San Jose, Palo Alto, Redwood City, San Mateo, Santa Clara, Sunnyvale. The wind dataset is publically available at \protect\url{https://github.com/Duvey314/austin-green-energy-predictor}. 

\begin{figure}[htbp]
    \centering
    \includegraphics[width=\linewidth]{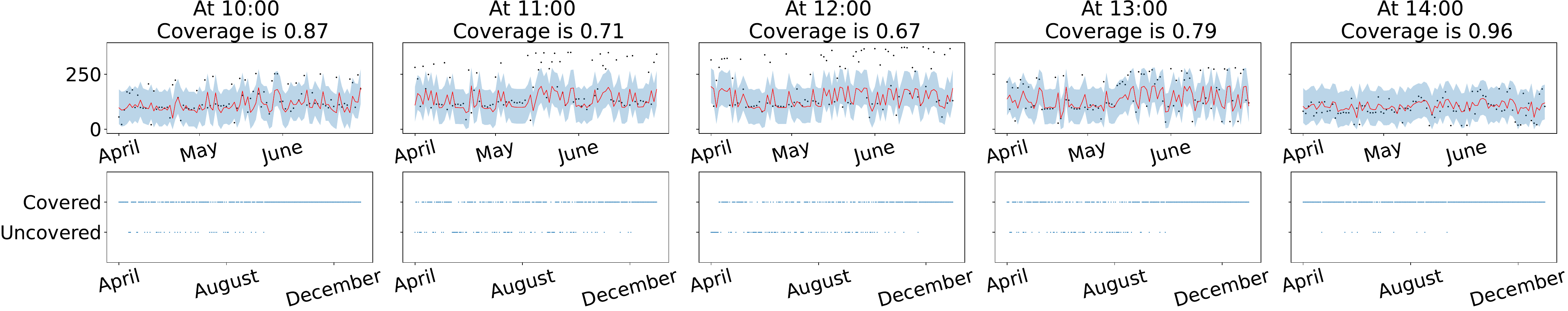}
    \vspace{-0.2cm}
    \caption{Solar power prediction in Atlanta when no feedback is available. The top figure at each hour visualizes observations in black, estimates in red, and prediction intervals in blue for three months (April-June). The bottom figure at each hour plots whether the prediction interval correctly covers the response during test (April---December). We observe clear decrease in conditional coverage even if no missing data is present and the same prediction models are used. This situation illustrates the necessity to updated past residuals based on new feedback.}
    \label{fig:plts_no_slide}
\end{figure}

\vspace{0.1in}
\noindent \textit{A. Multi-step ahead inference when $s=\infty$. } We train on the same set of data as in Figure \ref{fig:plts_daily_slide_nn}. Since change points are present in the data (that is, data near summer have very different radiation levels from the training data), we expect \EnbPI \ to perform less well if not slide. Indeed, Figure \ref{fig:plts_no_slide} shows poor conditional coverage, even if we train on the same set of data and further assume \textit{no missing data exist}.

\begin{table*}[!b]
    \centering
    \cprotect \caption{\magenta{Conditional coverage (abbreviated as Cov.) and width by \EnbPI, QOOB, and Adaptive-CI, where the setup is identical to Figure \ref{fig:plts_daily_slide_nn}. Comparing to \EnbPI, these methods produce prediction intervals that either fail to maintain coverage validity or remain slightly wider. Figure \ref{fig:plts_daily_slide_QOOB_Adaptive_CI} shows detailed results for QOOB and AdaptCI as those in Figure \ref{fig:plts_daily_slide_nn} for \EnbPI.}
    \label{tab:cond_cov_QOOB_AdaptiveCI}}
    \resizebox{\linewidth}{!}{%
    \def\arraystretch{1.5}
    \begin{tabular}{P{1.6cm} P{1.2cm}P{1.2cm}|P{1.2cm}P{1.2cm}|P{1.2cm}P{1.2cm}|P{1.2cm}P{1.2cm}|P{1.2cm}P{1.2cm}|P{1.2cm}P{1.2cm}|P{1.2cm}P{1.2cm}|P{1.2cm}P{1.2cm}}
    \toprule
   At Hour & \multicolumn{2}{c}{9:00} & \multicolumn{2}{c}{10:00} & \multicolumn{2}{c}{16:00} & \multicolumn{2}{c}{17:00} & \multicolumn{2}{c}{11:00} & \multicolumn{2}{c}{12:00} & \multicolumn{2}{c}{13:00} & \multicolumn{2}{c}{14:00} \\
    \hline
    Result & Cov. &   Width & Cov. &   Width & Cov. &   Width & Cov. &   Width & Cov. &   Width & Cov. &   Width & Cov. &   Width & Cov. &   Width \\
\hline
\EnbPI   &     0.87 &  254.08 &     0.87 &  253.47 &     0.90 &  253.82 &     0.89 &  253.92 &     0.87 &  388.29 &     0.88 &  387.55 &     0.89 &  387.98 &     0.85 &  387.75 \\
QOOB    &     0.78 &  184.85 &     0.79 &  183.86 &     0.82 &  183.44 &     0.75 &  184.20 &     0.74 &  325.14 &     0.75 &  324.82 &     0.73 &  324.41 &     0.74 &  325.57 \\
AdaptCI &     0.88 &  250.00 &     0.89 &  252.11 &     0.92 &  250.27 &     0.88 &  250.16 &     0.92 &  410.49 &     0.92 &  407.69 &     0.88 &  407.51 &     0.90 &  409.91 \\
    \bottomrule
    \end{tabular}
    }
\end{table*}

\begin{figure}[htbp]
    \centering
    \begin{minipage}[t]{\linewidth}
        \includegraphics[width=\textwidth]{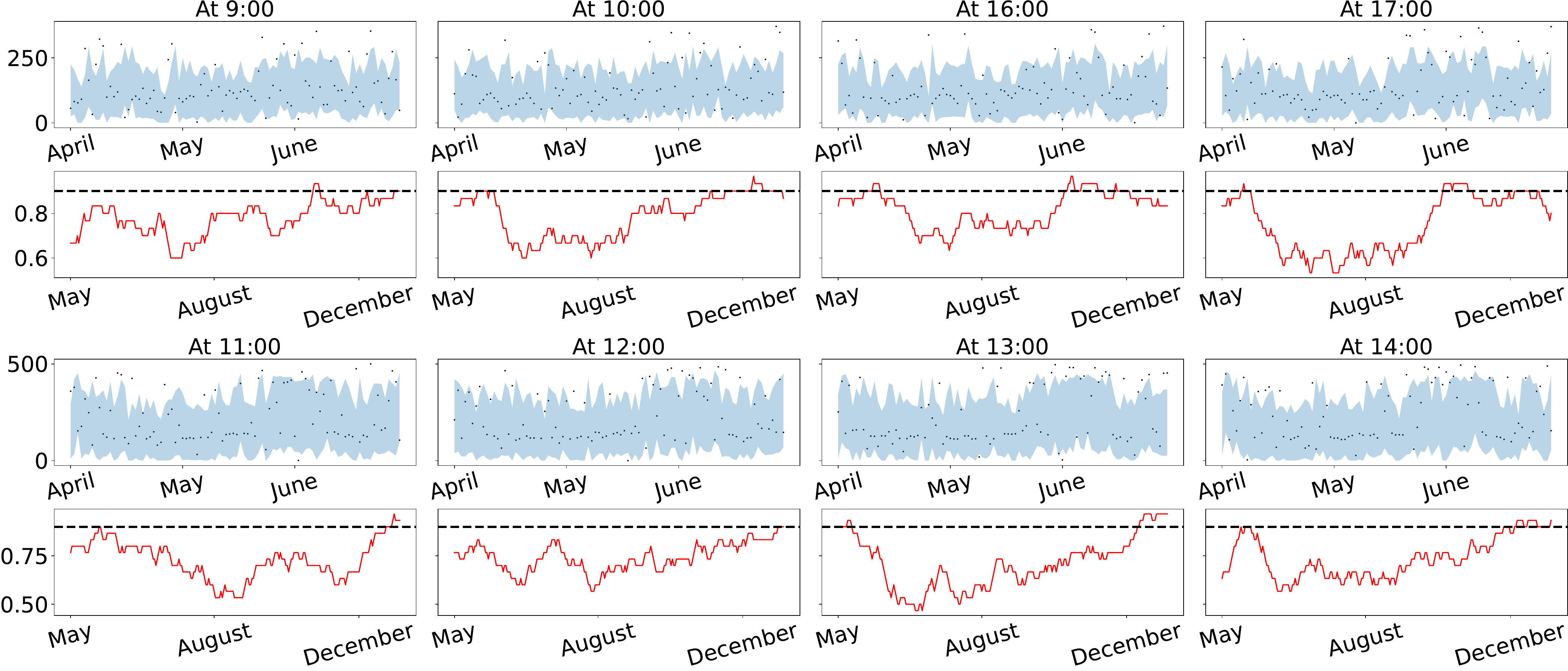}
        \subcaption{QOOB: Prediction intervals and sliding coverage by hour}
    \end{minipage}
    \begin{minipage}[t]{\linewidth}
        \includegraphics[width=\textwidth]{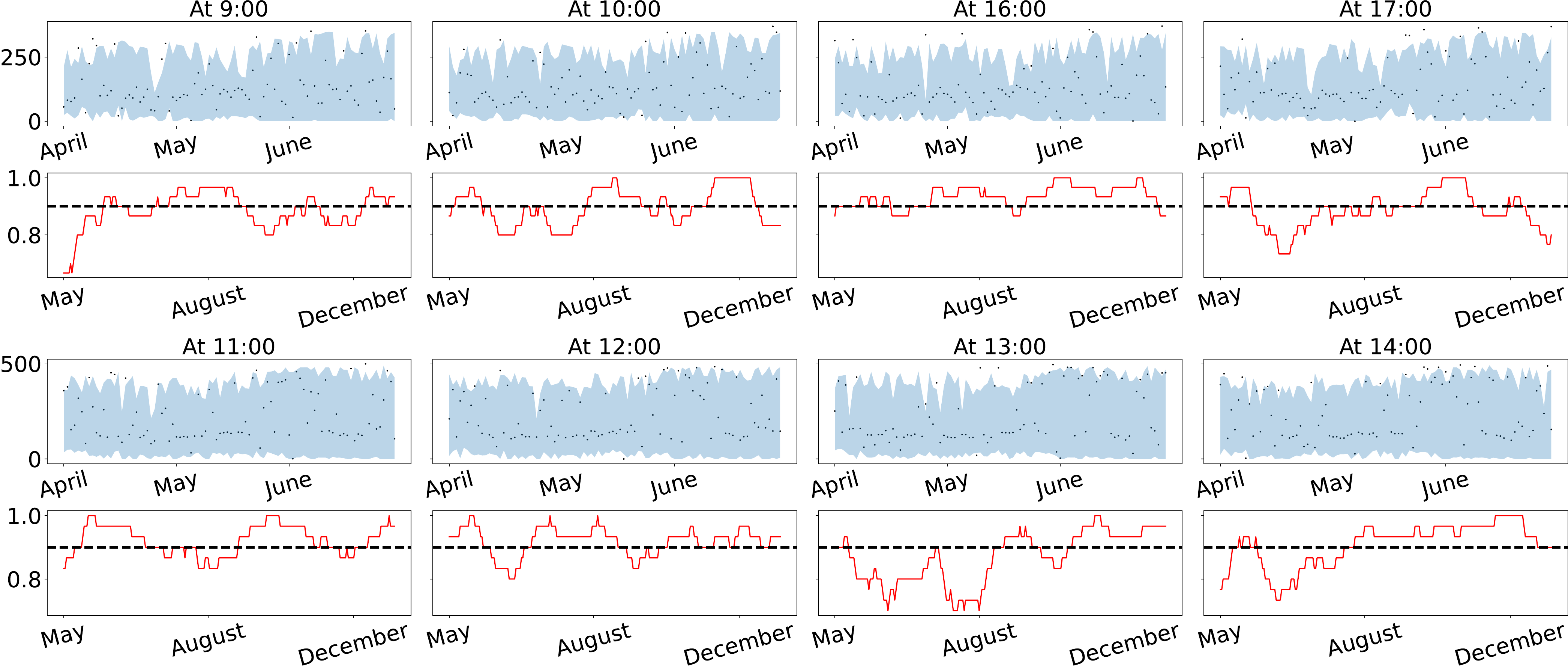}
        \subcaption{AdaptCI: Prediction intervals and sliding coverage by hour}
    \end{minipage}
    \cprotect \caption{\magenta{Solar power prediction in Atlanta, when QOOB or AdaptCI looks ahead beyond one step. A summary of results was presented in Table \ref{tab:cond_cov_QOOB_AdaptiveCI}. Figures on the top of each four rows visualize observations in black, estimates in red, and prediction intervals in blue for three months (April-June). Average coverage and width values are described in the caption. Figures on the bottom of each four rows visualize coverage under a sliding window of 30 days.}
    \label{fig:plts_daily_slide_QOOB_Adaptive_CI}}
\end{figure}

\vspace{0.1in}
\noindent \magenta{\textit{B. Conditional coverage of QOOB and Adaptive-CI. } Figure \ref{fig:plts_daily_slide_QOOB_Adaptive_CI} shows the detailed hourly conditional coverage by QOOB and AdaptCI, where a summary was presented in Table \ref{tab:cond_cov_QOOB_AdaptiveCI}.}
\vspace{0.1in}

\subsection{Results on solar---California and wind---Austin}\label{app_expr:CA_solar_wind}

\noindent \textit{\rev{Network prediction.}} Note that the whole California data constitute a network. \rev{We can use \EnbPI{} in this network prediction setting as follows.} Consider a network with $K$ nodes, so that the observations at node $k\in [K]$ are given by $\{(y_t^k,x_t^k)\}_{t\geq 1}$. To incorporate network temporal information, we define a new feature $\tilde{x}^k_t$ at node $k$ and time $t$ as a collection of features from neighbors of $k$ at time $t$ and earlier. By adopting this setup, we can perform nodal prediction (from history and neighboring information) by applying our conformal prediction scheme; theoretical guarantees equally hold at each node as long as we use \EnbPI{} for each node.

In general, \EnbPI \ on California solar data and on Austin wind data generates results very similar to those on Atlanta solar data. Therefore, we do not provide separate analyses of individual figures but highlight the overall pattern and differences in each part below.

\vspace{0.1in}
\noindent\textit{A. Marginal validity and the interval width}

\noindent\textit{On California solar network data.} In general, results on different Californian cities look very similar to each other so that we only provide plots on the Palo Alto solar data. 

We summarize the performance of traditional time series methods and CP methods on time series from all Californian cities. Details are in Table \ref{tab:table_solar_ca_ridge} for the ridge regression.
We only show results for ridge because results under different $\mathcal{A}$ for CP methods are similar. We can see from the table that \EnbPI \ performs similarly as the time series methods. On the other hand, the Winkler score\footnote{Let the upper and lower end of the prediction interval at time $t$ under level $\alpha$ be $L_t(\alpha)$, $U_t(\alpha)$, so width is $W_t(\alpha)=U_t(\alpha)-L_t(\alpha)$. Then, Winkler score (WS) is:\begin{equation*}
    (WS)_t= \begin{cases}
        W_t(\alpha), & \text{if } L_t(\alpha) \leq y_t\leq U_t(\alpha)\\
        W_t(\alpha)+2\cdot \frac{L_t(\alpha)-y_t}{\alpha}, & \text{if } y_t < L_t(\alpha)\\
        W_t(\alpha)+2\cdot \frac{y_t-U_t(\alpha)}{\alpha}, & \text{if } y_t > U_t(\alpha)
        \end{cases} 
\end{equation*} It was used in \citep{Kath_2021} as a quantitative measure of both coverage and width.} by \EnbPI \ using ridge regression can sometimes be the smallest so that it reaches a better balance between validity and efficiency. Meanwhile, ICP and WeightedICP can greatly lose coverage especially with multivariate $X_t$ so that they should not be used for dynamic time series data.

On the other hand, because \EnbPI \ performs very similarly on the wind data, we will only apply it on the more challenging multi-step ahead inference with missing data.

\vspace{0.1in}
\noindent\textit{B. Missing data, conditional coverage}

\begin{figure}[h!]
    \centering
    \begin{minipage}[t]{\linewidth}
        \includegraphics[width=\textwidth]{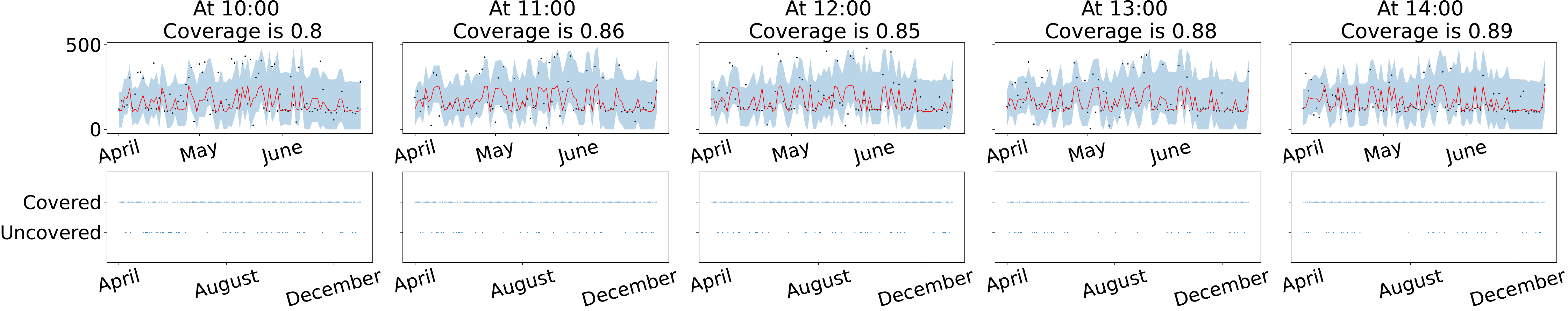}
        \cprotect \subcaption{Solar Palo Alto: \EnbPI \ under RF with missing data}
    \end{minipage}
    \begin{minipage}[t]{\linewidth}
        \includegraphics[width=\textwidth]{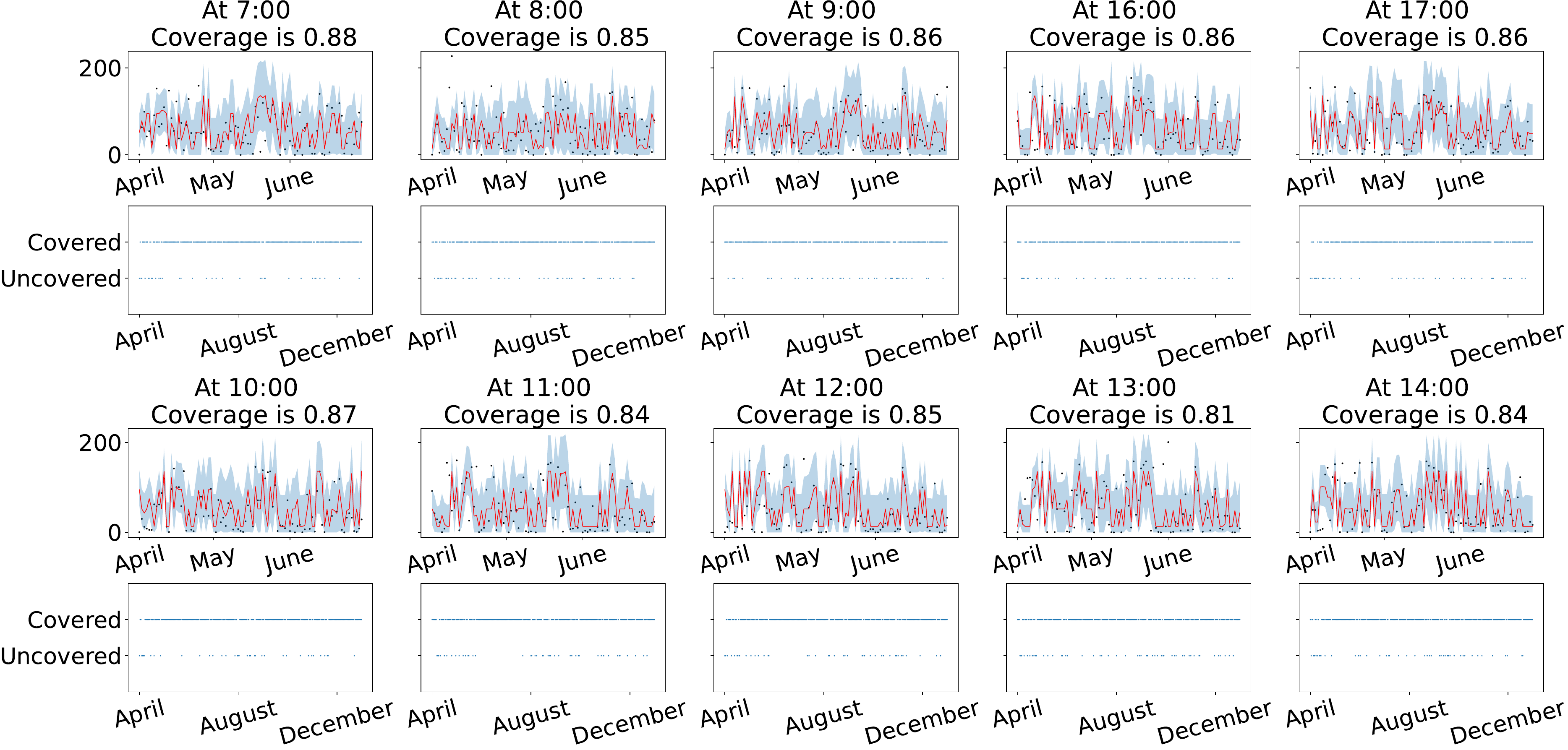}
        \cprotect \subcaption{Wind: \EnbPI \ under RF with missing data, $s=24$}
    \end{minipage}
    \cprotect\caption{Solar data in Palo Alto and wind data when \EnbPI \ looks ahead beyond one step. Results are similar to those in Figure \ref{fig:plts_daily_slide_nn} (Solar Atlanta).
    }
    \label{fig:plts_daily_slide_rf_append}
\end{figure}

\noindent\textit{On California solar network data:} Figure \ref{fig:plts_daily_slide_rf_append}(a) shows \textit{conditional} coverage of \EnbPI \ under RF at hours near noon, with the presence of missing data. 
The results look very similar to earlier ones, where conditional coverage by \EnbPI \ is still validly attained. 

\vspace{0.1in}
\noindent\textit{On wind power data:} Figure \ref{fig:plts_daily_slide_rf_append}(b) shows conditional coverage of multi-step ahead inference of \EnbPI \ under RF with missing data. No feature is available so we can only use past history of the wind power as response (that is, $X_t$ is the history of $Y_t$). Note, one difference from applying \EnbPI \ on earlier solar energy  results is that we do not choose $s=5$, but only train \EnbPI \ on the whole 24 hourly data (for example, $s=24$). We do so because this wind data do not exhibit clear differences at different hours of the day. Results show that \EnbPI \ can reach valid conditional coverage at these hours, even under the presence of missing data.


\begin{table*}[htbp]
\cprotect \caption{Results on all 9 cities in California. Smaller Winkler scores indicate a better balance between coverage and width. Bold cells indicate the smallest Winkler score}
\label{tab:table_solar_ca_ridge}
\centering
\resizebox{0.6\linewidth}{!}{\begin{tabular}{p{2.2cm}p{2.5cm}p{1.6cm}p{2cm}p{2cm}}
\hline
& & \multicolumn{3}{c}{Multivariate $X_t$}\\
Location & Method & Cov. & Width & Winkler Score\\
\hline
\multirow{6}{*}{Fremont} & \EnbPI & 0.93 & 259.31 & 2.56e+06 \\
  & ICP & 0.67 & 142.40 & 6.76e+06 \\
  & WeightedICP & 0.69 & 150.68 & 5.21e+06 \\
  & ARIMA & 0.93 & 106.76 & 1.61e+06 \\
  & ExpSmoothing & 0.93 & 128.55 & 1.62e+06 \\
  & DynamicFactor & 0.94 & 148.87 & 1.73e+06 \\
  \hline
\multirow{6}{*}{Milpitas} & \EnbPI & 0.93 & 257.14 & 2.53e+06 \\
  & ICP & 0.66 & 142.32 & 7.16e+06 \\
  & WeightedICP & 0.65 & 143.64 & 6.66e+06 \\
  & ARIMA & 0.92 & 99.71 & 1.70e+06 \\
  & ExpSmoothing & 0.94 & 126.03 & 1.62e+06 \\
  & DynamicFactor & 0.94 & 140.25 & 1.74e+06 \\
  \hline
\multirow{6}{*}{Mountain View} & \EnbPI & 0.93 & 261.92 & 2.57e+06 \\
  & ICP & 0.64 & 140.73 & 7.41e+06 \\
  & WeightedICP & 0.66 & 154.15 & 6.23e+06 \\
  & ARIMA & 0.92 & 93.55 & 1.50e+06 \\
  & ExpSmoothing & 0.94 & 116.52 & 1.48e+06 \\
  & DynamicFactor & 0.94 & 133.90 & 1.62e+06 \\
  \hline
\multirow{6}{*}{North San Jose} & \EnbPI & 0.93 & 261.04 & 2.55e+06 \\
  & ICP & 0.68 & 145.12 & 7.09e+06 \\
  & WeightedICP & 0.65 & 141.33 & 6.49e+06 \\
  & ARIMA & 0.92 & 101.12 & 1.60e+06 \\
  & ExpSmoothing & 0.94 & 120.17 & 1.54e+06 \\
  & DynamicFactor & 0.94 & 142.07 & 1.69e+06 \\
  \hline
\multirow{6}{*}{Palo Alto} & \EnbPI & 0.92 & 258.10 & 2.54e+06 \\
  & ICP & 0.65 & 143.62 & 7.06e+06 \\
  & WeightedICP & 0.66 & 152.01 & 5.77e+06 \\
  & ARIMA & 0.93 & 99.33 & 1.51e+06 \\
  & ExpSmoothing & 0.94 & 119.96 & 1.52e+06 \\
  & DynamicFactor & 0.93 & 137.79 & 1.65e+06 \\
  \hline
\multirow{6}{*}{Redwood City} & \EnbPI & 0.92 & 259.57 & 2.53e+06 \\
  & ICP & 0.68 & 151.59 & 6.44e+06 \\
  & WeightedICP & 0.64 & 149.69 & 6.47e+06 \\
  & ARIMA & 0.93 & 100.89 & 1.54e+06 \\
  & ExpSmoothing & 0.94 & 118.54 & 1.52e+06 \\
  & DynamicFactor & 0.94 & 142.87 & 1.64e+06 \\
  \hline
\multirow{6}{*}{San Mateo} & \EnbPI & 0.93 & 257.20 & 2.50e+06 \\
  & ICP & 0.68 & 147.86 & 6.03e+06 \\
  & WeightedICP & 0.65 & 151.80 & 6.18e+06 \\
  & ARIMA & 0.92 & 105.25 & 1.63e+06 \\
  & ExpSmoothing & 0.94 & 128.21 & 1.59e+06 \\
  & DynamicFactor & 0.94 & 153.77 & 1.72e+06 \\
  \hline
\multirow{6}{*}{Santa Clara} & \EnbPI & 0.93 & 251.93 & 2.47e+06 \\
  & ICP & 0.68 & 146.22 & 6.56e+06 \\
  & WeightedICP & 0.68 & 148.40 & 5.49e+06 \\
  & ARIMA & 0.93 & 101.71 & 1.55e+06 \\
  & ExpSmoothing & 0.94 & 117.94 & 1.53e+06 \\
  & DynamicFactor & 0.94 & 138.07 & 1.64e+06 \\
  \hline
\multirow{6}{*}{Sunnyvale} & \EnbPI & 0.92 & 261.17 & 2.56e+06 \\
  & ICP & 0.68 & 153.19 & 6.73e+06 \\
  & WeightedICP & 0.63 & 150.14 & 6.77e+06 \\
  & ARIMA & 0.93 & 100.63 & 1.46e+06 \\
  & ExpSmoothing & 0.94 & 114.14 & 1.47e+06 \\
  & DynamicFactor & 0.94 & 137.63 & 1.59e+06 \\
\hline
\end{tabular}}
\hspace{-0.1in}
\resizebox{0.332\linewidth}{!}{\begin{tabular}{p{1.6cm}p{2cm}p{2cm}}
\hline
\multicolumn{3}{c}{Univariate $X_t$}\\
Coverage & Width & Winkler Score\\
\hline
 0.95 & 135.64 & \textbf{1.43e+06} \\
0.91 & 100.35 & 1.69e+06 \\
0.89 & 103.41 & 1.80e+06 \\
0.93 & 107.32 & 1.61e+06 \\
0.94 & 124.93 & 1.60e+06 \\
0.94 & 149.25 & 1.73e+06 \\
\hline
0.95 & 128.25 & \textbf{1.41e+06} \\
0.91 & 100.49 & 1.63e+06 \\
0.91 & 101.13 & 1.69e+06 \\
0.92 & 99.88 & 1.67e+06 \\
0.95 & 122.90 & 1.59e+06 \\
0.94 & 140.51 & 1.74e+06 \\
\hline
0.95 & 126.76 & \textbf{1.41e+06} \\
0.90 & 98.68 & 1.65e+06 \\
0.91 & 105.05 & 1.66e+06 \\
0.92 & 94.76 & 1.48e+06 \\
0.94 & 116.84 & 1.48e+06 \\
0.94 & 134.04 & 1.62e+06 \\
\hline
0.95 & 129.01 & \textbf{1.38e+06} \\
0.90 & 103.55 & 1.65e+06 \\
0.90 & 100.84 & 1.69e+06 \\
0.92 & 101.34 & 1.59e+06 \\
0.95 & 120.72 & 1.55e+06 \\
0.94 & 142.32 & 1.69e+06 \\
\hline
0.95 & 129.69 & \textbf{1.43e+06} \\
0.90 & 100.40 & 1.66e+06 \\
0.91 & 106.87 & 1.66e+06 \\
0.93 & 99.74 & 1.52e+06 \\
0.95 & 117.99 & 1.49e+06 \\
0.93 & 137.98 & 1.65e+06 \\
\hline
0.95 & 132.31 & \textbf{1.43e+06} \\
0.92 & 115.10 & 1.63e+06 \\
0.93 & 124.41 & 1.69e+06 \\
0.93 & 101.72 & 1.54e+06 \\
0.94 & 118.49 & 1.52e+06 \\
0.94 & 143.14 & 1.64e+06 \\
\hline
0.95 & 139.29 & \textbf{1.45e+06} \\
0.92 & 127.50 & 1.77e+06 \\
0.93 & 126.55 & 1.72e+06 \\
0.92 & 105.49 & 1.63e+06 \\
0.94 & 125.95 & 1.56e+06 \\
0.94 & 154.09 & 1.72e+06 \\
\hline
0.95 & 128.49 & \textbf{1.38e+06} \\
0.91 & 105.98 & 1.64e+06 \\
0.92 & 110.49 & 1.66e+06 \\
0.93 & 103.44 & 1.54e+06 \\
0.95 & 117.64 & 1.51e+06 \\
0.94 & 138.31 & 1.64e+06 \\
\hline
0.95 & 131.52 & \textbf{1.42e+06} \\
0.91 & 102.26 & 1.69e+06 \\
0.92 & 113.00 & 1.61e+06 \\
0.93 & 98.32 & 1.47e+06 \\
0.94 & 114.69 & 1.44e+06 \\
0.94 & 137.79 & 1.59e+06 \\
\hline
\end{tabular}}
\end{table*} 


\subsection{Results on datasets in other domains}\label{app_expr:other_data}
\textit{Data Description.} We describe the additional three datasets being used, which are greenhouse gas emission data, air pollution data, and appliances energy data. The first dataset contains Greenhouse Gas observation (Greenhouse) \citep{greenhouse} from 5.10 till 7.31, 2010, with four samples every day and 6 hours apart between data points. The goal is to find the optimal weights for the 15 observation series to match the synthetic control series. The second dataset contains appliances energy usage (Appliances) \citep{appliances}. Consecutive data points are 10 minutes apart for about 4.5 months. We can use 27 different humidity and temperature indicators to predict the appliances' energy use in Wh. The third dataset on Beijing air quality (Beijing air) \citep{BJair} contains air pollutants data from 12 nationally-controlled air-quality monitoring sites. The data is from 3.1, 2013 to 2.28, 2017. The goal is to predict PM2.5 air pollutant levels using 10 different air pollutants and meteorological variables. We use the data from the Tiantan district.

\vspace{0.1in}
\noindent\textit{Results.} We first show additional average coverage and width versus $1-\alpha$ line plots. Then, we present grouped boxplots using both multivariate and univariate $X_t$. We do not examine conditional coverage on these dataset as we primarily aim to demonstrate the applicability of \EnbPI{}.

\vspace{0.1in}
\noindent\textit{(1) Observations from the other coverage/width versus $1-\alpha$ plots}

\begin{figure}[htbp]
    \centering
    \begin{minipage}[t]{\linewidth}
        \includegraphics[width=\textwidth]{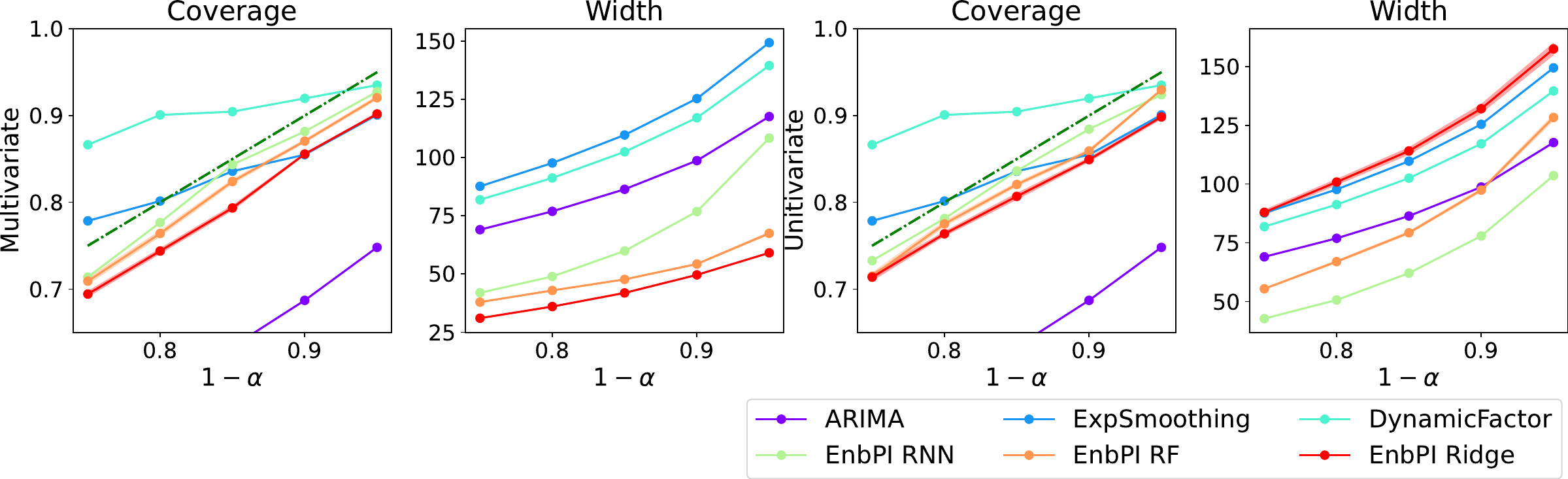}
        \subcaption{Greenhouse Gas}
    \end{minipage}
    \begin{minipage}[t]{\linewidth}
        \includegraphics[width=\textwidth]{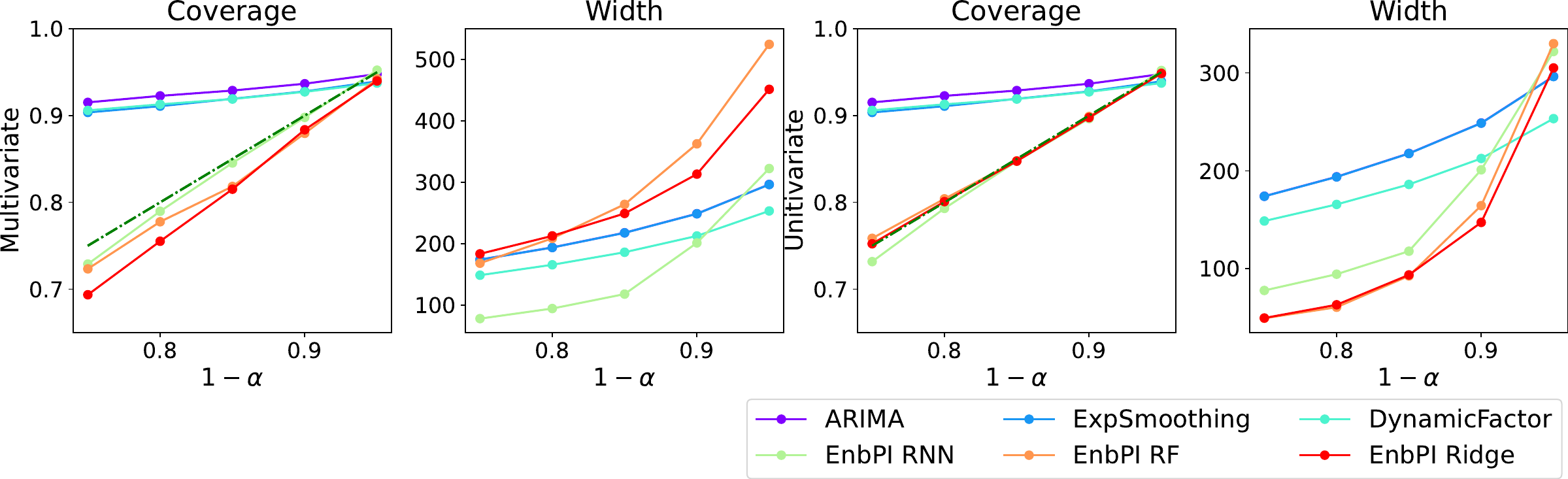}
        \subcaption{Appliances Energy}
    \end{minipage}
    \begin{minipage}[t]{\linewidth}
        \includegraphics[width=\textwidth]{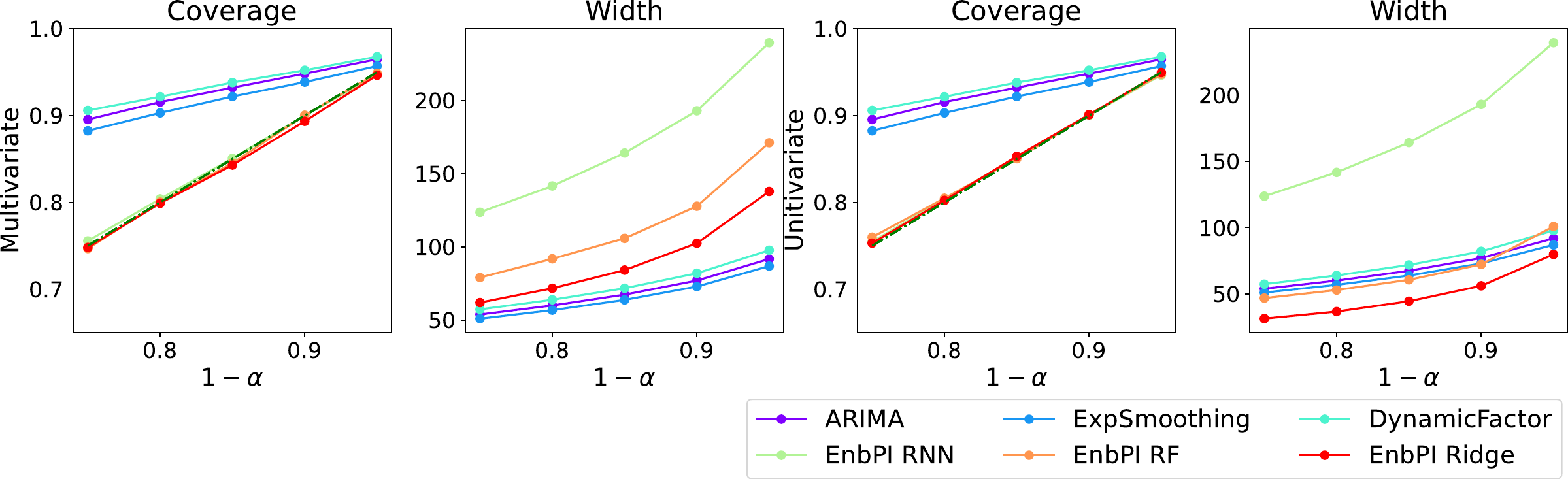}
        \subcaption{Beijing Air}
    \end{minipage}
    \cprotect \caption{Prediction on three other time series Average coverage and width versus $1-\alpha$ target coverage by \Verb|EnbPI| under different prediction algorithms and by ARIMA, Exponential Smoothing, and Dynamic Factor models. Five equally spaced $1-\alpha \in [0.75,0.95]$ are chosen. The green dash-dotted line at 0.9 represents the target coverage.}
    \label{fig:real_others_alpha}
\end{figure}

\begin{itemize}[itemsep=0em]
    \item Figure \ref{fig:real_others_alpha} (a) (Greenhouse): Except the Dynamic Factor model, all methods tend to lose coverage; however, \EnbPI \ under RNN tends to reach better coverage than other methods with much shorter interval widths. Therefore, we still favor \EnbPI, although one needs to be more selective with the regression model $\mathcal{A}$.
    \item Figure \ref{fig:real_others_alpha} (b) (Appliances Energy): All time series methods no longer lose coverage, but \EnbPI \ under RNN yields shortest intervals without coverage losses when $X_t$ is multivariate. When $X_t$ is univariate, \EnbPI \ almost always maintains coverage and yields much shorter intervals than time series methods.
    \item Figure \ref{fig:real_others_alpha} (c) (Beijing Air): Time-series methods do not lose coverage. However, \EnbPI \ under RF or Ridge with univariate $X_t$ can produce shorter intervals with exact coverage guarantee.
\end{itemize}

\begin{figure}[htbp]
    \centering
    \begin{minipage}[t]{\linewidth}
        \includegraphics[width=0.49\textwidth]{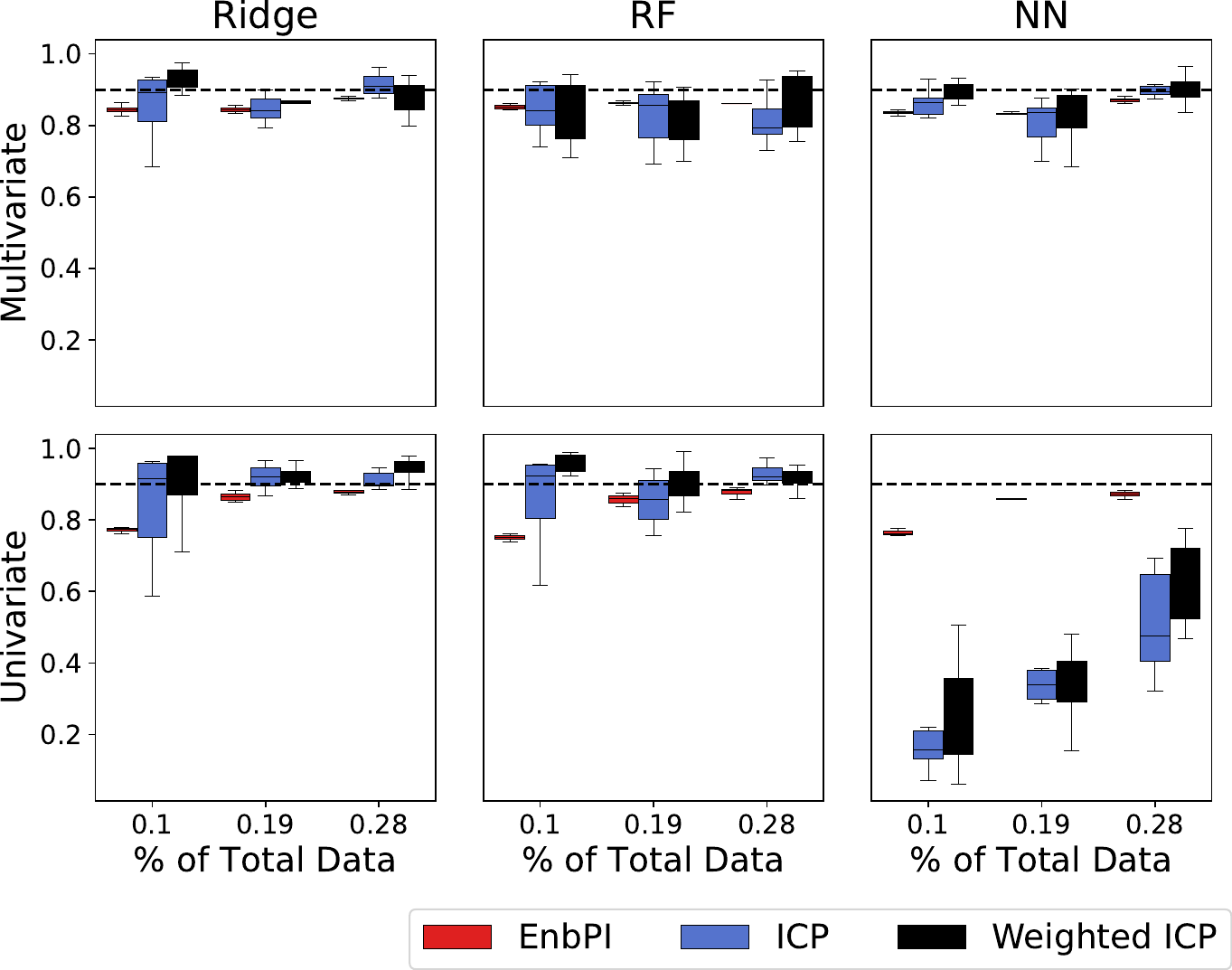}
        \includegraphics[width=0.49\textwidth]{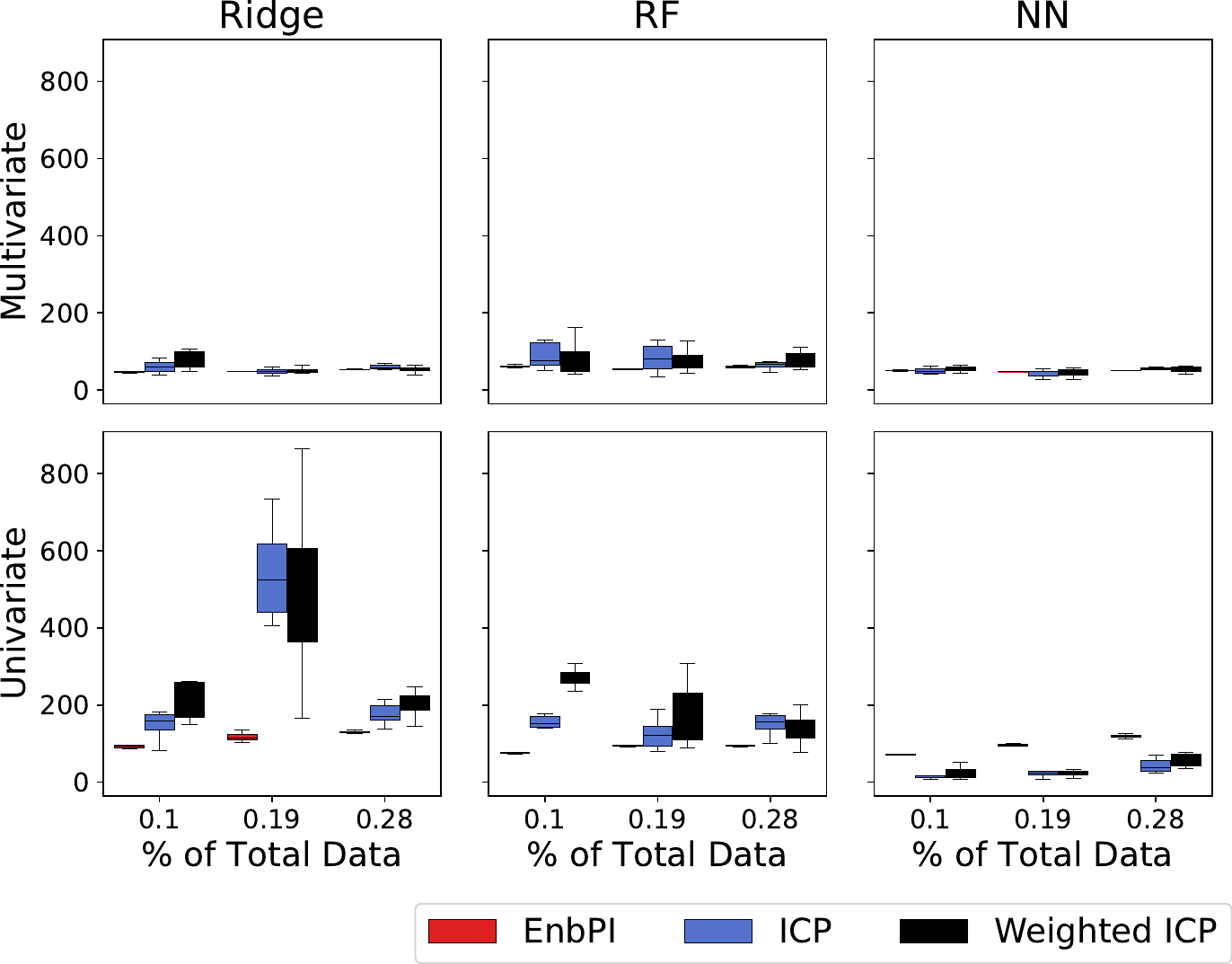}
        \subcaption{Greenhouse Gas}
    \end{minipage}
    \begin{minipage}[t]{\linewidth}
        \includegraphics[width=0.49\textwidth]{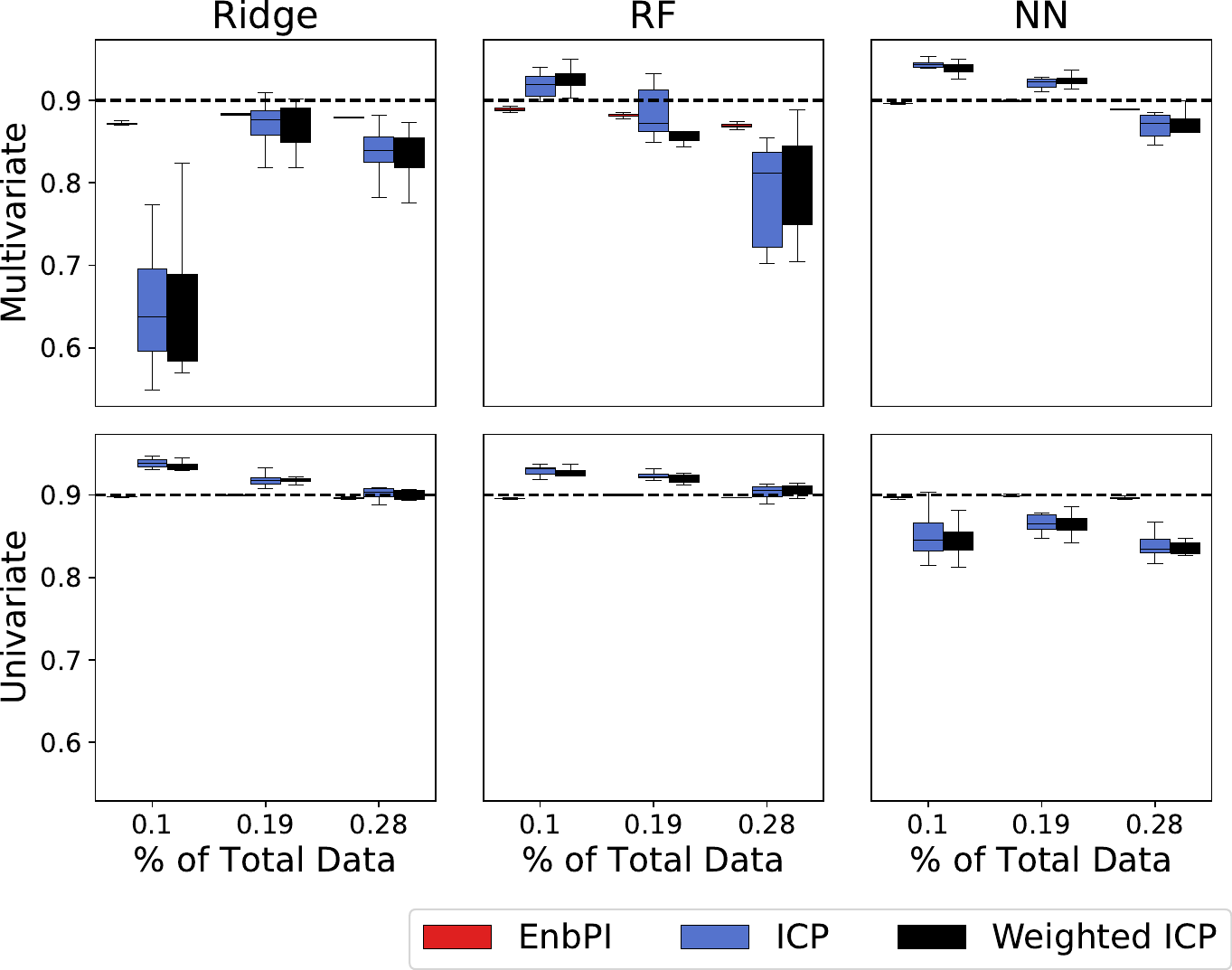}
        \includegraphics[width=0.49\textwidth]{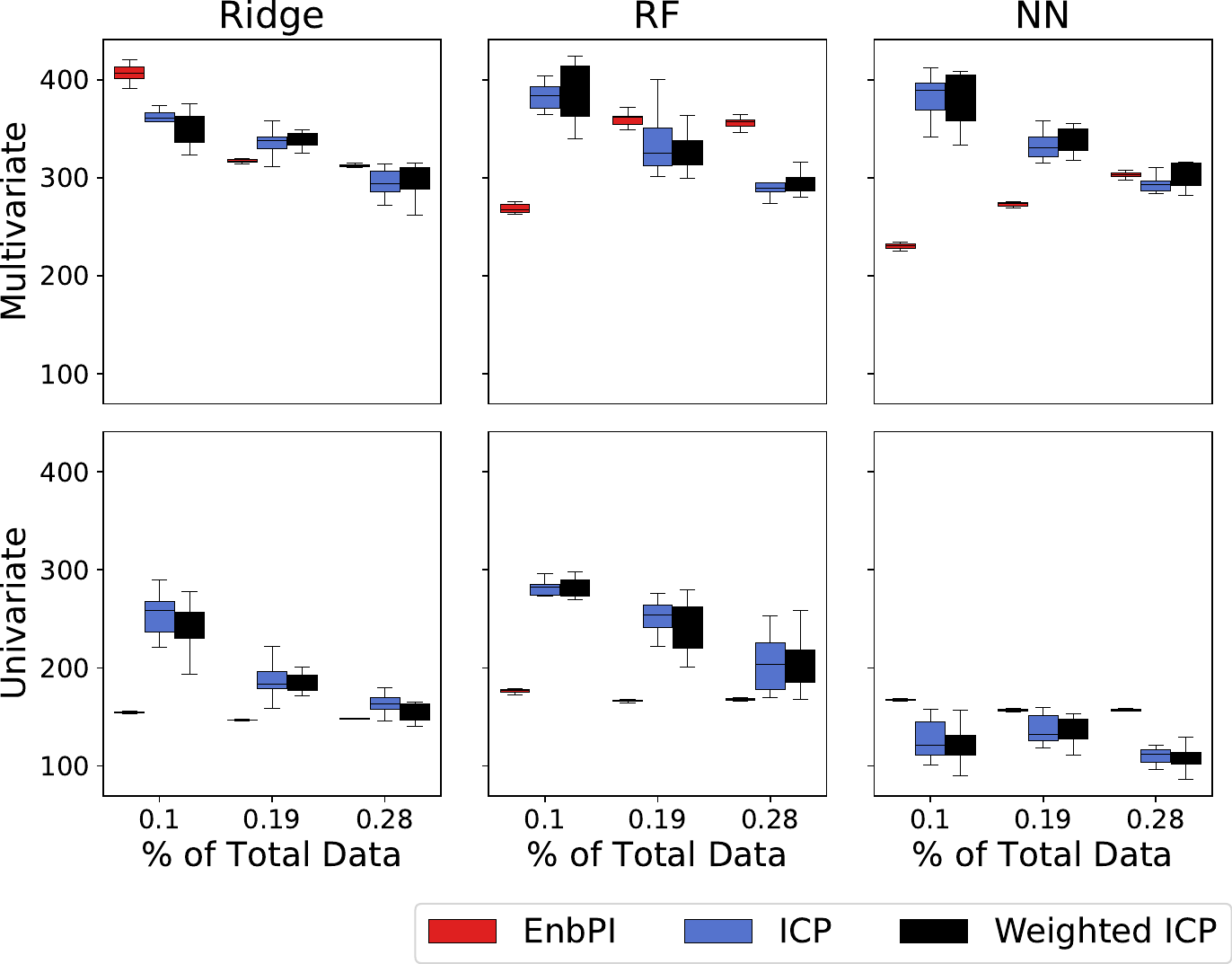}
        \subcaption{Appliances Energy}
    \end{minipage}
    \begin{minipage}[t]{\linewidth}
        \includegraphics[width=0.49\textwidth]{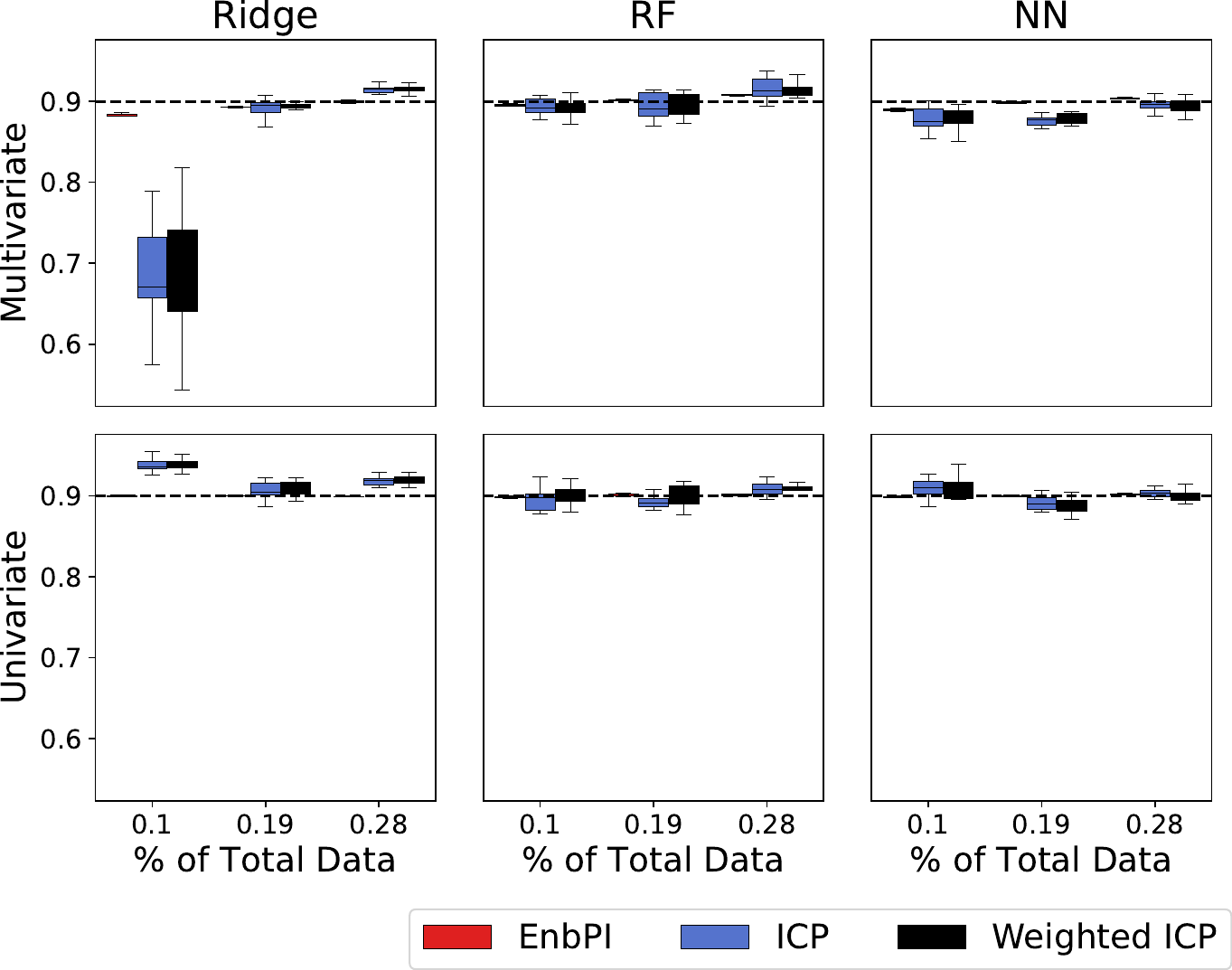}
        \includegraphics[width=0.49\textwidth]{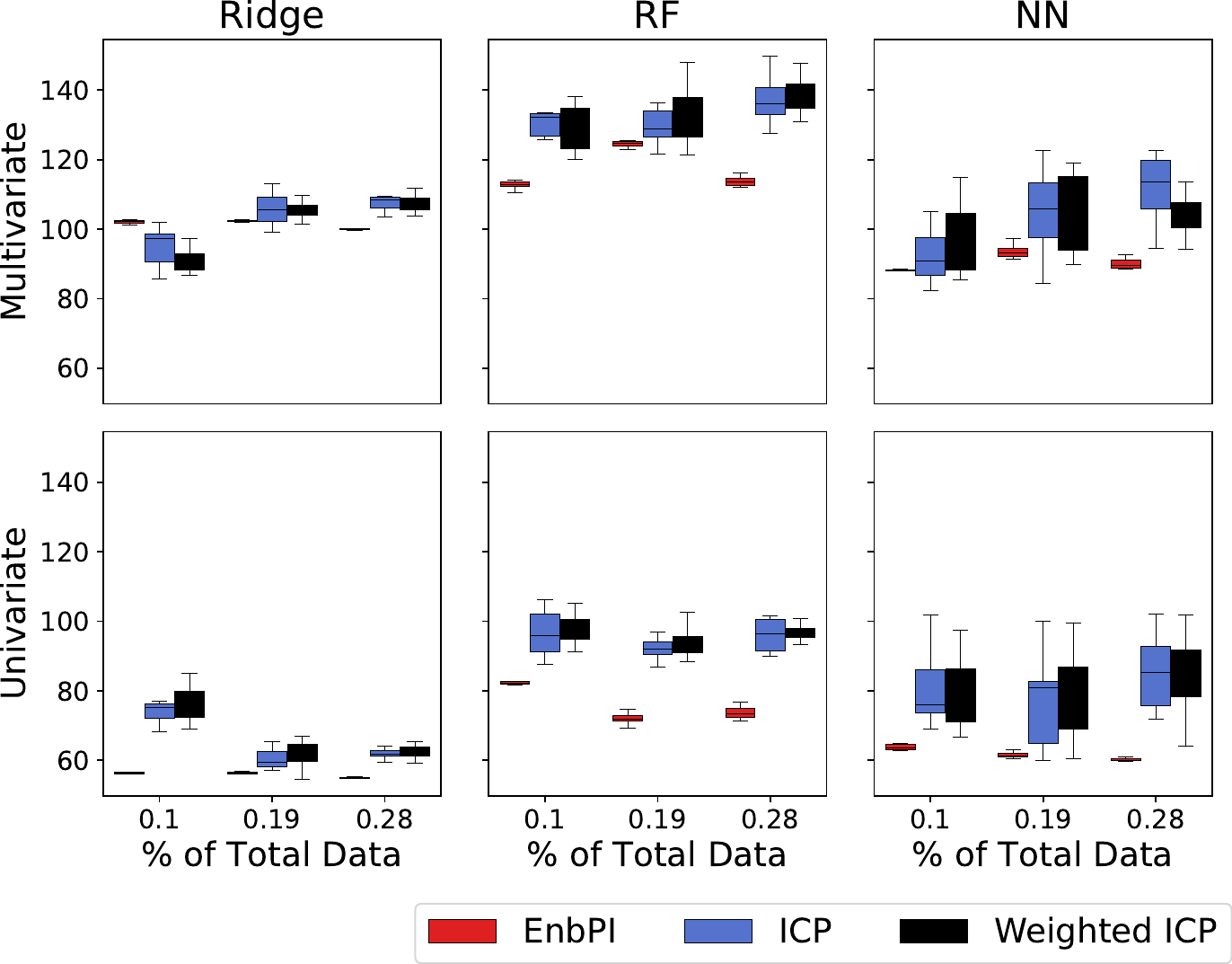}
        \subcaption{Beijing Air}
    \end{minipage}
    \cprotect \caption{Prediction on three other time series. Each box contains results from 10 independent trials. The black dash-dotted line at 0.9 indicates target coverage.}
    \label{fig:real_box_width}
\end{figure}

\noindent\textit{(2) Observations from the other grouped boxplots}
\begin{itemize}[itemsep=0em]
    \item Figure \ref{fig:real_box_width} (a) (Greenhouse): \Verb|EnbPI| almost never loses coverage, whereas ICP and WeightedICP can significantly under-cover (for example, see NN on univariate $X_t$). Moreover, \Verb|EnbPI| coverage and widths have much less variance than the other ones. 
    \item Figure \ref{fig:real_box_width} (b) (Appliances Energy) reveals similar patterns. In particular, ICP and WeightedICP can significantly lose coverage significantly (for example, see ridge on multivariate $X_t$). Moreover, ICP and WeightedICP also have higher widths with much larger variances than \verb|EnbPI| (for example, see RNN on multivariate $X_t$). Overall, we notice that intervals on univariate versions are much shorter than those on multivariate versions, likely because the past history of energy use is more predictive of future energy use than the exogenous variables such as humidity and temperature of the surrounding (for example, kitchen, bathroom, living room, etc.).
    \item Figure \ref{fig:real_box_width} (c) (Beijing Air) shows similar performances by all three CP methods, although ICP and Weighted ICP may greatly lose coverage (for example, see ridge on multivariate $X_t$).
\end{itemize}
In general, we think \Verb|EnbPI| is stable across different combinations of prediction algorithms and datasets. Since other CP methods such as ICP and WeightedICP can severely lose coverage, we advocate the use of \Verb|EnbPI| for time series predictive inference. Regarding interval width, using the history of the response (univariate version) to predict its future values tends to yield shorter intervals.

\subsection{Details on unsupervised anomaly detection} \label{append:anomaly_detection}

\vspace{0.1in}
\noindent {\it A. Data Description.} The raw bi-hourly traffic flow data are from the California Department of Transportation. The same source of data was also used in \citep{xu2021conformal}, but sensors are chosen differently, and some competing methods are also different. Thus, we follow how \citep{xu2021conformal} define anomalies with minor differences; the definitions are emphasized here for self-contained exposition. Suppose $K$ sensors are in the network and $T$ observations are available at each sensor. Denote $Y_{tk}$ as the traffic flow observation of sensor $k$ at time $t$. The symbol $\boldsymbol Y=[Y_{tk}]$ denotes the whole data matrix with no missing entries. 
Then for each $k \in \{1,\ldots,10\}$ and $t\in \{1,\ldots,17520\}$ (48$\times$365 bi-hourly data in 2020), define $Y_{tk}$ as an anomaly if 
\begin{equation*}
    Y_{tk} \geq q_{1-\alpha}(Y^{-d}_{t,N_k}) \text{ or } Y_{tk} \leq q_{\alpha}(Y^{-d}_{t,N_k}),
\end{equation*}
where $q_{\alpha}(\cdot)$ is the $\alpha$ percentile of its input vector, $N_k$ contains sensors closest to $k$ (including itself), and $Y^{-d}_{t,N_k}$ contains past $d$ hourly flows from sensors in $N_k$. Let $\alpha=0.01$, $d=|N_k|=5$, all of which are unknown.

We remark that the training data in $\boldsymbol Y$ is given to us with 30$\%$ missing entries in each sensor column. Thus, we use the \Verb|IterativeImputer| from the Python \Verb|sklearn| package to denote the imputed matrix. Denote $\boldsymbol{\hat{Y}}$ as the imputed data matrix. To apply \EnbPI, we first define $X_{tk}:=\hat{Y}^{-m}_{t,\hat{N}_k} \in \mathbb R^{m|\hat{N}_k|}$ as the past $m$ hourly flows from $\hat{N}_k$ closest sensors to $k$. Let $m=|\hat{N}_k|=8$. Then, the test data $Y_{tk}$ is predicted as an anomaly if its residual is either higher than $1-\alpha+\betaReal$ or lower than $\betaReal$ quantile of residuals on $Y^{-m}_{t,\hat{N}_k}$.

\vspace{0.1in}
\noindent {\it B. Setup and comparison methods.} Our goal is to identify binary anomalies at each sensor, defined as traffic flow observations with extremely large or small magnitude compared to those from its neighbor and/or in the past. It is natural to use our conformal prediction method as the whole data constitute a traffic network, and \EnbPI \ can easily capture the spatio-temporal information. We compute the standard precision, recall, and $F_1$ score at each node; methods with higher $F_1$ scores are preferable for performance metrics. We build 15 pre-trained bootstrap models in \EnbPI, fix the significance level $\alpha$ at 0.05, and use the mean aggregation function to build LOO ensemble predictors. The four regression algorithms are used in Section \ref{expr:first_part} and \ref{expr:second_part}. We compare \EnbPI \ against eight competing anomaly detectors, four of which are unsupervised (for example, IForest, PCA, OCSVM, and HBOS), and the other four are supervised (for example, MLPClassifier, GBoosting, KNN, SVC).
\begin{itemize}
    \item \textit{Four unsupervised methods.} All the unsupervised methods are implemented in the \verb|pyod| library in Python. We consider IForest, PCA,OCSVM, and HBOS and descriptions below mostly come from the package description with minor changes:
\begin{itemize}[noitemsep]
    \item The IsolationForest (IForest) ``isolates'' observations $x_t$ by randomly selecting a feature of $x_t$ and then randomly selecting a split value between the maximum and minimum values of the selected feature. See \citep{IForest}.
    \item In the Principal Component Analysis (PCA) for anomaly detection, covariance matrix of the data is first decomposed to orthogonal vectors, which are eigenvectors. Then, outlier scores are obtained as the sum of the projected distance of a sample on all eigenvectors. See \citep{PCA}.
    \item The one-class support vector machine (OCSVM) is a wrapper of scikit-learn one-class SVM Class with more functionalities. See \url{https://scikit-learn.org/stable/modules/svm.html#svm-outlier-detection} for detailed descriptions.
    \item The Histogram-based Outlier Detection (HBOS) assumes feature independence in $x_t$ and calculates the degree of outlyingness by building histograms. See \citep{HBOS}.
\end{itemize} 
\item \textit{Four supervised methods.} All the supervised methods are taken as binary classification methods from the \verb|sklearn| package in Python. We take descriptions of methods from the package and specify the following parameters for each method
\begin{itemize}[noitemsep]
    \item The Gradient Boosting Classifier (GBoosting) builds an additive model in a forward stage-wise fashion; it allows for the optimization of arbitrary differentiable loss functions. We build 100 estimators, pick a learning rate of 1, and let maximum depth be 1.
    \item The Multi-layer Perceptron classifier (MLPClassifier) optimizes the log-loss function. We use LBFGS for optimization, let $l_2$ penalty $\alpha$ be 1e-5, and pick two hidden layers with 5 neurons in the first and 2 in the second.
    \item The $k$-nearest neighbor (KNN) algorithm is specified with $k=20$ and weights=``distance'', so that closer neighbors of a query point will have a greater influence than neighbors which are further away.
    \item The support vector classification (SVC) uses all the default settings except with gamma=``auto'', which uses 1 / $\#$ features as the kernel coefficient.
\end{itemize}
\end{itemize}

\end{document}